\documentclass[12pt,onecolumn]{IEEEtran}
\usepackage[cmex10]{amsmath}
\usepackage{amsfonts}
\usepackage[pdftex]{color,graphicx}
\usepackage{subfig}
\usepackage{rotating}

\newtheorem{thm}{Theorem}
\newtheorem{prop}[thm]{Proposition}

\newcommand{\defineqq}{\ensuremath{\stackrel{\text{def}}{=}}}
\newcommand{\wtilde}{\widetilde}
\newcommand{\what}{\widehat}
\newcommand{\PVTCOOP}{\text{PVT-COOP}}

\begin{document}

\title{Interference Channels with Source Cooperation}
\author{Vinod M. Prabhakaran and Pramod Viswanath\thanks{The authors are with
the Coordinated Science Laboratory, 
University of Illinois at Urbana-Champaign,
Urbana, IL 61801. Email: \{vinodmp,pramodv\}@illinois.edu}}

\maketitle

\begin{abstract}
The role of cooperation in managing interference -- a fundamental feature
of the wireless channel -- is investigated by studying the two-user
Gaussian interference channel  where the source nodes can {\em both}
transmit and receive in full-duplex. The sum-capacity of this channel is
obtained within a gap of a constant number of bits. The coding scheme used
builds up on the superposition scheme of Han and Kobayashi for the two-user
interference channel without cooperation. New upperbounds on the
sum-capacity are also derived. The same coding scheme is shown to obtain
the sum-capacity of the symmetric two-user Gaussian interference channel
with noiseless feedback within a constant gap.
\end{abstract}


\section{Introduction}

The standard engineering approach to dealing with interference in wireless
systems is to orthogonalize the transmissions and/or treat interference as
noise at the receivers. However, these strategies can be far from optimal
in several scenarios, including the classical two user Gaussian
interference channel~\cite{Carleial78,EtkinTseWang08}. Superposition coding
and interference alignment have been shown to perform well in interference
channels (where the sources only transmit and destinations only receive)
\cite{EtkinTseWang08,CadambeJafar08}.

A potentially more effective approach to interference management is
available when the wireless nodes can cooperate among themselves (this
situation is not feasible in the classical interference channel when
sources only transmit and destinations only receive). A coarse result
(scaling of symmetric capacity as the number of radios grows) derived in
\cite{OLT07} shows that distributed cooperation (a so-called {\em
hierarchical MIMO} strategy) can manage interference between the different
traffic flows so well as to get near the performance of (co-located) MIMO
operation among the nodes. While this is a strong result, it is coarse --
in the asymptotic regime of a very large number of radios.

Our goal in this paper is to better understand the role of cooperation in
providing interference management gains. We study a simple wireless
network: we start with the classical two-user Gaussian interference
channel, but endow the sources with the capability to receive as well as
transmit. This sets up the possibility of cooperation between the sources.
The cooperative links between the sources are over the same frequency band
as the rest of the links. In this paper, we treat only the sum-rate under a
full-duplex mode of operation. The main result is a characterization of the
sum-rate within 20 bits. In specific instances our characterization is
readily sharpened -- we provide an example where the gap reduces to 6 bits,
for instance. We also provide recipes to improve the lower and upper
bounds. The two-user Gaussian interference channel with destination
cooperation where the destinations have receive and transmit capabilities
is investigated in a companion paper~\cite{DestCoop} where the sum-capacity
is obtained within a constant gap. In section~\ref{sec:reversibility}, we
present a reversibility property between the two results.  As we discuss
there, one setting can be viewed as being obtained from the other by (a)
reversing the roles of sources and destinations and (b) changing the
directions of the links while preserving the channel coefficients. We show
that the sum-capacities of the two settings connected by this
transformation are within a constant gap.

One approach to using the cooperative links is to employ
orthogonalization to emulate essentially noise-free bit-pipes between the
sources which they can use to conference. The capacity region of the
two-user Gaussian interference channels with sources conferencing over
orthogonal links has been characterized within a constant gap
in~\cite{WangConference}.  However, as we will see, the
orthogonalization approach does not lead to an efficient
solution for our problem in general.

Han and Kobayashi~\cite{HanKobayashi} proposed a coding scheme for the
two-user interference channel based on superposition coding of
Cover~\cite{CoverBroadcast}. It involves the two destinations partially
decoding the interference they receive. In order to facilitate this, the
sources encode their messages as a superposition of two partial messages.
One of these partial messages, termed the {\sf public} message, is decoded
by the destination where it appears as interference along with the two
partial messages which are meant for this destination. The other partial
message, called the {\sf private} message, from the interfering source is
treated as noise. Our coding scheme use two additional types of messages:
\begin{itemize}
\item  a {\sf
cooperative-public} message which is decoded not only by both the
destinations, but also by the other source which aids its own destination
in decoding it, and
\item  a {\sf cooperative-private} message which is
decoded by the destination to which it is intended and by the other source
which cooperates with the original source in its transmission.
\end{itemize}
These two messages have similarities to the suggestions of
\cite{Tuninetti07,CaoChen07,YangTuninetti08} -- works which proposed
achievable strategies for the two-user interference channel with generalized
feedback. But they differ in the details of implementation from our scheme
in ways which have a bearing on the rates achieved in the Gaussian setting.

Closely related to the problem of Gaussian interference channel with source
cooperation is that of Gaussian interference channel with feedback. As we
discuss in section~\ref{sec:feedback}, our coding strategy can be directly
employed to obtain the sum-capacity of the two-user symmetric Gaussian
interference channel with noiseless output feedback within a constant gap
of 19 bits. \cite{SuhTse09} independently investigated this problem and
obtained a better bound.

Other related works include~\cite{HostMadsen06} which studied the same
model, but did not provide a constant-gap result,  and
\cite{Mohajer08,Thejaswi08} which studied a two-stage, two-source
interference network. Also related are the works on the so-called cognitive
interference channel~\cite{MaricPartial,WuCognitive,
JovicicCognitive,RiniCognitive} and interference channel with
unidirectional cooperation~\cite{BagheriUnidirectional}.

\section{Problem Statement}
\label{sec:model}

\begin{figure*}[!t]
\centerline{{\scalebox{0.5}{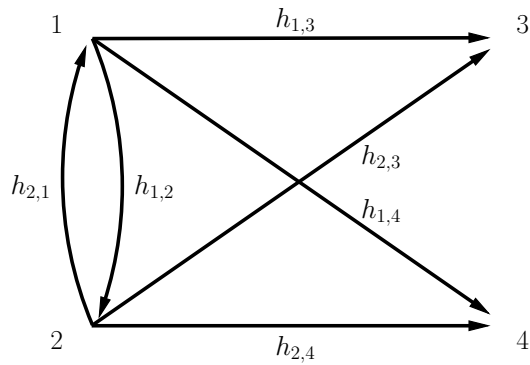}%
}
}
\caption{Problem Setup}
\label{fig:setup}
\end{figure*}

We consider the following channel model for source cooperation (see
Figure~\ref{fig:setup}). At each discrete-time instant -- indexed by
$t=1,2,\ldots$ -- the source nodes 1 and 2 send out, respectively, $X_1(t)$
and $X_2(t)\in{\mathbb C}$. The source nodes 1 and 2, and the destination
nodes 3 and 4 receive respectively
\begin{align*}
Y_1(t)&=h_{2,1}X_2(t)+N_1(t),\\
Y_2(t)&=h_{1,2}X_1(t)+N_2(t),\\
Y_3(t)&=h_{1,3}X_1(t)+h_{2,3}X_2(t)+N_3(t),\\
Y_4(t)&=h_{2,4}X_2(t)+h_{1,4}X_1(t)+N_4(t),
\end{align*}
where the channel coefficients $h$'s are complex numbers and
$N_k(t),\;k=1,2,3,4,\;t=1,2,\ldots$ are independent and identically
distributed (i.i.d.) zero-mean Gaussian random variables with unit
variance. It is easy to see that, without loss of generality, we may consider
a channel where the channel coefficients $h_{1,3},h_{1,2},h_{2,4},h_{2,1}$ are
replaced by their magnitudes $|h_{1,3}|, |h_{1,2}|, |h_{2,4}|, |h_{2,1}|$,
and the channel coefficient $h_{1,4}$ is replaced by
$|h_{1,4}|e^{j\theta/2}$ and $h_{2,3}$ is replaced by
$|h_{2,3}|e^{j\theta/2}$, where $\theta \defineqq \arg(h_{1,4}) +
\arg(h_{2,3}) - \arg(h_{1,3}) - \arg(h_{2,4})$. We will consider this
channel. We will also assume that $|h_{1,2}|=|h_{2,1}|=h_C$, say, which
models the reciprocity of the link between nodes 1 and 2. Further, we
consider unit power constraints which is without loss of generality when
both sources have the same power constraint.

There is a causality restriction on what the sources are
allowed to transmit: it can only depend on the message it sends and
everything it has heard up to the previous time instance, {\em i.e.},
\[X_k(t)=f_{k,t}(M_k,Y_k^{t-1}),\; k=1,2,\] where $M_k$ is the message to
be conveyed by source~$k$ and $f$ is a (deterministic) encoding function.
A blocklength-$T$ codebook of rate $(R_1,R_2)$ is (for each $k=1,2$) a
sequence of encoding functions, $f_{k,t},\;t=1,2,\ldots,T$ such that
\[{\mathbb E}\left[\frac{1}{T}\sum_{t=1}^T|X_k(t)|^2\right]\leq 1,\] with
message alphabets ${\mathcal M}_k=\{1,2,\ldots,2^{TR_k}\}$ over which the
messages $M_k$ are uniformly distributed, and decoding functions
$g_{k+2}:{\mathcal C}^{T}\rightarrow{\mathcal M}_k$. We say that a rate
$(R_1,R_2)$ is achievable if there is sequence of rate $(R_1,R_2)$
codebooks such that as $T\rightarrow \infty$,
\[{\mathbb P}\left(g_{k+2}(Y_{k+2}^T)\neq M_k\right)\rightarrow 0,\;k=1,2.\]

As in aid in describing our solution, we would also like to consider a linear
deterministic model (introduced in \cite{Avestimehr07}) for the above channel.
In order to treat both models together, we will adopt the following notation.
\begin{align*}
Y_1(t)&=h_{2,1}^\ast(X_2(t)),\\
Y_2(t)&=h_{1,2}^\ast(X_1(t)),\\
Y_3(t)&=h_{1,3}(X_1(t))+h_{2,3}^\ast(X_2(t)),\\
Y_4(t)&=h_{2,4}(X_2(t))+h_{1,4}^\ast(X_1(t)).
\end{align*}
Here the functions with a $\ast$ are potentially random functions and the
others are deterministic functions.

{\em Gaussian case:} In the Gaussian case, we specialize to (with some
abuse of notation\footnote{The correct notation would be
$Y_1(t)=h_{2,1}^{\ast(t)}(X_2(t))=h_{2,1}X_2(t)+N_1(t)$. This $t$-index in
the notation for random functions like $h_{2,1}^{\ast(t)}$ will be
suppressed.  We will tacitly assume that application of $\ast$-ed functions
for different values of $t$ result in independent realizations of $N$'s.}):
\begin{align*}
h_{2,1}^\ast(X_2)&=h_{2,1}X_2+N_1,\\
h_{1,2}^\ast(X_1)&=h_{1,2}X_1+N_2,\\
h_{1,3}(X_1)&=h_{1,3}X_1,\\
h_{2,4}(X_2)&=h_{2,4}X_2,\\
h_{2,3}^\ast(X_2)&=h_{2,3}X_2+N_3,\\
h_{1,4}^\ast(X_1)&=h_{1,4}X_1+N_4.
\end{align*}
We will further assume that $h_{1,2}=h_{2,1}$ which is justified by the
reciprocity of the links between the sources.

{\em Linear deterministic case:}
Let $n_{1,2},n_{1,3}$, $n_{1,4},n_{2,1}$, $n_{2,3},n_{2,4}$ be non-negative
integers and $n\defineqq\max(n_{1,2},n_{1,3}, n_{1,4},n_{2,1},
n_{2,3},n_{2,4})$. The inputs to the channel $X_1$ and $X_2$ are $n$-length
vectors over a finite field ${\mathbb F}$. We define
\begin{align*}
h_{2,1}^\ast(X_2)&={\bf S}^{n-n_{2,1}}X_2,\\
h_{1,2}^\ast(X_1)&={\bf S}^{n-n_{1,2}}X_1,\\
h_{1,3}(X_1)&={\bf S}^{n-n_{1,3}}X_1,\\
h_{2,4}(X_2)&={\bf S}^{n-n_{2,4}}X_2,\\
h_{2,3}^\ast(X_2)&={\bf S}^{n-n_{2,3}}X_2,\\
h_{1,4}^\ast(X_1)&={\bf S}^{n-n_{1,4}}X_1,
\end{align*}
where
\[{\bf S}=\left(\begin{array}{ccccc}
 0&0&0&\ldots&0\\
 1&0&0&\ldots&0\\
 0&1&0&\ldots&0\\
 \vdots&\ddots&\ddots&\ddots&\\
 0&\ldots&0&1&0
 \end{array}\right)_{n\times n}\]
is the $n\times n$ shift matrix.
Further, to model the reciprocity of the links between the sources, we set
$n_{1,2}=n_{2,1}=n_C$, say.

\section{Main Results}

\subsection{Sum-rate Characterization}\label{sec:results}

The following theorems characterize the sum-rates of the interference channels
with source cooperation introduced in the previous section.

\begin{thm}{\em Linear deterministic case.}\label{thm:sourcecoopLD}
The sum-capacity of the linear deterministic channel with source
cooperation is the minimum of the following:
\begin{align}
u_1&=\max(n_{1,3}-n_{1,4}+n_C,n_{2,3},n_C) +
\max(n_{2,4}-n_{2,3}+n_C,n_{1,4},n_C),\label{eq:LDu1}\\
u_2&=
\max(n_{1,3},n_{2,3}) + \left(\max(n_{2,4},n_{2,3},n_C)-n_{2,3}\right),
\label{eq:LDu2}\\
u_3&=
\max(n_{2,4},n_{1,4}) + \left(\max(n_{1,3},n_{1,4},n_C)-n_{1,4}\right),
\label{eq:LDu3}\\
u_4&=\max(n_{1,3},n_C)+\max(n_{2,4},n_C),\text{ and}\label{eq:LDu4}\\
u_5&=\left\{\begin{array}{ll}
\max(n_{1,3}+n_{2,4},n_{1,4}+n_{2,3}),
 &\text{ if }n_{1,3}-n_{2,3}\neq n_{1,4}-n_{2,4},\\
\max(n_{1,3},n_{2,4},n_{1,4},n_{2,3}),
 &\text{ otherwise}.
\end{array}\right.\label{eq:LDu5}
\end{align}
\end{thm}
The achievability of the above theorem in proved in
Appendix~\ref{app:LDachieve}, and the upperbounds are derived in
Appendix~\ref{app:sourcecoopupperbounds}.

\begin{thm}{\em Gaussian case.}\label{thm:sourcecoopG}
The sum-capacity of the Gaussian channel with source cooperation is at most
the minimum of the following five quantities and a sum-rate can be achieved
to within a constant (20 bits) of this minimum.
\begin{align}
u_1&=
\log\left(1+\left(\frac{\left|h_{1,3}\right|}{\max\left(1,\left|h_{1,4}\right|\right)}+\frac{\left|h_{2,3}\right|}{\max\left(1,\left|h_C\right|\right)}\right)^2\right)(1+|h_C|^2) \\ &\quad +
\log\left(1+\left(\frac{\left|h_{2,4}\right|}{\max\left(1,\left|h_{2,3}\right|\right)}+\frac{\left|h_{1,4}\right|}{\max\left(1,\left|h_C\right|\right)}\right)^2\right)(1+|h_C|^2),\label{eq:Gaussu1}\\ 
u_2&=
\log2\left(1+\left(|h_{1,3}|+|h_{2,3}|\right)^2\right)
\left(1+\frac{\max\left(|h_{2,4}|^2,|h_{2,3}|^2,|h_C|^2\right)}{\max\left(1,|h_{2,3}|^2\right)}\right),
\label{eq:Gaussu2}\\ 
u_3&=
\log2\left(1+\left(|h_{2,4}|+|h_{1,4}|\right)^2\right)
\left(1+\frac{\max\left(|h_{1,3}|^2,|h_{1,4}|^2,|h_C|^2\right)}{\max\left(1,|h_{1,4}|^2\right)}\right),
\label{eq:Gaussu3}\\ 
u_4&=
\log\left(1+|h_{1,3}|^2+|h_C|^2\right) +
\log\left(1+|h_{2,4}|^2+|h_C|^2\right),\text{ and}\label{eq:Gaussu4}\\ 
u_5&=
\log\Bigg( 1 +
2 \left(|h_{1,3}|^2+|h_{2,4}|^2+|h_{1,4}|^2+|h_{2,3}|^2\right) \notag\\
&\qquad\qquad+4 \left( |h_{1,3}h_{2,4}|^2 + |h_{1,4}h_{2,3}|^2 -
2|h_{1,3}h_{2,4}h_{1,4}h_{2,3}|\cos\theta\right) \Bigg). \label{eq:Gaussu5}
\end{align}
\end{thm}
The achievability proof is presented in Appendix~\ref{app:Gachieve} and the
converse in Appendix~\ref{app:sourcecoopupperbounds}.

\subsection{Gains from cooperation}

To illustrate the gains from cooperation, in this section we will explore a
specific instance of the symmetric interference channel:
$|h_{1,3}|=|h_{2,4}|=h_D$, $|h_{1,4}| = |h_{2,3}| = h_I = \sqrt{h_D}$,
and $\theta=0$. In appendix~\ref{app:Example} we prove the following
proposition.
\begin{prop}
Let $C$ be the minimum of the following
\begin{align*}
&2\log2h_D\left(1+h_C^2\right),\\
&\log2h_D^2\left(1+\frac{\max(h_D^2,h_C^2)}{h_D}\right),\\
&\log\left(4h_D^4\right).
\end{align*}
For the symmetric channel described above, for any $\epsilon>0$, the
sum-capacity lies in $(C-6-\epsilon,C+\epsilon)$, for sufficiently large
$h_D$.
\end{prop}
We plot in Figure~\ref{fig:plot}, as a function of $\log |h_C|^2/\log |h_D|^2$,
the sum-rate $C$ normalized by the capacity of the direct link, in the limit
of $|h_D|\rightarrow \infty$ while keeping the ratios
$\log|h_C|^2/\log|h_D|^2$ and $\log|h_I|^2/\log|h_D|^2$
constants. Since $C$ is achievable within a constant gap (6 bits), this
plot is also that of the sum-capacity in this limit. This reveals three
regimes of operation:
\begin{figure}
\centerline{\scalebox{0.5}{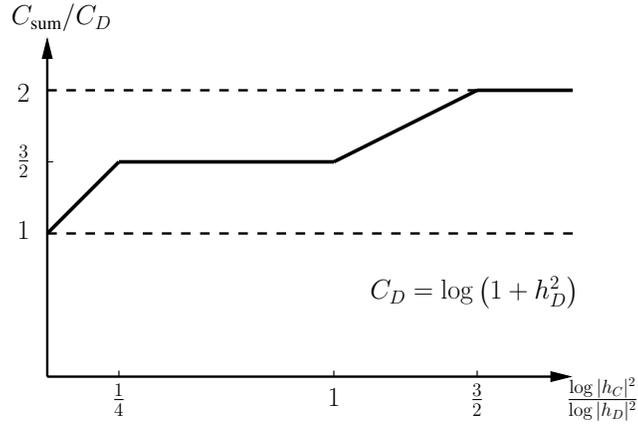}%
}
\caption{Normalized sum-capacity of the symmetric interference channel with
$h_I=\sqrt{h_D}$ under source cooperation in the limit of $h_D \rightarrow
\infty$ keeping $\log|h_I|^2/\log|h_D|^2$ and $\log|h_C|^2/\log|h_D|^2$
fixed.}
\label{fig:plot}
\end{figure}

\begin{itemize}

\item $\log|h_C|^2/\log|h_D|^2 \leq 1/2.$ In this regime, the plot shows
that the capacity increases linearly with the strength of the cooperation
link (measured in the dB scale). For every 3dB increase in link strength
the sum-capacity increases by 2 bits. Note that without cooperation, the
sum-capacity is essentially achieved by time-sharing. Thus, cooperation
can be seen to deliver significant gains.

\item $1/2 < \log|h_C|^2/\log|h_D|^2 \leq 1.$ The linear gain saturates
when the cooperation link strength is half the direct link strength. No
further gains are accrued until the cooperation link is as strong as the
direct link.

\item $1 < \log|h_C|^2/\log|h_D|^2 \leq 3/2.$ The capacity again increases
linearly with the cooperation link strength, but here an increase in
capacity by 2 bits requires a 6dB increase in the cooperation channel
strength. This linear increase continues until the cooperation capacity is
approached when the cooperation link is 3/2 times as strong as the direct
link, after which the capacity is flat.

\end{itemize}

\section{Coding Scheme: Illustrative Examples}

In this section, our cooperative coding scheme is illustrated through a few
simple examples. These examples have been hand-picked so that uncoded
(signal processing) strategies are enough to achieve the sum-capacity.
However, they cover the key novel aspects of our coding strategy.  In
particular, the first example shows how the sources cooperate in conveying
{\sf cooperative-public} messages, which are messages decoded by both
sources and which benefit from cooperation. Example~1 also involves the use
of {\sf private} messages which are decoded only by the destinations to
which they are intended and get conveyed without the benefits of
cooperation. Example~2 shows how {\sf cooperative-public} messages can
occur together with the two types of messages in Han-Kobayashi's scheme,
namely, {\sf private} messages and {\sf public} messages ({\sf public}
messages are decoded by both the destinations and do not benefit from
cooperation). Example~3 illustrates a second kind of cooperative message:
{\sf cooperative-private} message which benefits from cooperation and is
decoded only by the destination which is interested in it.

\begin{sidewaysfigure}
\centering
\subfloat[]{\scalebox{0.55}{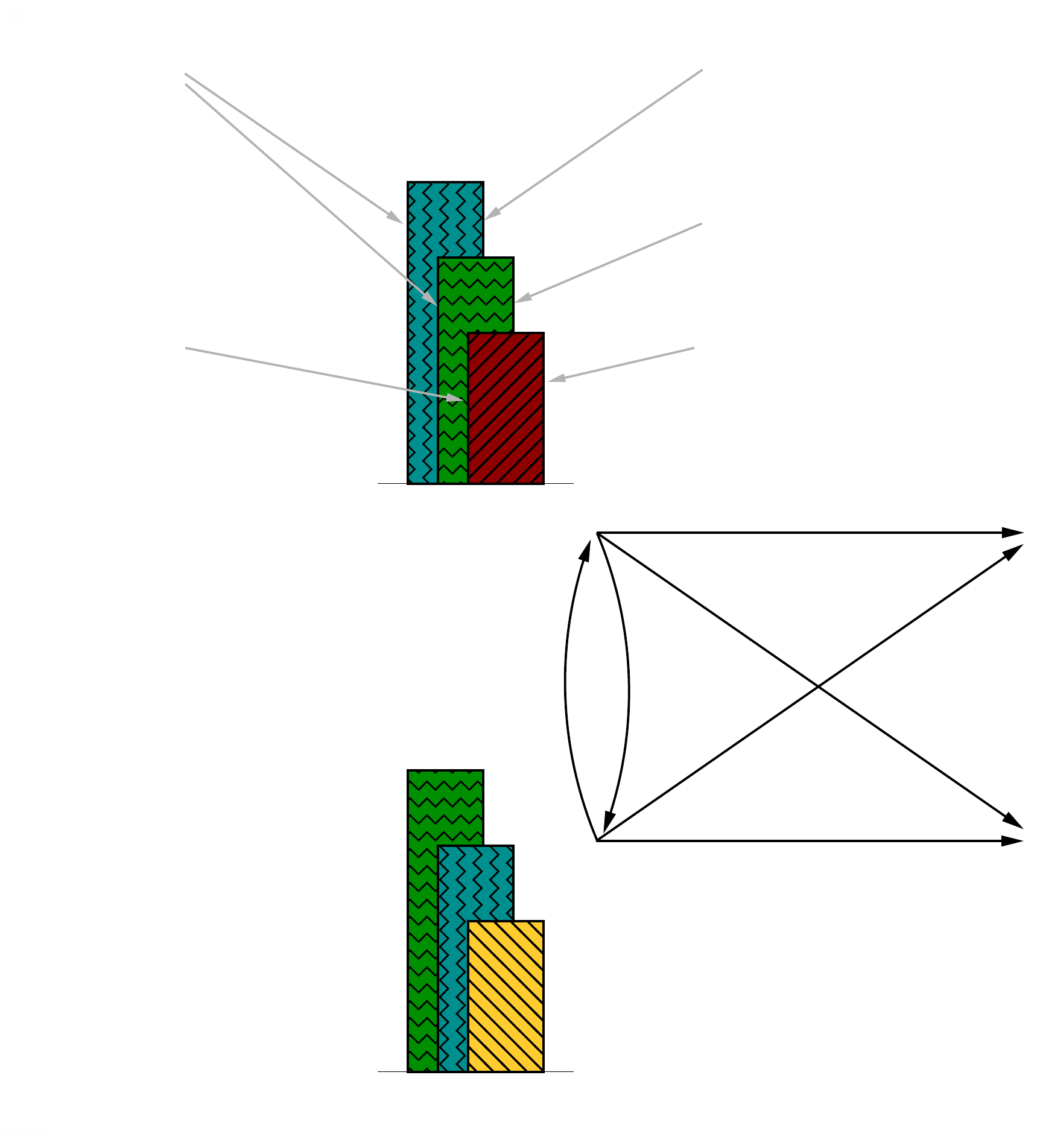}}%
\subfloat[]{\scalebox{0.55}{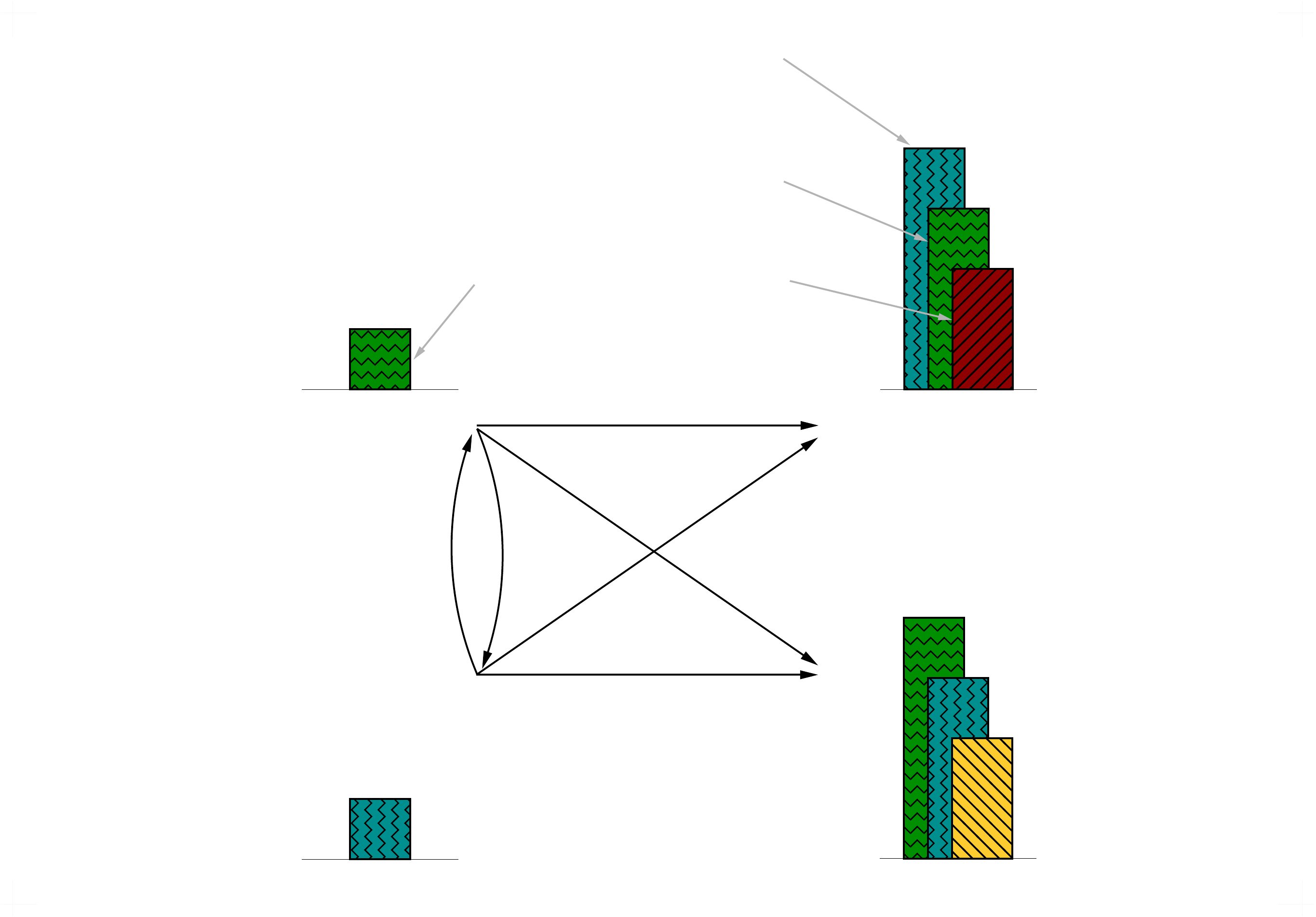}}
\caption{Example 1. (a) Transmitted signals. (b) Received signals. The
transmitted signals can be interpreted as a superposition of three
codewords which leads to a generalization as discussed at the end of the
section.}
\label{fig:example1}
\end{sidewaysfigure}

\noindent{\em Example 1:} For the linear deterministic case,
let us consider the symmetric channel with direct links
$n_{1,3}=n_{2,4}=n_D$, say, and interference links $n_{1,4}=n_{2,3}=n_I$,
say, such that $n_D=2n_I$. When source cooperation is absent, {\em i.e.},
$n_C=0$, the sum capacity turns out to be $n_D$ and it can be achieved
simply by time-sharing.  Now, let us consider $n_C=n_D/4$, and in
particular, $n_D=4, n_I=2, n_C=1$. See Figure~\ref{fig:example1}.
The sources transmit
\[ x_1(t)=\left(\begin{array}{c}
v_1(t)+v_1(t-1)\\v_2(t-1)\\z_{1a}(t)\\z_{1b}(t)\end{array}\right)\text{ and  }
x_2(t)=\left(\begin{array}{c}
v_2(t)+v_2(t-1)\\v_1(t-1)\\z_{2a}(t)\\z_{2b}(t)\end{array}\right).\]
Note that this is possible because
\[y_1(t)=\left(\begin{array}{c}
0\\0\\0\\v_2(t)+v_2(t-1)\end{array}\right)\text{ and  }
y_2(t)=\left(\begin{array}{c}
0\\0\\0\\v_1(t)+v_1(t-1)\end{array}\right),\]
which means that before time $t$, source~1 knows
$v_2(t-1)$ and source~2 knows $v_1(t-1)$. The transmissions are over a long
block of length $T$ with the signals at time $T$ such that $z_{ka}(T)=
z_{kb}(T)=0,\;k=1,2$. Also, we interpret $v_k(0)=0$. Then, destination~3
receives
\begin{align*}
y_3(t)&=\left(\begin{array}{c}
v_1(t)+v_1(t-1)\\v_2(t-1)\\z_{1a}(t)+v_2(t)+v_2(t-1)\\z_{1b}(t)+v_1(t-1)
\end{array}\right),\;t=1,2,\ldots,T-1,\text{ and}\\
y_3(T)&=\left(\begin{array}{c}
v_1(T)+v_1(T-1)\\v_2(T-1)\\v_2(T)+v_2(T-1)\\v_1(T-1)\end{array}\right).
\end{align*}
At the end of time $T$, destination~3 starts reading off the signals
backwards starting from what it received at time $T$. From $y_3(T)$ it can
recover $v_1(T), v_2(T), v_1(T-1), v_2(T-1)$. Making use of the latter two,
{\em i.e.}, $v_1(T-1), v_2(T-1)$, it can recover $z_{1a}(T-1), z_{1b}(T-1),
v_1(T-2), v_2(T-2)$ from $y_3(T-1)$. Then, employing its knowledge of
$v_1(T-2), v_2(T-2)$, it recovers  $z_{1a}(T-2), z_{1b}(T-2), v_1(T-3),
v_2(T-3)$ from $y_3(T-2)$, and so on.
Thus, destination~3 can
recover $\{(v_1(t),z_{1a}(t),z_{1b}(t),v_2(t)): t=1,2,\ldots,T\}$. By
symmetry, destination~4 can also recover its messages.
Thus, a rate of $R_1=3,\;R_2=3$ can be achieved (asymptotically as
$T\rightarrow\infty$). Thus we obtain a sum-rate of 6 which is in fact the
sum-capacity of this channel with cooperation.

The above scheme has two kinds of signals:
\begin{itemize}

\item {\sf Private} signals: $z_{1a},z_{1b},z_{2a},z_{2b}$ are recovered
only by the destination to which it is intended. Note that these signals
occupy the lower levels of the transmitted vector such that they do not
appear at the destination where they may act as interference.

\item {\sf Cooperative-public} signals: $v_1, v_2$. These signals are read
off by the other source at the end of each time $t$ and then incorporated
into the transmission by both the sources at the time $t+1$. Thus, the
transmission of these signals exploits the possibility of cooperation among
the two sources. Destinations perform ``backwards decoding.'' They recover
the cooperative-public signals sent cooperatively starting from the final
received vector and proceeding backwards. Hence, the initial transmission
used by the sources to convey the signals to each other, being already
available, does not act as interference at the destinations. In order to
facilitate the recovery of these signals at the sources, they occupy the
higher levels of transmitted vector.

\end{itemize}

\noindent{\em Example 2:} Let us consider the following asymmetric
linear deterministic case,
$n_{1,3}=6, n_{1,4}=3, n_{2,4}=4, n_{2,3}=3$, and $n_C=1$. The capacity as
given by Theorem~\ref{thm:sourcecoopLD} is 7. To achieve this,
the sources transmit
\[ x_1(t)=\left(\begin{array}{c}
v_1(t)+v_1(t-1)\\u_1(t)+v_2(t-1)\\u_1(t)\\z_{1a}(t)\\z_{1b}(t)\\z_{1c}(t)
\end{array}\right)\text{ and  }
x_2(t)=\left(\begin{array}{c}
v_2(t)+v_2(t-1)\\v_1(t-1)\\0\\z_{2}(t)\\0\\0\end{array}\right).\]
Note that this is again possible because
\[y_1(t)=\left(\begin{array}{c}
0\\0\\0\\v_2(t)+v_2(t-1)\end{array}\right)\text{ and  }
y_2(t)=\left(\begin{array}{c}
0\\0\\0\\v_1(t)+v_1(t-1)\end{array}\right).\]
and hence, source~1 knows
$v_2(t-1)$ and source~2 knows $v_1(t-1)$ before time $t$.
The transmissions are again over a long
block of length $T$ with the signals at time $T$ such that at time $T$ all
the $u,v,z$ signals are 0. Also, we interpret $v_k(0)=0, \; k=1,2$. Then,
destination~3 receives
\begin{align*}
y_3(t)&=\left(\begin{array}{c}
v_1(t)+v_1(t-1)\\u_1(t)+v_2(t-1)\\u_1(t)\\z_{1a}(t)+v_2(t)+v_2(t-1)\\z_{1b}(t)+v_1(t-1)\\z_{1c}(t)\end{array}\right),\;t=1,2,\ldots,T-1,
\text{ and }
y_3(T)=\left(\begin{array}{c}
v_1(T-1)\\v_2(T-1)\\0\\v_2(T-1)\\v_1(T-1)\\0\end{array}\right),
\end{align*}
and destination~4 receives
\begin{align*}
y_4(t)&=\left(\begin{array}{c}
0\\0\\v_2(t)+v_2(t-1)\\v_1(t)+2v_1(t-1)\\u_1(t)+v_2(t-1)\\u_1(t)+z_{2}(t)\end{array}\right),\;t=1,2,\ldots,T-1,
\text{ and }
y_3(T)=\left(\begin{array}{c}
0\\0\\v_2(T-1)\\2v_1(T-1)\\v_2(T-1)\\0\end{array}\right),
\end{align*}
Now it is easy to verify that if the destinations read off the signals
starting from the vectors they received at time $T$ and proceeding
backwards as in the previous examples, destination~3 can recover
$\{(v_1(t),u_1(t),z_{1a}(t),z_{1b}(t),z_{1c}(t),v_2(t)): t=1,2,\ldots,T\}$,
and destination~4
$\{(v_2(t),z_{2}(t),v_1(t),u_1(t)): t=1,2,\ldots,T\}$, and hence achieve a
sum-rate of 7.

This example involved a new type of signal apart from the {\sf private} and
{\sf cooperative-public} types of the previous example.
\begin{itemize}
\item {\sf Public} signal: $u_1$. This signal is decoded by both the
destinations. However, note that unlike the {\sf cooperative-public}
signal, the other source does not participate in its transmission. Indeed,
this signal is transmitted in such a way that it is not visible to the
other source.
\end{itemize}

Both the examples above involved cooperative links which are weaker than
the direct and interfering links. The mode of cooperation involved the
sources cooperating in aiding the destinations recover an interfering
signal. When the cooperative link is strong, yet another possible form of
cooperation becomes feasible. The next example illustrates this.

\begin{sidewaysfigure}
\centering
\subfloat[]{\scalebox{0.5}{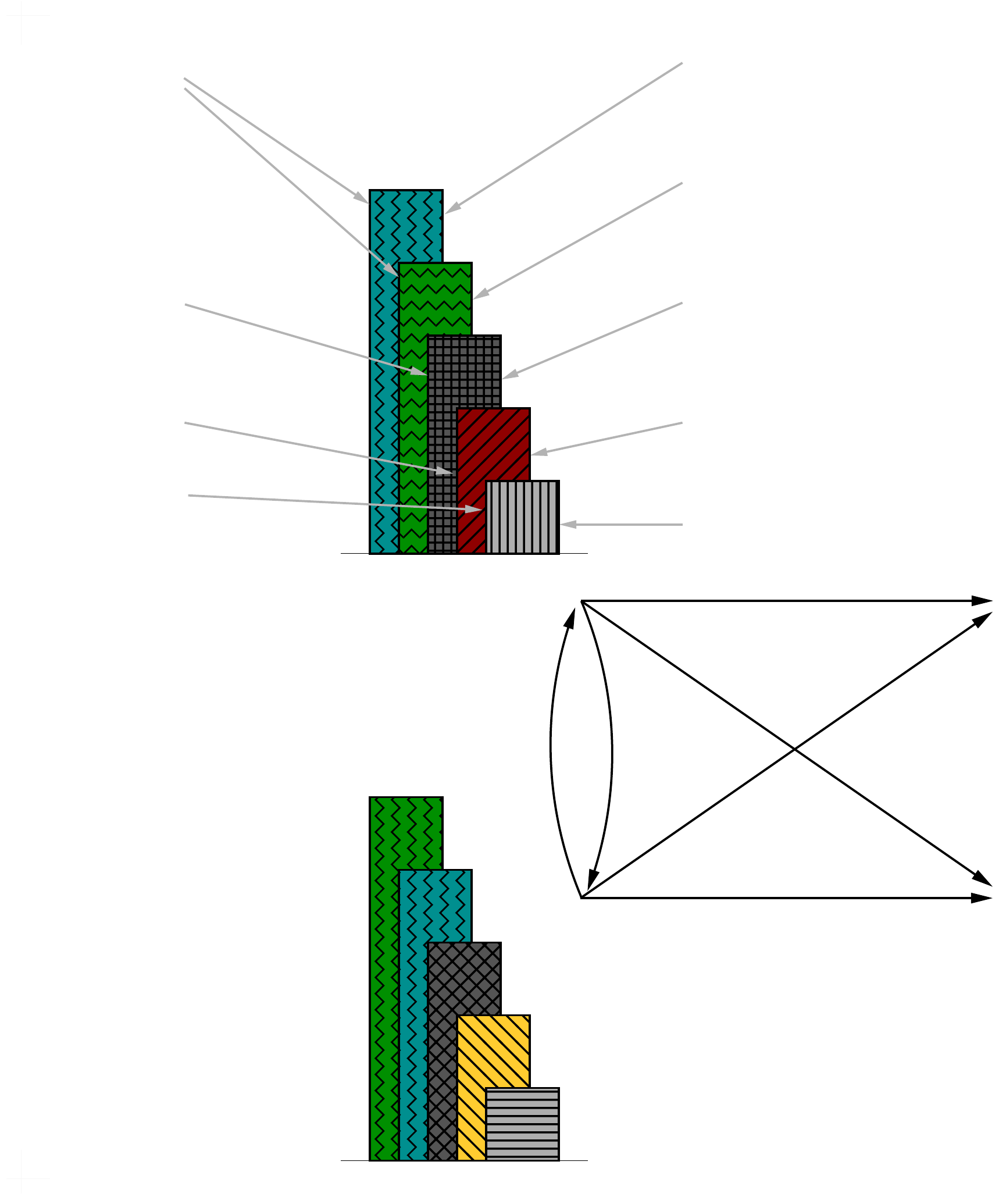}}%
\subfloat[]{\scalebox{0.5}{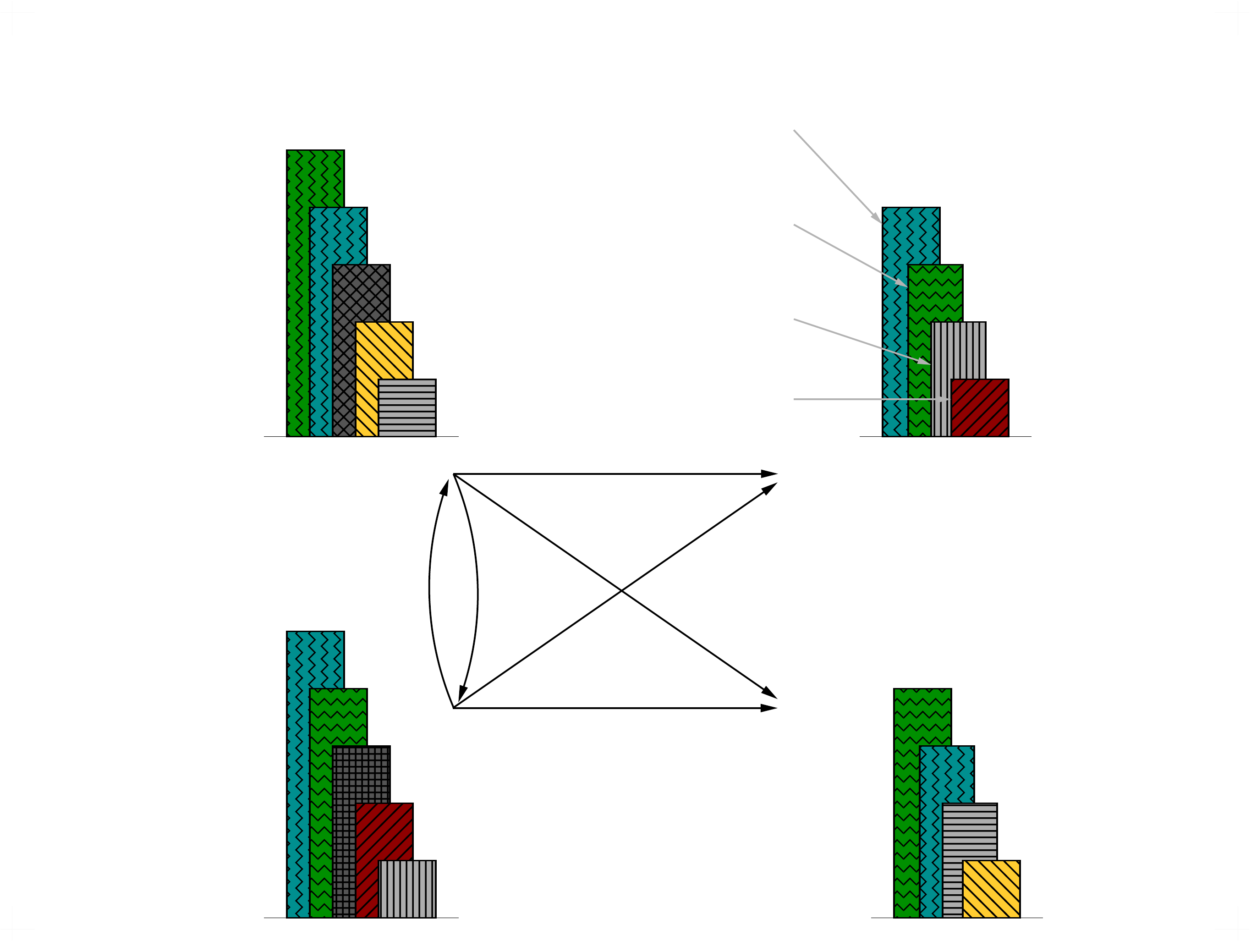}}%
\caption{Example 3. (a) Transmitted signals. (b) Received signals.}
\label{fig:example3}
\end{sidewaysfigure}

\noindent{\em Example 3:} Let us again consider the symmetric case, but
now with $n_D=4, n_I=3, n_C=5$. Note that, the cooperation link is now
stronger than both the direct and interference links. Without cooperation,
the sum capacity is 5; but with cooperation, we will show that a sum rate
of 6 can be achieved. See Figure~\ref{fig:example3}.
The sources transmit
\[
x_1(t)=\left(\begin{array}{c}
v_1(t)+v_1(t-1)\\v_2(t-1)\\s_{1}(t)\\z_{1}(t)-s_2(t)\\s_1(t+1)\end{array}\right)\text{ and  }
x_2(t)=\left(\begin{array}{c}
v_2(t)+v_2(t-1)\\v_1(t-1)\\s_{2}(t)\\z_{2}(t)-s_1(t)\\s_2(t+1)\end{array}\right).
\]
Note that this transmission scheme is possible since $y_1(t)=x_2(t),
\;y_2(t)=x_1(t)$. This allows the sources to exchange their $s(t)$ signals
one time step in advance over the lowest level. Also, we set $s_k(1)=0,
v_k(T)=0$, and interpret $v_k(0)=0, s_k(T+1)=0,\;k=1,2$.
Destinations now receive
\begin{align*}
y_3(t)&=\left(\begin{array}{c}
0\\v_1(t)+v_1(t-1)\\v_2(t)+2v_2(t-1)\\s_1(t)+v_1(t-1)\\z_{1}(t)
\end{array}\right),\;t=1,2,\ldots,T-1,\text{ and}\\
y_3(T)&=\left(\begin{array}{c}
0\\v_1(T-1)\\2v_2(T-1)\\s_1(T)+v_1(T-1)\\z_1(T)\end{array}\right).
\end{align*}
Recovery of signals proceeds backwards from the last received vector as in
Example~1. Assuming that the characteristic of the field ${\mathbb F}$ is
not 2, destination~3 can recover the signals
$\{(v_1(t),s_1(t),z_1(t),v_2(t)): t=1,2,\ldots,T\}$. This gives a rate of
$R_1=3$, and by symmetry a sum-rate of 6.

In addition to the previously encountered {\sf private} and {\sf
cooperative-public} types of signals, another kind of signal plays an
important role in this example. The strong cooperation link allows the
sources to share with each other signals which are eventually only
recovered by the destination it is intended for. {\em i.e.,}
\begin{itemize}

\item {\sf Cooperative-private} signals: $s_1,s_2$. The sources learn these
signals from each other one time step ahead, and in the next time step they
cooperate with each other to convey these signals only to the destination
it is intended for. Note that the collaboration between the sources in this
example is a rudimentary form of precoding. The two sources can be thought
of as two antennas of a broadcast transmitter when they cooperate to
transmit these signals. Only the destination to which the signal is
intended for recovers it and the precoding ensures that no interference is
caused at the other destination. The main differences from {\sf
cooperative-public} signals are: (1) the sources convey the {\sf
cooperative-private} signals to each other through the lowest levels of
their transmission vector in such a way that it is not visible to the
destinations, whereas the {\sf cooperative-public} signals are conveyed
over the top-most levels of the transmission vector, and (2) the sources
collaborate in sending the {\sf cooperative-private} signals by precoding
the signals to ensure that no interference results at the destinations,
while {\sf cooperative-public} signals are visible to both the destinations
which end up recovering them. Thus, while the role of cooperation is to aid
both the destinations in recovering the {\sf cooperative-public} signals,
it aims to conceal the {\sf cooperative-private} from the destination it is
not intended for.

\end{itemize}

In general, such uncoded schemes are not sufficient to cover all possible
linear deterministic channels (indeed, even in the above example, we relied
on the characteristic of the field not being 2), and more importantly the
Gaussian channels. But the basic intuition can be used to build coding
schemes which do. These schemes are presented in detail in the next
section.

To roughly see how the intuition extends, consider again example~1 in
Figure~\ref{fig:example1}. Consider a coding scheme where the transmitted
signal from each source is a superposition of three codewords (as shown in
the figure). The codewords encode the two {\sf cooperative-public} messages
from the two sources and the {\sf private} message of the transmitting
source. As the figure suggests, these codewords have individual ``power
allocations'' in the sense that the codeword for the {\sf private} message
has zeros in the top two rows while the codeword for the other source
sources {\sf cooperative-public} message has zeros in the top-most row. The
specific uncoded scheme of the example clearly satisfies these ``power
allocations.'' But we could also consider a coding scheme where the
non-zero rows of the codewords carry linear equations on the messages
expressed as variables in the field. As long as the rates at which the
message variables are introduced into the linear equations are as in the
example, a random coding argument can be invoked to show the existence of
such a linear code (over a large enough extension field, or equivalently
over a long enough block of symbols) which will allow the sources to
successfully decode each other's {\sf cooperative-public} messages and the
destinations to successfully decode all the messages. In the next section
we follow this direction and present coding schemes for discrete-memoryless
interference channels with source cooperation.

\section{Coding schemes}
We first present our key coding theorem (in Theorem~\ref{thm:sourcecoopgenericschemes}).
It is a block-Markov (in the sense of \cite{ElGamalCoverRelay}) coding scheme
 which builds on Han and
Kobayashi's classical superposition coding scheme~\cite{HanKobayashi} for the two-user
interference channel and has elements of decode-and-forward
strategy~\cite{ElGamalCoverRelay} and
backwards-decoding~\cite{Willems} for relay
channels. The schemes are generic in the sense that they apply to any
memoryless interference channel with source cooperation
$p_{Y_1,Y_2,Y_3,Y_4|X_1,X_2}$. Then, we apply these schemes to the problems
at hand to obtain the achievability part of Theorems~\ref{thm:sourcecoopLD}
and~\ref{thm:sourcecoopG}. We would like to point out that
Theorem~\ref{thm:sourcecoopgenericschemes}(a) is identical to the one which
appears in~\cite{Tuninetti07}.

\begin{thm}\label{thm:sourcecoopgenericschemes}
(a) Given a joint distribution $p_W$ $p_{V_1,U_1,X_1|W}$
$p_{V_2,U_2,X_2|W}$, the rate pair $(R_1,R_2)$ is achievable if there are
non-negative $r_{V_1},r_{V_2},r_{U_1},r_{U_2},r_{X_1},r_{X_2}$ such that
$R_1=r_{V_1}+r_{U_1}+r_{X_1}$, $R_2=r_{V_2}+r_{U_2}+r_{X_2}$, and
\begin{align*}
r_{V_1}&\leq I(V_1;Y_2|W)\\
\\
r_{X_1}&\leq I(X_1;Y_3|V_1,V_2,W,U_1,U_2),\\
r_{U_1}+r_{X_1}&\leq I(U_1,X_1;Y_3|V_1,V_2,W,U_2),\\
r_{U_2}+r_{X_1}&\leq I(U_2,X_1;Y_3|V_1,V_2,W,U_1),\\
r_{U_1}+r_{U_2}+r_{X_1}&\leq I(U_1,U_2,X_1;Y_3|V_1,V_2,W),\\
(r_{V_1}+r_{V_2})+r_{U_1}+r_{U_2}+r_{X_1}&\leq
I(W,V_1,V_2,U_1,U_2,X_1;Y_3),
\end{align*}
and the corresponding inequalities with subscripts 1 and 2 exchanged,
and 3 replaced with 4.\\
(b) Given a joint distribution $p_{W}$
$p_{V_1,U_1|W}p_{V_2,U_2|W}$ $p_{S_1|W}p_{S_2|W}$
$p_{Z_1|W,V_1,U_1,S_1}p_{Z_2|W,V_2,U_2,S_2}$
$p_{X_1|W,V_1,U_1,Z_1,S_1,S_2}$ $p_{X_2|W,V_2,U_2,Z_2,S_1,S_2}$, the rate
pair $(R_1,R_2)$ is achievable if there are non-negative
$r_{V_k},r_{U_k},r_{Z_k},r_{S_k},\; k=1,2$ such that
$R_k=r_{V_k} + r_{U_k} + r_{Z_k}+r_{S_k}$, $k=1,2$ and
\begin{align*}
r_{S_1}&\leq I(X_1;Y_2|W,S_1,S_2,Z_1,U_1,V_1)\\
r_{Z_1}+r_{S_1}&\leq I(Z_1,X_1;Y_2|W,S_1,S_2,U_1,V_1)\\
r_{U_1}+r_{Z_1}+r_{S_1}&\leq I(U_1,Z_1,X_1;Y_2|W,S_1,S_2,V_1)\\
r_{V_1}+r_{U_1}+r_{Z_1}+r_{S_1}&\leq
I(V_1,U_1,Z_1,X_1;Y_2|W,S_1,S_2),\\
\\
r_{Z_1}&\leq I(Z_1;Y_3|V_1,V_2,W,U_1,U_2,S_1),\\
r_{U_1}+r_{Z_1}&\leq I(U_1,Z_1;Y_3|V_1,V_2,W,U_2,S_1),\\
r_{S_1}+r_{Z_1}&\leq I(S_1,Z_1;Y_3|V_1,V_2,W,U_1,U_2),\\
r_{S_1}+r_{U_1}+r_{Z_1}&\leq I(S_1,U_1,Z_1;Y_3|V_1,V_2,W,U_2),\\
r_{U_2}+r_{Z_1}&\leq I(U_2,Z_1;Y_3|V_1,V_2,W,U_1,S_1),\\
r_{U_2}+r_{U_1}+r_{Z_1}&\leq I(U_2,U_1,Z_1;Y_3|V_1,V_2,W,S_1),\\
r_{U_2}+r_{S_1}+r_{Z_1}&\leq I(U_2,S_1,Z_1;Y_3|V_1,V_2,W,U_1),\\
r_{U_2}+r_{S_1}+r_{U_1}+r_{Z_1}
&\leq I(U_2,S_1,U_1,Z_1;Y_3|V_1,V_2,W),\\
(r_{V_1}+r_{V_2})+r_{U_1}+r_{U_2}+r_{S_1}+r_{Z_1}
&\leq
I(W,V_1,V_2,U_1,U_2,S_1,Z_1;Y_3),
\end{align*}
and the corresponding inequalities with subscripts 1 and 2 exchanged,
and 3 replaced by 4.\\
(c) Given a joint distribution $p_{W}$
$p_{V_1,U_1|W}p_{V_2,U_2|W}$ $p_{S_1|W}$
$p_{Z_1|W,V_1,U_1,S_1}p_{Z_2|W,V_2,U_2}$
$p_{X_1|W,V_1,U_1,Z_1,S_1}$ $p_{X_2|W,V_2,U_2,Z_2,S_1}$, the rate
pair $(R_1,R_2)$ is achievable if there are non-negative
$r_{V_k},r_{U_k},r_{Z_k},
\; k=1,2$, and $r_{S_1}$, such that
$R_1=r_{V_1} + r_{U_1} + r_{Z_1} + r_{S_1}$,
$R_2=r_{V_1} + r_{U_1} + r_{Z_1}$, and
\begin{align*}
r_{S_1}&\leq I(X_1;Y_2|W,S_1,Z_1,U_1,V_1)\\
r_{Z_1}+r_{S_1}&\leq I(Z_1,X_1;Y_2|W,S_1,U_1,V_1)\\
r_{U_1}+r_{Z_1}+r_{S_1}&\leq I(U_1,Z_1,X_1;Y_2|W,S_1,V_1)\\
r_{V_1}+r_{U_1}+r_{Z_1}+r_{S_1}&\leq I(V_1,U_1,Z_1,X_1;Y_2|W,S_1),\\
\\
r_{V_2}&\leq I(V_2;Y_1|W,S_1),\\
\\
r_{Z_1}&\leq I(Z_1;Y_3|V_1,V_2,W,U_1,U_2,S_1),\\
r_{U_1}+r_{Z_1}&\leq I(U_1,Z_1;Y_3|V_1,V_2,W,U_2,S_1),\\
r_{S_1}+r_{Z_1}&\leq I(S_1,Z_1;Y_3|V_1,V_2,W,U_1,U_2),\\
r_{S_1}+r_{U_1}+r_{Z_1}&\leq I(S_1,U_1,Z_1;Y_3|V_1,V_2,W,U_2),\\
r_{U_2}+r_{Z_1}&\leq I(U_2,Z_1;Y_3|V_1,V_2,W,U_1,S_1),\\
r_{U_2}+r_{U_1}+r_{Z_1}&\leq I(U_2,U_1,Z_1;Y_3|V_1,V_2,W,S_1),\\
r_{U_2}+r_{S_1}+r_{Z_1}&\leq I(U_2,S_1,Z_1;Y_3|V_1,V_2,W,U_1),\\
r_{U_2}+r_{S_1}+r_{U_1}+r_{Z_1}
&\leq I(U_2,S_1,U_1,Z_1;Y_3|V_1,V_2,W),\\
(r_{V_1}+r_{V_2})+r_{U_1}+r_{U_2}+r_{S_1}+r_{Z_1}
&\leq
I(W,V_1,V_2,U_1,U_2,S_1,Z_1;Y_3),\\
\\
r_{Z_2}&\leq I(Z_2;Y_4|V_1,V_2,W,U_1,U_2),\\
r_{U_2}+r_{Z_2}&\leq I(U_2,Z_2;Y_4|V_1,V_2,W,U_1),\\
r_{U_1}+r_{Z_2}&\leq I(U_1,Z_2;Y_4|V_1,V_2,W,U_2),\\
r_{U_1}+r_{U_2}+r_{Z_2}&\leq I(U_1,U_2,Z_2;Y_4|V_1,V_2,W),\\
(r_{V_1}+r_{V_2})+r_{U_1}+r_{U_2}+r_{Z_2}
&\leq I(W,V_1,V_2,U_1,U_2,Z_2;Y_4).
\end{align*}
\end{thm}

We prove this theorem in Appendix~\ref{app:sourcecoopgeneric}. Here, we
interpret these theorems in the context of the examples and discussion in
the previous section:

Scheme (a) involves the source nodes aiding their respective
destination nodes decode part of the interference (the {\sf
cooperative-public} message from the interfering source) by essentially
retransmitting the part of the interference observed by the source. In
order to facilitate this, the {\sf cooperative-public} message is coded
separately (in $V_1$ and $V_2$) which are decoded by the sources and
retransmitted (in $W$). There are further {\sf public} messages ($U_1$ and
$U_2$) which are not aided by the other source, and {\sf private} messages as
well. This scheme is closely related to Example~1 of the previous section.

In scheme (b), in addition to the above, the sources collaborate in
sending private messages by sharing with each other in advance the part of
the message on which they want to collaborate. Thus, the connection is to
Example~3 of the previous section.
The auxiliary random variables in scheme (b) have the following
interpretation:\\
\begin{center}\begin{tabular}{ccl}
Aux.&Decoding&Remarks\\
 RV&destinations&\\
\hline\\
$V_1$&3,4&{\sf cooperative-public} message from source~1\\
$V_2$&3,4&{\sf cooperative-public} message from source~2\\
$W$&3,4&used by the sources to cooperatively send the two {\sf
cooperative-public} messages\\
$U_1$&3,4& {\sf public} message from source~1\\
$U_2$&3,4& {\sf public} message from source~2\\
$S_1$&3& carries the {\sf cooperative-private} message to destination~3\\
$S_2$&4& carries the {\sf cooperative-private} message to destination~4\\
$Z_1$&3&{\sf private} message from source~1\\
$Z_2$&4&{\sf private} message from source~2
\end{tabular}
\end{center}
Note that scheme (a) is not a special case of (b) as it might appear. The
key difference is that while in scheme (b), the sources perform a joint
decoding of all the messages sent by their counterparts (including the {\sf
public} and {\sf private} messages which they do not aid in the
transmission of), in scheme (a), only the part of the {\sf
cooperative-public} message meant to be used for collaboration is decoded
while treating all the other messages as noise. At low strengths for the
cooperative link, scheme (a) can perform better than scheme (b) specialized
to not include a {\sf cooperative-private} message. On the other hand, for
strong cooperative links, adopting a joint decoding scheme at the sources
can lead to better overall rates for other messages.

Scheme (c) combines these two schemes in a limited manner. Only
source node~1's transmission benefits from transmission of some {\sf
cooperative-public} and some {\sf cooperative-private} messages while
source node~2's transmission benefits only from collaborative transmission
of some {\sf cooperative-public} message. Source node~1 adopts a decoding
strategy similar to that adopted by the sources in scheme (a) whereas
source node~2's decoding strategy is similar to the one in scheme (b).
The three schemes (a), (b), and (c) can be easily combined into a single
scheme (at the expense of considerably more involved notation) which may
improve the achievable region in general, but since the focus here is on
the sum-rate of the Gaussian case (within a constant gap) and linear
deterministic cases which are obtained by considering the schemes
separately, we do not explore this here.

Intuitively, a source employs a {\sf cooperative-private} message only when
the cooperative link is stronger than the direct link from this source, and
the source shares this message with the other source in advance by having
it ``ride below'' the other messages it sends. Thus, one of the differences
of our scheme from the proposals of~\cite{CaoChen07,YangTuninetti08} is
that when sharing the {\sf cooperative-private} message in advance, the
decoding source performs a joint decoding of all the messages including
those messages it does not aid in the transmission of, rather than treat
these other messages as noise.
Note that in our schemes~(b) and (c), the sources cooperate in sending the
{\sf cooperative-private} messages by employing a simple form of precoding
along the lines of Example~3 in the last section. Use of dirty-paper
coding~\cite{GP,CostaDPC} instead may lead to an improved gap.

The achievability proofs presented in Appendix~\ref{app:LDachieve}
and~\ref{app:Gachieve} of our main theorems (Theorems~\ref{thm:sourcecoopLD}
and~\ref{thm:sourcecoopG}) make use of these schemes.

\section{Interference Channel with Feedback}\label{sec:feedback}

A closely related problem is that of the interference channel with feedback.
Let us consider the symmetric Gaussian interference channel with noiseless
feedback. As in the model we considered earlier,
\begin{align*}
Y_3(t)&=h_{1,3}X_1(t)+h_{2,3}X_2(t)+N_3(t),\\
Y_4(t)&=h_{2,4}X_2(t)+h_{1,4}X_1(t)+N_4(t).
\end{align*}
However, instead of receiving signals through the cooperation link, the
sources now receive noiseless feedback from their respective
destinations. {\em i.e.},
\begin{align*}
Y_1(t)&=Y_3(t),\\
Y_2(t)&=Y_4(t).
\end{align*}
As before, the transmissions from the sources are deterministic functions of
their messages and their observations (here, the feedback received) in the
past. Since the sources have access to the symbols they transmitted in the
past, it is clear that the above problem is equivalent to one where the
sources observe
\begin{align*}
Y_1(t)&=h_{2,3}X_2(t)+N_3(t),\\
Y_2(t)&=h_{1,4}X_1(t)+N_4(t).
\end{align*}
We can rewrite these as
\begin{align*}
Y_1(t)&=h_{2,1}X_2(t)+N_1(t),\\
Y_2(t)&=h_{1,2}X_1(t)+N_2(t).
\end{align*}
Here $h_{2,1}=h_{2,3}$, $h_{1,2}=h_{1,4}$,
$N_1(t)=N_3(t)$, and $N_2(t)=N_4(t)$.
This is identical to the channel model stated at the beginning of
Section~\ref{sec:model} except for the fact that now $N_1,N_2,N_3,N_4$ are
not independent, but $N_1$ and $N_3$ are identical, and so are $N_2$ and
$N_4$.

The above difference notwithstanding, we will argue below that the results
presented in Section~\ref{sec:results} on the sum-capacity of the
interference channel with source cooperation have a bearing on this model
as well.  Note, however, that our source cooperation result was proved
under the restriction that $|h_{1,2}|=|h_{2,1}|$. In general, this may not
hold true for interference channels with feedback, but a range of channels
including, most importantly, the symmetric interference channel is covered.
We will focus our attention on the symmetric channel, where
$|h_{1,3}|=|h_{2,4}|=h_D$, and $|h_{1,4}|=|h_{2,3}|=h_I$.

Let us begin by noting that a linear deterministic formulation of the above
feedback problem is identical to the one with source cooperation, and
hence, Theorem~\ref{thm:sourcecoopLD} applies directly. Turning to the
Gaussian case, let us note that the achievability proof for the source
cooperation case depended only on the marginal distributions of the noises
and not on their correlation. Hence the achievability proof also holds
directly. We only need to argue that the converse also applies. In
appendix~\ref{app:feedback}, we will show that the biting upperbound for
the symmetric channel with noiseless feedback indeed holds to give us the
following proposition:
\begin{prop}
The sum-capacity of the symmetric, Gaussian interference channel with output
feedback is within a constant (19 bits) of 
\begin{align}
\log2\left(1+\left(|h_D|+|h_I|\right)^2\right)
\left(1+\frac{\max\left(|h_D|^2,|h_I|^2\right)}
 {\max\left(1,|h_I|^2\right)}\right).\label{eq:u2GaussFeedback}
\end{align}
\end{prop}
This case was studied independently in~\cite{SuhTse09} which also
characterizes the sum-capacity and obtains a better constant than we do
here.

\section{Discussion}

\subsection{Reversibility}\label{sec:reversibility}

A related setting to the one studied in this paper is the interference
channel with destination cooperation. This case will be presented in a
companion paper~\cite{DestCoop}. An interesting {\em reversibility}
property connects the two settings. We briefly discuss it here.

\begin{figure*}[!t]
\centerline{\subfloat[Source
Cooperation]{\scalebox{0.5}{\input{sourcecoop.pdf_t}}%
\label{fig:sourcecoop}}
\hfil
\subfloat[Destination Cooperation]{
\scalebox{0.5}{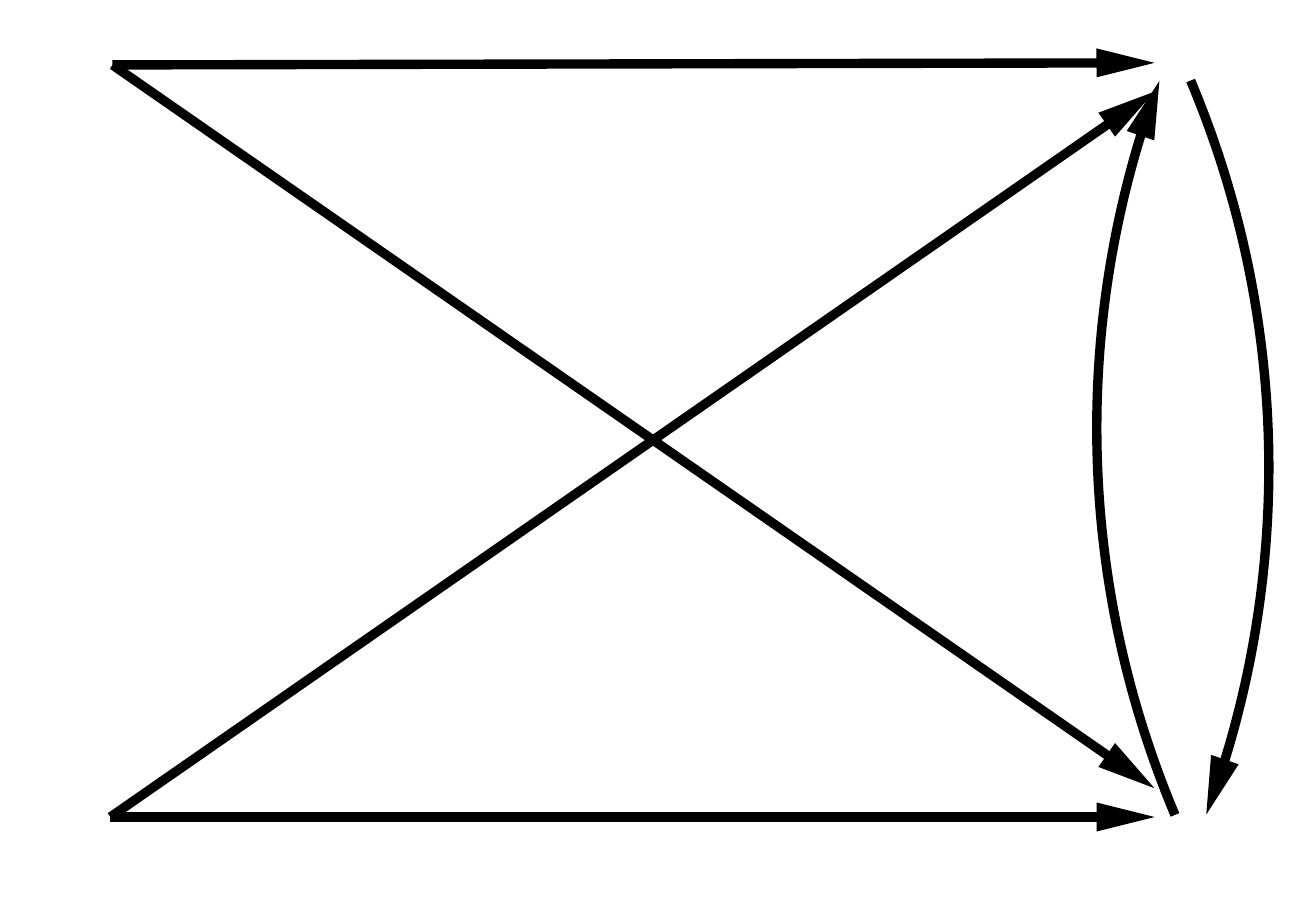}
\label{fig:destinationcoop}}}
\caption{Source Cooperation and Destination Cooperation. Note that the
destination cooperation channel (b) is obtained from the source cooperation
channel (a) by: (i) reversing the roles of the sources and destinations,
{\em i.e.}, nodes~1 and~2 are sources in (a) and destinations in (b);
nodes~3 and 4 are destinations in (a) and sources in (b), and (ii)
reversing the directions of all the links while maintaining the link
coefficients to be the same. The noise variances and transmitter power
constraints are normalized to unity. The sum-capacities of the two channels
are within a constant gap (50 bits) of each other irrespective of the
channel coefficients. For the linear deterministic model, both channels
have the same sum-capacity.}
\label{fig:reversibility setup}
\end{figure*}
In the destination cooperation case, the destinations can not only receive,
but they can also transmit. But these transmissions have to satisfy a
causality constraint -- the transmissions from each destination is a
function of everything it has received up to the previous time instant. In
order to illustrate the reversibility between destination and source
cooperation scenarios, we will number the nodes as shown in the
Fig.~\ref{fig:reversibility setup}(b): 3 and 4 are the source nodes now
which want to communicate to destination nodes~1 and 2 respectively, and
there is a cooperation link between the destination nodes. The destination
nodes receive
\begin{align*}
Y_1(t)&=h_{1,3}(X_3(t))+h_{1,4}^\ast(X_4(t))+h_{1,2}(X_2(t)),\\
Y_2(t)&=h_{2,4}(X_4(t))+h_{2,3}^\ast(X_3(t))+h_{2,1}(X_1(t)).
\end{align*}
where the (deterministic) encoding functions at the sources are of the form
\begin{align*}
X_k(t)&=f_{k,t}(M_k),\;k=3,4,
\end{align*}
and the (deterministic) relaying functions at the destinations are of the form
\begin{align*}
X_k(t)&=f_{k,t}(Y_k^{t-1}),\;k=2,1.
\end{align*}
For the Gaussian channel,
\begin{align*}
h_{1,3}(X_3)&=h_{1,3}X_3,\\
h_{2,4}(X_4)&=h_{2,4}X_4,\\
h_{1,4}^\ast(X_4)&=h_{1,4}X_4+N_1,\\
h_{2,3}^\ast(X_3)&=h_{2,3}X_3+N_2,\\
h_{1,2}(X_2)&=h_{1,2}X_2,\\
h_{2,1}(X_1)&=h_{2,1}X_1,
\end{align*}
where $N_1$ and $N_2$ are independent white Gaussian noise processes with
zero mean and unit variance. The encoding and relaying maps must
satisfy average power constraints of unity. The reciprocity between the
destinations is modeled by setting $|h_{2,1}|=|h_{1,2}|=h_C$.

We may also define a linear deterministic analog. Let
$n_{1,3},n_{2,3}$, $n_{1,4},n_{2,4}$, $n_{2,1},n_{1,2}$ be non-negative
integers and $n\defineqq\max(n_{1,3},n_{2,3}, n_{1,4},n_{2,4}
,n_{2,1},n_{1,2})$. The inputs to the channel $X_3$ and $X_4$ are
$n$-length vectors over a finite field ${\mathbb F}$. We define
\begin{align*}
h_{1,3}(X_3)&={\bf S}^{n-n_{1,3}}X_3,\\
h_{2,4}(X_4)&={\bf S}^{n-n_{2,4}}X_4,\\
h_{1,4}^\ast(X_4)&={\bf S}^{n-n_{1,4}}X_4,\\
h_{2,3}^\ast(X_3)&={\bf S}^{n-n_{2,3}}X_3,\\
h_{1,2}(X_2)&={\bf S}^{n-n_{1,2}}X_2,\\
h_{2,1}(X_1)&={\bf S}^{n-n_{2,1}}X_1.
\end{align*}
Further, to model the reciprocity of the links between the two receivers,
we set $n_{2,1}=n_{1,2}=n_C$.

The reversibility property in the context of the linear deterministic
channels is that the sum-capacity expression in
Theorem~\ref{thm:sourcecoopLD} is also the sum-capacity of the above
channel with destination cooperation. This turns out to be a feature of our
achievable strategy which holds in more general cases (larger networks,
more number of sources-destination pairs etc.) as discussed further
in~\cite{RPV_ITW09}. It would be interesting to investigate whether the
optimality results presented here also extend to more general settings.

In the Gaussian case, the sum-capacities are within a universal constant.
\begin{thm} \label{thm:reversibility}
The sum-capacities of the two-user Gaussian interference channels in
Figures~\ref{fig:sourcecoop} and~\ref{fig:destinationcoop} with source
cooperation and destination cooperation, respectively, are within a
constant gap of at most 50 bits.
\end{thm}
The proof is presented in Appendix~\ref{app:reversibility}.

\subsection{Dependence on channel state information}

Throughout the paper we assumed that full channel state information is
available at both the sources and the destinations. However, this can be
relaxed. The application of the scheme in
Theorem~\ref{thm:sourcecoopgenericschemes}(a) only requires the sources to
know the channel strengths to the two destinations, and not their phases.
Note that Theorem~\ref{thm:sourcecoopgenericschemes}(a) caters to the case
where the strength of the cooperative link is weaker than that of all the
links to the two destinations. But, to apply the schemes in
Theorem~\ref{thm:sourcecoopgenericschemes}(b) and~(c), we do require the
sources to have full channel state information. This is not surprising
since the analogous setting of a multiantenna broadcast channel also
requires full channel state information.

\appendices

\section{Proof of Theorem~\ref{thm:sourcecoopgenericschemes}}
\label{app:sourcecoopgeneric}

\noindent{(a)} We present a block-Markov scheme with backwards decoding.
Given $p_W p_{V_1,U_1,X_1|W} p_{V_2,U_2,X_2|W}$, we construct the following
blocklength-$T$ codebooks:
\begin{itemize}

\item {$W$ codebook:} We create a $W$-codebook ${\mathcal C}_W$ of size
$2^{Tr_W}$ with codewords of length $n$ by choosing the elements
independently according to the distribution $p_W$. We will denote the
codewords by $c_W(m_W)$ where $m_W\in\{1,\ldots,2^{T(r_W-\epsilon)}\}$,
where $\epsilon>0$.

\item {$V$ codebooks:} For each codeword $c_W(m_W)\in{\mathcal C}_W$, and
for each $k=1,2$, we create $V_k$-codebook ${\mathcal C}_{V_k}(m_W)$ of
size $2^{T(r_{V_k}-\epsilon)}$ respectively, by choosing elements
independently according to $p_{V_k|W}(.|w)$ where $w$ is set to the
respective element of the $c_W(m_W)$ codeword. These codewords will be
denoted by $c_{V_k}(m_{V_k},m_W)$ where
$m_{V_k}\in\{1,\ldots,2^{T(r_{V_k}-\epsilon)}\}$.  Moreover, we set
\[r_W=r_{V_1}+r_{V_2}.\]

\item {$U$ codebooks:} For each codeword $c_{V_k}(m_{V_k},m_W)$, we create
a $U_k$-codebook ${\mathcal C}_{U_k}(m_{V_k},m_W)$ of size
$2^{T(r_{U_k}-\epsilon)}$ by choosing elements according to
$p_{U_k|V_k,W}(.|v_k,w)$ by setting $v_k$ and $w$ to be the respective
elements of the $c_{V_k}(m_{V_k},m_W)$ and $c_W(m_W)$ codewords.

\item {$X$ codebooks:} Finally, for each codeword
$c_{U_k}(m_{U_k},m_{V_k},m_W)$, we similarly create a $X_k$-codebook
${\mathcal C}_{X_k}(m_{U_k},m_{V_k},m_W)$ of size $2^{T(r_{X_k}-\epsilon)}$ using
$p_{X_k|U_k,V_k,W}$.

\end{itemize}

\noindent {\em Encoding}: For block-$j$, we will assume for the moment that the
source nodes have successfully decoded the messages $m_{V_k}(j-1)$. Then
the encoding proceeds as follows. Both encoders set
$m_W(j)=(m_{V_1}(j-1),m_{V_2}(j-1))$. They then proceed to choose the
codewords $c_{W}(m_W(j))$, $c_{V_k}(m_{V_k}(j),m_W(j))$,
$c_{U_k}(m_{U_k}(j),m_{V_k}(j),m_W(j))$, and
$c_{X_k}(m_{X_k}(j),m_{U_k}(j),m_{V_k}(j),m_W(j))$. The $X$-codewords are
transmitted. For the first block, $j=1$, we set $m_W(1)=1$ and for the last
block $J$, we set $m_{V_1}(J)=m_{V_2}(J)=1$.

\noindent {\em Decoding at the sources}: At the end of block-$j$, source~1 will
try to decode the message $m_{V_2}(j)$ from source~2 and {\em vice versa}.
Using standard arguments, we can show that for joint-typical decoding 
to succeed (with probability
approaching 1 as the blocklength $n$ approaches $\infty$), it is enough to
ensure that
\begin{align*}
r_{V_1}&\leq I(V_1;Y_2|W),\text{ and}\\
r_{V_2}&\leq I(V_2;Y_1|W).
\end{align*}
When this decoding fails, we will say that ``encoding at the sources has
failed at block-$j$,'' and declare an error.

\noindent {\em Decoding at the destinations}: Destinations perform {\em backwards
decoding}.
We will assume that before destination~3 processes
block-$j$, it has already successfully decoded
$m_{W}(j+1)=(m_{V_1}(j),m_{V_2}(j))$. This is true for $j=J$ since
$m_{V_1}(J)=m_{V_2}(J)=1$. And, for each $j$, we will ensure that from
block-$j$, destination~3 decodes $m_{W}(j)$ successfully thereby ensuring
that the above assumption holds true. Assuming that $m_{V_1}(j),m_{V_2}(j)$ is
available at destination~3, we will try to ensure that from block-$j$, the
messages $m_W(j)$, $m_{U_1}(j)$, $m_{X_1}(j)$ are successfully decoded. In
trying to decode these messages, destination~3 will also try to jointly
decode the message $m_{U_2}(j)$. It can be shown that a high probability of decoding success can
be ensured ({\em i.e.}, the probability of failure in decoding the messages
$m_W(j)$, $m_{U_1}(j)$, and $m_{X_1}(j)$ from what destination~3 receives
in block-$j$ assuming $m_W(j+1)$ is available, goes to 0 as blocklength $n$
goes to $\infty$) if the following conditions are met. 
\begin{align*}
r_{X_1}&\leq I(X_1;Y_3|V_1,V_2,W,U_1,U_2),\\
r_{U_1}+r_{X_1}&\leq I(U_1,X_1;Y_3|V_1,V_2,W,U_2),\\
r_{U_2}+r_{X_1}&\leq I(U_2,X_1;Y_3|V_1,V_2,W,U_1),\\
r_{U_1}+r_{U_2}+r_{X_1}&\leq I(U_1,U_2,X_1;Y_3|V_1,V_2,W),\text{ and}\\
(r_{V_1}+r_{V_2})+r_{U_1}+r_{U_2}+r_{X_1}&\leq
I(W,V_1,V_2,U_1,U_2,X_1;Y_3).
\end{align*}
A similar set of conditions ensure success of decoding at destination~4. If
decoding fails for block-$j$ for either of the destinations, we will say
that ``decoding failed at block-$j$'' and declare an error.

Overall, an error results if for at least one block-$j$, either encoding
fails or decoding fails. Since there are a finite number $J$ of blocks, by
union bound, the above discussion implies that the probability of error
goes to 0 as the blocklength goes to $\infty$ when the above conditions are
met. This completes the random coding argument.

\noindent{(b)} We present a block-Markov scheme with backwards decoding at
the destinations.
Given $p_{W}$
$p_{V_1,U_1|W}$ $p_{V_2,U_2|W}$ $p_{S_1|W}p_{S_2|W}$
$p_{Z_1|W,V_1,U_1,S_1}$ $p_{Z_2|W,V_2,U_2,S_2}$
$p_{X_1|W,V_1,U_1,Z_1,S_1,S_2}$ $p_{X_2|W,V_2,U_2,Z_2,S_1,S_2}$,
we construct the following blocklength-$n$ codebooks:
\begin{itemize}

\item {$W$, $V$, and $U$ codebooks:} These codebooks are constructed as in
scheme~(a) above. We create a $W$-codebook ${\mathcal C}_W$ of size
$2^{Tr_W}$ with codewords of length $n$ by choosing the elements
independently according to the distribution $p_W$. We will denote the
codewords by $c_W(m_W)$, where $m_W\in\{1,\ldots,2^{T(r_W-\epsilon)}\}$,
where $\epsilon>0$.

For each codeword $c_W(m_W)\in{\mathcal C}_W$, and for each $k=1,2$, we
create $V_k$-codebook ${\mathcal C}_{V_k}(m_W)$ of size
$2^{T(r_{V_k}-\epsilon)}$ respectively, by choosing elements independently
according to $p_{V_k|W}(.|w)$ where $w$ is set to the respective element of
the $c_W(m_W)$ codeword. These codewords will be denoted by
$c_{V_k}(m_{V_k},m_W)$ where
$m_{V_k}\in\{1,\ldots,2^{T(r_{V_k}-2\epsilon)}\}$.  Moreover, we set
\[r_W=r_{V_1}+r_{V_2}.\]

For each codeword $c_{V_k}(m_{V_k},m_W)$, we create a $U_k$-codebook
${\mathcal C}_{U_k}(m_{V_k},m_W)$ of size $2^{T(r_{U_k}-\epsilon)}$ by
choosing elements according to $p_{U_k|V_k,W}(.|v_k,w)$ by setting $v_k$
and $w$ to be the respective elements of the $c_{V_k}(m_{V_k},m_W)$ and
$c_W(m_W)$ codewords. These codewords will be denoted by
$c_{U_k}(m_{U_k},i_{V_k},m_W)$ where
$m_{U_k}\in\{1,\ldots,2^{T(r_{U_k}-\epsilon)}\}$.

\item {$S$ codebooks:} For each $c_W(m_W)\in{\mathcal C}_W$, and
for each $k=1,2$, we create an $S_k$-codebook ${\mathcal C}_{S_k}(m_W)$ of
size $2^{T(r_{S_k}-\epsilon)}$ respectively, by choosing elements
independently according to $p_{S_k|W}(.|w)$ where $w$ is set to the
respective element of the $c_W(m_W)$ codeword. These codewords will be
denoted by $c_{S_k}(m_{S_k},m_W)$ where
$m_{S_k}\in\{1,\ldots,2^{T(r_{S_k}-2\epsilon)}\}$.

\item {$Z$ codebooks:} For each pair of codewords
$(c_{U_k}(m_{U_k},m_{V_k},m_W),c_{S_k}(m_{S_k},m_W))$ $\in$\\ ${\mathcal
C}_{U_k}(m_{V_k},m_W)$ $\times {\mathcal C}_{S_k}(m_W)$, and for each
$k=1,2$, we create a $Z_k$-codebook\\ ${\mathcal
C}_{Z_k}(m_{U_k},m_{V_k},m_W,i_{S_k})$ of size $2^{T(r_{Z_k}-\epsilon)}$
respectively, by choosing elements independently according to
$p_{Z_k|W,V_k,U_k,S_k}(.|w,v_k,u_k,s_k)$ where $w$, $v_k$, $u_k$ and $s_k$
are set to the respective elements of the $c_W(m_W)$,
$c_{V_k}(m_{V_k},m_W)$, $c_{U_k}(m_{U_k},m_{V_k},m_W)$, and
$c_{S_k}(m_{S_k},m_W)$ codewords, respectively. The codewords so generated
will be denoted by\\ $c_{Z_k}(m_{Z_k},m_{U_k},m_{V_k},m_W,m_{S_k})$ where
$i_{Z_k}\in\{1,\ldots,2^{T(r_{Z_k}-\epsilon)}\}$.

\item {$X$ codebooks:} Finally, consider pairs of codewords\\
$(c_{Z_k}(m_{Z_k},m_{U_k},m_{V_k},m_W,m_{S_k}),
c_{S_{\bar{k}}}(m_{S_{\bar{k}}},m_W))$ $\in {\mathcal
C}_{Z_k}(m_{U_k},m_{V_k},m_W,m_{S_k})\times {\mathcal
C}_{S_{\bar{k}}}(m_W)$, for each $k=1,2$, where $\bar{k}=2$, if $k=1$, and
$\bar{k}=1$, if $k=2$. For each pair, we create a $X_k$-codebook ${\mathcal
C}_{X_k}(m_{Z_k},m_{U_k},m_{V_k},m_W,i_{S_k},i_{S_{\bar{k}}})$ of size
$2^{T(r_{S_k}-\epsilon)}$ by choosing elements independently according to
$p_{X_k|W,V_k,U_k,Z_k,S_k,S_{\bar{k}}}(.|w,v_k,u_k,s_k,s_{\bar{k}})$ where
$w$, $v_k$, $u_k$, $z_k$, $s_k$, and $s_{\bar{k}}$ are set to the
respective elements of the $c_W(m_W)$, $c_{V_k}(i_{V_k},m_W)$,
$c_{U_k}(i_{U_k},i_{V_k},m_W)$, $c_{Z_k}(i_{Z_k},i_{U_k},i_{V_k},m_W)$,
$c_{S_k}(i_{S_k},m_W)$, and $c_{S_{\bar{k}}}(i_{S_{\bar{k}}},m_W)$
codewords, respectively.  The codewords so generated will be denoted by\\
$c_{X}(m_{\PVTCOOP_k},m_{Z_k},m_{U_k},m_{V_k},m_W,i_{S_{\bar{k}}},i_{S_k})$
where $m_{\PVTCOOP_k}\in\{1,\ldots,2^{T(r'_{S_k}-\epsilon)}\}$.

\end{itemize}

\noindent {\em Encoding}: For block-$j$, we will assume for the moment that the
source nodes have successfully decoded the messages $m_{V_k}(j-1),
m_{\PVTCOOP_k}(j-1)$. Then the encoding proceeds as follows.  Both encoders
set $m_W(j)=(m_{V_1}(j-1),m_{V_2}(j-1))$. Then, encoder-$k$ proceeds to
select the codewords $c_{W}(m_W(j))$, $c_{V_k}(m_{V_k}(j),m_W(j))$,
$c_{U_k}(m_{U_k}(j),m_{V_k}(j),m_W(j))$, $c_{S_k}(m_{S_k}(j),m_W(j))$, and
$c_{Z_k}(m_{Z_k}(j),m_{U_k}(j),m_{V_k}(j),m_W(j),i_{S_k}(j))$. It transmits
the $X$-codeword\\
$c_{X}(m_{\PVTCOOP_k}(j),m_{Z_k}(j),m_{U_k}(j),m_{V_k}(j),m_W(j),m_{S_{\bar{k}}}(j),m_{S_k}(j))$.
For the first block, $j=1$, we set $m_W(1)=m_{S_1}(1)=m_{S_2}(1)=1$ and for
the last block $J$, we set
$m_{V_1}(J)=m_{V_2}(J)=m_{\PVTCOOP_1}(J)=m_{\PVTCOOP_2}(J)=1$.

\noindent {\em Decoding at the sources}: At the end of block-$j$, source~2 will
try to jointly decode \begin{align*}
m_{V_1}(j),m_{U_1}(j),m_{Z_1}(j),m_{\PVTCOOP_1}(j)
\end{align*}
 from
source~1 and {\em vice versa}. Note that both sources have access to the
$W$, $S_1$, and $S_2$-codewords. For joint-typical decoding to
succeed (with probability approaching 1 as the blocklength $T$ approaches
$\infty$), we can show that it is enough to ensure that
\begin{align*}
r_{S_1}&\leq I(X_1;Y_2|W,S_1,S_2,Z_1,U_1,V_1)\\
r_{Z_1}+r_{S_1}&\leq I(Z_1,X_1;Y_2|W,S_1,S_2,U_1,V_1)\\
r_{U_1}+r_{Z_1}+r_{S_1}&\leq I(U_1,Z_1,X_1;Y_2|W,S_1,S_2,V_1)\\
r_{V_1}+r_{U_1}+r_{Z_1}+r_{S_1}&\leq
I(V_1,U_1,Z_1,X_1;Y_2|W,S_1,S_2).
\end{align*}
When this decoding fails, we will say that ``encoding at the sources has
failed at block-$j$,'' and declare an error.

\noindent {\em Decoding at the destinations}: Destinations perform {backwards
decoding}. We will assume that before destination~3 processes
block-$j$, it has already successfully decoded
$m_{W}(j+1)=(m_{V_1}(j),m_{V_2}(j))$. This is true for $j=J$ since
$m_{V_1}(J)=m_{V_2}(J)=1$. And, for each $j$, we will ensure that from
block-$j$, destination~3 decodes $m_{W}(j)$ successfully thereby ensuring
that the above assumption holds true. Assuming that $m_{V_1}(j),m_{V_2}(j)$
is available at destination~3, we will try to ensure that from block-$j$,
the messages $m_W(j)$, $m_{U_1}(j)$, $m_{Z_1}(j)$, and $m_{S_1}(j)$ are
successfully decoded. In trying to decode these messages, destination~3
will also try to jointly decode the message $m_{U_2}(j)$. The decoding is
performed by looking for a unique tuple of $W,V_1,V_2,U_2,U_1,Z_1,S_1$
codewords consistent with the information already known (namely,
$m_{V_1}(j),m_{V_2}(j)$) and which are jointly (strongly) typical 
with the ($T$-length) block of signal $Y_3$ received by destination~3
corresponding to the block-$j$. Using standard arguments, a high probability
of decoding success can be ensured ({\em i.e.}, the probability of failure
in decoding the messages $m_W(j)$, $m_{U_1}(j)$, and $m_{Z_1}(j)$ and
$i_{S_1}(j)$ from what destination~3 receives in block-$j$ assuming
$m_W(j+1)$ is available, goes to 0 as blocklength $T$ goes to $\infty$) if
the following conditions are met.
\begin{align*}
r_{Z_1}&\leq I(Z_1;Y_3|V_1,V_2,W,U_1,U_2,S_1),\\
r_{U_1}+r_{Z_1}&\leq I(U_1,Z_1;Y_3|V_1,V_2,W,U_2,S_1),\\
r_{S_1}+r_{Z_1}&\leq I(S_1,Z_1;Y_3|V_1,V_2,W,U_1,U_2),\\
r_{S_1}+r_{U_1}+r_{Z_1}&\leq I(S_1,U_1,Z_1;Y_3|V_1,V_2,W,U_2),\\
r_{U_2}+r_{Z_1}&\leq I(U_2,Z_1;Y_3|V_1,V_2,W,U_1,S_1),\\
r_{U_2}+r_{U_1}+r_{Z_1}&\leq I(U_2,U_1,Z_1;Y_3|V_1,V_2,W,S_1),\\
r_{U_2}+r_{S_1}+r_{Z_1}&\leq I(U_2,S_1,Z_1;Y_3|V_1,V_2,W,U_1),\\
r_{U_2}+r_{S_1}+r_{U_1}+r_{Z_1}
&\leq I(U_2,S_1,U_1,Z_1;Y_3|V_1,V_2,W),\text{ and}\\
(r_{V_1}+r_{V_2})+r_{U_1}+r_{U_2}+r_{S_1}+r_{Z_1}
&\leq
I(W,V_1,V_2,U_1,U_2,S_1,Z_1;Y_3).
\end{align*}
A similar set of conditions ensure success of decoding at destination~4. If
decoding fails for block-$j$ for either of the destinations, we will say
that ``decoding failed at block-$j$'' and declare an error.

For the purposes of illustration, let us see how one of the above
conditions is arrived at. One of the possible error-events under which
decoding at destination~3 fails is when, for block-$j$, only the following
decoding errors occur: $\what{m}_{Z_1} \neq m_{Z_1}$, $\what{m}_{U_2} \neq
m_{U_2}$, and $\what{m}_{S_1} \neq m_{S_1}$, where the $\what{}$'s indicate
the decoded values. The probability of this (under random coding as
described above) is
\begin{align*}
\sum_{\what{m}_{Z_1},\what{m}_{U_2},\what{m}_{S_1}} &&
{\mathbb P}\Bigg(
(c_W,c_{V_1},c_{V_2},{c}_{U_1},\what{c}_{Z_1},\what{c}_{S_1},\what{c}_{U_2},Y_3^T)
\in {\mathcal T}^\delta \Bigg|
m_W,m_{V_1},m_{V_2},m_{U_1},m_{U_2},m_{Z_1},m_{Z_2}, \\
&&\quad\quad\quad m_{S_1},i_{S_2},m_{\PVTCOOP_1},m_{\PVTCOOP_2}\Bigg),
\end{align*}
where we suppressed the indices for the codewords, and the time-index $j$
for the codeword indices of the conditioning event. The unhatted codewords
have indices from the messages of the conditioning event, while the hatted
codewords are short-hand notations for the codewords with the corresponding
indices replaced by their hatted forms:
\begin{align*}
\what{c}_{Z_1}&=c_{Z_1}(\what{m}_{Z_1},m_{U_1},m_{V_1},m_W,\what{m}_{S_1}),\\
\what{c}_{S_1}&=c_{S_1}(\what{m}_{S_1},m_W),\text{ and}\\
\what{c}_{U_2}&=c_{U_2}(\what{m}_{U_2},m_{V_2},m_W).
\end{align*}
We also suppressed the subscript for the $\delta$-typical set ${\mathcal
T}_{W,V_1,V_2,U_1,U_2,Z_1,S_1,Y_3}$. We will continue to do that in the
sequel; the appropriate subscripts will be clear from the context.
Below, we will also suppress the conditioning event. Then,
\begin{align*}
&{\mathbb P}\Bigg(
(c_W,c_{V_1},c_{V_2},{c}_{U_1},\what{c}_{Z_1},\what{c}_{S_1},\what{c}_{U_2},Y_3^T)
\in {\mathcal T}^\delta \Bigg)\\
&\leq 
\sum_{(\wtilde{c}_W,\wtilde{c}_{V_1},\wtilde{c}_{V_2},\wtilde{c}_{U_1})
\in{\mathcal T}^\delta}
{\mathbb
P}\left((c_W,c_{V_1},c_{V_2},{c}_{U_1})=(\wtilde{c}_W,\wtilde{c}_{V_1},\wtilde{c}_{V_2},\wtilde{c}_{U_1})\right)\\
&\qquad\qquad\qquad\cdot{\mathbb P}\Bigg(
(c_W,c_{V_1},c_{V_2},{c}_{U_1},\what{c}_{Z_1},\what{c}_{S_1},\what{c}_{U_2},Y_3^T)
\in {\mathcal T}^\delta \Bigg|
(c_W,c_{V_1},c_{V_2},{c}_{U_1})=(\wtilde{c}_W,\wtilde{c}_{V_1},\wtilde{c}_{V_2},\wtilde{c}_{U_1}).
\Bigg)
\end{align*}
Further,
\begin{align*}
&{\mathbb P}\left(
(c_W,c_{V_1},c_{V_2},{c}_{U_1},\what{c}_{Z_1},\what{c}_{S_1},\what{c}_{U_2},Y_3^T)
\in {\mathcal T}^\delta \middle|
(c_W,c_{V_1},c_{V_2},{c}_{U_1})=(\wtilde{c}_W,\wtilde{c}_{V_1},\wtilde{c}_{V_2},\wtilde{c}_{U_1})\right)\\
&= \sum_{(\wtilde{c}_{Z_1},\wtilde{c}_{S_1},\wtilde{x}_{U_2},\wtilde{Y}_3^T)
 \in{\mathcal T}^\delta}
{\mathbb P}\left((\what{c}_{Z_1},\what{c}_{S_1},\what{c}_{U_2},{Y}_3^T)
= (\wtilde{c}_{Z_1},\wtilde{c}_{S_1},\wtilde{c}_{U_2},\wtilde{Y}_3^T) \middle|
(c_W,c_{V_1},c_{V_2},{c}_{U_1})=(\wtilde{c}_W,\wtilde{c}_{V_1},\wtilde{c}_{V_2},\wtilde{c}_{U_1})
\right),
\end{align*}
where ${\mathcal T}^\delta$ in the summation index is the set of
conditionally $\delta$-typical $Z_1,S_1,U_2,Y_3$ sequences conditioned on the
$(W,V_1,V_2,U_1)$-typical sequence
$(\wtilde{c}_W,\wtilde{c}_{V_1},\wtilde{c}_{V_2},\wtilde{c}_{U_1})$. Note
that the cardinality of this set is upperbounded by
$2^{T(H(Z_1,S_1,U_2,Y_3|W,V_1,V_2,U_1)+\delta)}$.
\begin{align*}
&{\mathbb P}\left((\what{c}_{Z_1},\what{c}_{S_1},\what{c}_{U_2},{Y}_3^T)
= (\wtilde{c}_{Z_1},\wtilde{c}_{S_1},\wtilde{c}_{U_2},\wtilde{Y}_3^T) \middle|
(c_W,c_{V_1},c_{V_2},{c}_{U_1})=(\wtilde{c}_W,\wtilde{c}_{V_1},\wtilde{c}_{V_2},\wtilde{c}_{U_1})\right)\\
&=
{\mathbb P}\left((\what{c}_{Z_1},\what{c}_{S_1},\what{c}_{U_2}) =
(\wtilde{c}_{Z_1},\wtilde{c}_{S_1},\wtilde{c}_{U_2}) \middle|
(c_W,c_{V_1},c_{V_2},{c}_{U_1}) =
(\wtilde{c}_W,\wtilde{c}_{V_1},\wtilde{c}_{V_2},\wtilde{c}_{U_1})
\right) \cdot \\ 
& \quad\quad \quad \quad {\mathbb P}\left({Y}_3^T=\wtilde{Y}_3^T \middle|
(c_W,c_{V_1},c_{V_2},{c}_{U_1}) =
(\wtilde{c}_W,\wtilde{c}_{V_1},\wtilde{c}_{V_2},\wtilde{c}_{U_1})
\right) \\
&\leq
2^{-T(H(Z_1,S_1,U_2|W,V_1,V_2,U_1)-\delta)}2^{-T(H(Y_3|W,V_1,V_2,U_1)-\delta)},
\end{align*}
where the first step follows from the independence of the hatted-codewords
and what destination~3 receives conditioned on the unhatted-codewords.
Combining everything, we can conclude that the probability of the
error-event under consideration is less than or equal to
\[2^{T(r_{U_2}+r_{S_1}+r_{Z_1}-I(U_2,S_1,Z_1;Y_3|W,V_1,V_2,U_1)
-3\epsilon+3\delta)},\]
which goes to zero as the blocklength $T$ goes to $\infty$ if we choose
$0<\delta<\epsilon$, and the rates satisfy the condition
\[ r_{U_2} + r_{S_1} + r_{Z_1} \leq I(U_2,S_1,Z_1;Y_3|W,V_1,V_2,U_1).\]
Overall, an error results if for at least one block-$j$, either encoding
fails or decoding fails. Since there are a finite number $J$ of blocks, by
union bound, the above discussion implies that the probability of error
goes to 0 as the blocklength goes to $\infty$ when the above conditions are
met. This completes the random coding argument.

\noindent{(c)} This scheme is a combination of the two schemes above. Now,
only destination~3 receives a private message sent cooperatively by the two
sources. Hence, only an $S_1$ codebook is present and there is no $S_2$
codebook. The $W,V$ and $U$ codebooks are exactly as in~(a) and (b). The
$S_1$ codebook is identical to scheme~(b). The $Z$ and $X$ codebooks are also
similar and differ only in that the distribution used to construct them has
no $S_2$, and there is no $m_{\PVTCOOP_2}$. Hence, we may set $Z_2=X_2$, and
set $X_2$ codeword to be identical to the $Z_2$ codeword.

The encoding at node~1 proceeds exactly as in scheme~(b) above except that,
since there is no $S_2$ codeword, node~1 need only decode the $V_2$-codeword.
Unlike in scheme~(b), node~1 treats all the other codewords from node~2 as
noise when decoding the $V_2$ codeword. Thus, the only condition imposed by
decoding at node~1 is \[ r_{V_2}\leq I(V_2;Y_1|W,S_1).\] Encoding at node~2 is
exactly as in scheme~(b) except for the fact that there is no $S_2$ codeword.
Decoding at the destination~3 is identical to that in scheme~(b) while that at
destination~4 is identical to that in scheme~(a).\\

\section{ Proof of achievability of Theorem~\ref{thm:sourcecoopLD}}
\label{app:LDachieve}

If we fix $n_{1,3}$, $n_{1,4}$, $n_{2,3}$, and $n_{2,4}$, and consider the
$u_i$'s in \eqref{eq:LDu1}-\eqref{eq:LDu4} as functions of $n_C$, the
sum-rate expression in Theorem~\ref{thm:sourcecoopLD} (as a function of
$n_C$) breaks up into four natural regimes. We use different strategies to
achieve the sum-capacity in different regimes. The regimes are:
\begin{enumerate}
\item[(i)] $n_C\leq
n_\text{min}\defineqq\min(n_{1,3},n_{1,4},n_{2,3},n_{2,4})$.
It can be shown that for $n_C\geq n_\text{min}$,
\[u_1(n_C) \geq \min(u_2(n_C), u_3(n_C), u_4(n_C), u_5).\]
Hence, we need consider $u_1$ only in the regime $n_C\leq
n_\text{min}$. Moreover, in this regime, $u_2(n_C)$ through $u_4(n_C)$ are
constants ({\em i.e.}, they do not depend on $n_C$ and their values are the
same as when $n_C=0$). Since $u_1(n_C)$ is monotonically increasing in
$n_C$, this means that we need to employ cooperation only when $u_1(0) <
\min(u_2(0),u_3(0),u_4(0),u_5)$, {\em i.e.}, when
\begin{align}
\max&(n_{1,3}-n_{1,4},n_{2,3})+\max(n_{2,4}-n_{1,4},n_{1,4})\notag\\ &\quad<
\min(\max(n_{1,3},n_{2,3}) + \left(\max(n_{2,4},n_{2,3})-n_{2,3}\right),
 \notag\\ &\qquad\qquad\quad
\max(n_{2,4},n_{1,4})+ \left(\max(n_{1,3},n_{1,4})-n_{1,4}\right),
n_{1,3}+n_{2,4}).\label{eq:LDu1cond}
\end{align}
When the above condition is not true, the sum-rate expression reduces to
the sum-capacity without cooperation. We show below how
Theorem~\ref{thm:sourcecoopgenericschemes}(a) implies that the sum-rate
expression is achievable in this region, both when cooperation is required
and not.

\item[(ii)] $n_\text{min}< n_C \leq \min(n_{1,3},n_{2,4})$. In this regime,
we can observe that the sum-rate expression takes on a constant value since
$u_2(n_C)$, $u_3(n_C)$, and $u_4(n_C)$ are still constants. Hence, the
achievability here is implied by the achievability in regime (i).

\item[(iii)] $\min(n_{1,3},n_{2,4})< n_C \leq \max(n_{1,3},n_{2,4})$. In
this regime, we use Theorem~\ref{thm:sourcecoopgenericschemes}(c).

\item[(iv)] $\max(n_{1,3},n_{2,4}) < n_C$. The sum-capacity is achieved in this
regime by applying Theorem~\ref{thm:sourcecoopgenericschemes}(b).

\end{enumerate}
For integer $q$ satisfying $1\leq q\leq n$, we define
\begin{align*}
{\mathcal F}_{q}\stackrel{\text{def}}{=}\left\{ x\in
{\mathbb F}^n: x_i=0,\; i\leq q\right\},
\end{align*}
{\em i.e.}, all vectors in ${\mathbb F}^q$ such that their components
in the range $1,\ldots,q$ are zeros. We take the indexing of the
elements of vectors to start from the top as usual. For example, for binary field
and $n=4$,
\[{\mathcal F}_2=\left\{
\left[\begin{array}{c}0\\0\\0\\0\end{array}\right],
\left[\begin{array}{c}0\\0\\0\\1\end{array}\right],
\left[\begin{array}{c}0\\0\\1\\0\end{array}\right],
\left[\begin{array}{c}0\\0\\1\\1\end{array}\right]\right\}.\]

\noindent{\em Regime (i):} When the condition \eqref{eq:LDu1cond} holds, we
consider a restricted regime of $n_C$ where
\[ u_1(n_C) \leq \min(u_2(n_C),u_3(n_C),u_4(n_C),u_5).\]
Since $u_1(n_C)$ is monotonic in $n_C$, it is enough to prove achievability
in this regime to obtain achievability in regime~(i) when
\eqref{eq:LDu1cond} holds. We make the following choices for the auxiliary
random variables in Theorem~\ref{thm:sourcecoopgenericschemes}(a):
$W,  V_1,U_1,Z_1,V_2,U_2,Z_2$ are
independent of each other and uniformly distributed over their alphabets
which are as follows -- $V_1,V_2\in {\mathbb F}^n$, $U_1,U_2\in {\mathcal
F}_{n_C}$, $Z_1\in {\mathcal F}_{n_{1,4}}$, and $Z_2\in {\mathcal
F}_{n_{2,3}}$. $W$ is independent of all these and has the same cardinality
as $(V_1,V_2)$. $X_1$ and $X_2$ are defined as
\begin{align*}
X_1&=V_1+U_1+Z_1,\\
X_2&=V_2+U_2+Z_2.
\end{align*}
This defines $p_Wp_{V_1,U_1,X_1|W}p_{V_2,U_2,X_2|W}$. These choices are
such that destination~4's observation does not depend on the ``private''
signal $Z_1$, and, similarly, destination~3's observation does not depend
on $Z_2$. With these choices, the conditions on the non-negative rates
$r_{V_1},r_{V_2},r_{U_1},r_{U_2},r_{Z_1},r_{Z-2}$ after removing redundant
conditions are
\begin{align*}
r_{V_1} &\leq n_C\\
r_{Z_1} &\leq [n_{1,3}-n_{1,4}]_+,\\
r_{U_1}+r_{Z_1} &\leq n_{1,3}-n_C,\\
r_{U_2}+r_{Z_1} &\leq \max(n_{1,3}-n_{1,4},n_{2,3}-n_C),\\
r_{U_1}+r_{U_2}+r_{Z_1} &\leq \max(n_{1,3}-n_C,n_{2,3}-n_C),\\
r_{V_1}+r_{V_2}+r_{U_1}+r_{U_2}+r_{Z_1} &\leq \max(n_{1,3},n_{2,3}),
\end{align*}
and the corresponding inequalities with subscripts 1 and 2 exchanged, and 3
and 4 exchanged, where $[x]_+=\max(x,0)$. Further, we make the following
choices for the rates.
\begin{align*}
r_{Z_1}&=[n_{1,3}-n_{1,4}]_+,\\
r_{Z_2}&=[n_{2,4}-n_{2,3}]_+,\\
r_{V_1}=r_{V_2}&=n_C,\\
r_{U_1}&=\max(n_{2,4}-n_{2,3},n_{1,4}-n_C)-[n_{2,4}-n_{2,3}]_+,\text{ and}\\
r_{U_2}&=\max(n_{1,3}-n_{1,4},n_{2,3}-n_C)-[n_{1,3}-n_{1,4}]_+.
\end{align*}
It can be shown that under the restricted regime
of $n_C$, these choices satisfy all the conditions above. The resulting
sum-rate is $u_1(n_C)$ as required.

When condition \eqref{eq:LDu1cond} does not hold, as we mentioned earlier,
it is enough to prove that the sum-rate at $n_C=0$ is achievable. We apply
Theorem~\ref{thm:sourcecoopgenericschemes}(a) where we set $W,V_1,V_2$ to
be constants, and $U_1,U_2,Z_1,Z_2$ to be independent and uniformly
distributed over their alphabets. $U_1,U_2\in {\mathbb F}^{n}$, $Z_1\in
{\mathcal F}_{n_{1,4}}$, $Z_2\in {\mathcal F}_{n_{2,3}}$, and $X_1$ and
$X_2$ are defined as
\begin{align*}
X_1&=V_1+U_1+Z_1,\\
X_2&=V_2+U_2+Z_2.
\end{align*}
The conditions on the non-negative rates
$r_{U_1},r_{U_2},r_{Z_1},r_{Z-2}$ after removing redundant conditions are
\begin{align*}
r_{Z_1} &\leq [n_{1,3}-n_{1,4}]_+,\\
r_{U_1}+r_{Z_1} &\leq n_{1,3},\\
r_{U_2}+r_{Z_1} &\leq \max(n_{1,3}-n_{1,4},n_{2,3}),\\
r_{U_1}+r_{U_2}+r_{Z_1} &\leq \max(n_{1,3},n_{2,3}),\\
\end{align*}
and the corresponding inequalities with subscripts 1 and 2 exchanged, and 3
and 4 exchanged. Applying Fourier-Motzkin elimination, we can show that a
sum-rate of\\ $\min(u_1(0),u_2(0),u_3(0),u_4(0),u_5)$ is achievable.

\noindent{\em Regime (iii):} Without loss of generality, let us assume that
$n_{1,3}\leq n_C \leq n_{2,4}$. We will apply
Theorem~\ref{thm:sourcecoopgenericschemes}(c) in two different ways to show
achievability in this regime. The first application covers (1) $n_{2,4}
\geq n_{1,4}$, while the second covers (2) $n_{2,4} < n_{1,4}$.

For case~(1), $n_{2,4} \geq n_{1,4}$, we use the following choices for the
auxiliary random variables: $U_1$ is a constant. $U_2,V_1,V_2,Z_1,Z_2,
\wtilde{S}_{1,3}$, $S_{2,3}^\perp$, and $S'_1$ are chosen to be independent
and uniformly distributed over their alphabets. The alphabets are: $V_1,V_2
\in {\mathbb F}^n$, $Z_1\in {\mathcal F}_{n_{1,4}}$, $Z_2\in {\mathcal
F}_{n_{2,3}}$, $\wtilde{S}_{1,3} \in {\mathbb F}^n$, $S'_1 \in {\mathcal
F}_{\max(n_{1,3},n_{1,4})}$, $S_{2,3}^\perp \in {\mathcal F}_{n_{2,4}}$ and
$U_2\in {\mathcal F}_{n'_C}$, where $n'_C$ $(n'_C \leq n_C)$ is to be
specified.  $W$ is independent of all these and has the same cardinality as
$(V_1,V_2)$. We define $X_1$ and $X_2$ as follows
\begin{align*}
X_1&=V_1+U_1+Z_1+\wtilde{S}_{1,3}+S'_1,\\
X_2&=V_2+U_2+Z_2-{\bf S}^{n_{2,4}-n_{1,4}}\wtilde{S}_{1,3}+S_{2,3}^\perp
\end{align*}
with the result that
\begin{align*}
Y_3&={\bf S}^{n-n_{1,3}}(V_1+Z_1) + {\bf S}^{n-n_{2,3}}(V_2+U_2) + S_1,\\
Y_4&={\bf S}^{n-n_{2,4}}(V_2+U_2+Z_2) + {\bf S}^{n-n_{2,3}}(V_1),
\end{align*}
where we defined
\[ S_1 \defineqq ({\bf S}^{n-n_{1,3}}-{\bf
S}^{n-n_{2,3}+n_{2,4}-n_{1,4}})\wtilde{S}_{1,3} + {\bf
S}^{n-n_{2,3}}S_{2,3}^\perp.\]
We set $r_{U_1}=0$. The conditions on the other rates (after removing the
redundant ones) are
\begin{align*}
r_{S_1}&\leq [n_C-\max(n_{1,3},n_{1,4})]_+,\\
r_{V_1}+r_{U_1}+r_{Z_1}+r_{S_1}&\leq n_C,\\
\\
r_{V_2}&\leq n'_C,\\
\\
r_{Z_1}&\leq [n_{1,3}-n_{1,4}]_+,\\
r_{U_2}+r_{Z_1}&\leq \max([n_{1,3}-n_{1,4}]_+,n_{2,3}-n'_C),\\
r_{S_1}+r_{Z_1}&\leq \left\{\begin{array}{ll}
\max([n_{1,3}-n_{1,4}]_+,n_{2,3}-(n_{2,4}-n_{1,4})),
&\text{ if } n_{1,3}+n_{2,4}\neq n_{1,4}+n_{2,3}\\
\max([n_{1,3}-n_{1,4}]_+,n_{2,3}-n_{2,4}),
&\text{ otherwise},
\end{array}\right.
\end{align*}
\begin{align*}
r_{U_2}+r_{S_1}+r_{Z_1}&\leq
\left\{\begin{array}{ll}
\max([n_{1,3}-n_{1,4}]_+,n_{2,3}-n'_C,n_{2,3}-(n_{2,4}-n_{1,4})),
&\text{ if } n_{1,3}+n_{2,4}\neq n_{1,4}+n_{2,3}\\
\max([n_{1,3}-n_{1,4}]_+,n_{2,3}-n'_C,n_{2,3}-n_{2,4}),
&\text{ otherwise},
\end{array}\right.
\end{align*}
\begin{align*}
(r_{V_1}+r_{V_2})+r_{U_2}+r_{S_1}+r_{Z_1}
&\leq \max(n_{1,3},n_{2,3}),\\
\\
r_{Z_2}&\leq [n_{2,4}-n_{2,3}]_+,\\
r_{U_2}+r_{Z_2}&\leq \max([n_{2,4}-n_{2,3}]_+,n_{2,4}-n'_C),\\
(r_{V_1}+r_{V_2})+r_{U_2}+r_{Z_2} &\leq \max(n_{2,4},n_{1,4}).
\end{align*}
By Fourier-Motzkin elimination, we may conclude that an achievable
$R_1+R_2$ is given by the smaller of $u_2(n_C),u_3(n_C),u_4(n_C),u_5$ and
\[ \frac{[n_C-\max(n_{1,3},n_{1,4})]_+ + \max(n_{2,4},n_{1,4}) +
\max([n_{1,3}-n_{1,4}]_+,n_{2,3}-n'_C) + n_C + n'_C +
[n_{2,4}-n_{2,3}]_+}{2}.\]
The last term above can be shown to be not smaller than the minimum of
$u_2(n_C),u_3(n_C),u_4(n_C)$ and $u_5$ if $n'_C$ is chosen to be such that
$u_1(n'_C) = \min(u_2(n_C),u_3(n_C),u_4(n_C),u_5)$ if
$u_1(0)<\min(u_2(0),u_3(0),u_4(0),u_5)$, and $n'_C=0$
otherwise. Note that from earlier discussion, we know that this choice of
$n'_C$ must be less than or equal to $n_C$.

For (2)~$n_{2,4} < n_{1,4}$, we apply
Theorem~\ref{thm:sourcecoopgenericschemes}(c) with the same auxiliary
random variables $W,U_1,U_2,V_1,V_2,Z_1,Z_2,S_{2,3}^\perp,S'_1$ as in case~(1)
above. But instead of $\wtilde{S}_{1,3}$, we define $\wtilde{S}_{2,3}$
which is independent of all these random variables and distributed
uniformly over ${\mathbb F}^n$. We define $X_1$, and $X_2$ as follows
\begin{align*}
X_1&=V_1+Z_1+{\bf S}^{n_{1,4}-n_{2,4}}\wtilde{S}_{2,3}+S'_1,\\
X_2&=V_2+U_2+Z_2-\wtilde{S}_{2,3}+S_{2,3}^\perp
\end{align*}
with the result that
\begin{align*}
Y_3&={\bf S}^{n-n_{1,3}}(V_1+Z_1) + {\bf S}^{n-n_{2,3}}(V_2+U_2) + S_1,\\
Y_4&={\bf S}^{n-n_{2,4}}(V_2+U_2+Z_2) + {\bf S}^{n-n_{1,4}}(V_1),
\end{align*}
where we define $S_1$ as
\[ S_1 \defineqq ({\bf S}^{n-n_{1,3}+n_{1,4}-n_{2,4}} - {\bf
S}^{n-n_{2,3}})\wtilde{S}_{2,3} + {\bf S}^{n-n_{2,3}}S_{2,3}^\perp.\]
The conditions on the rates (after removing the redundant ones) are the
same as in case~(1) except for the following two
\begin{align*}
r_{S_1}+r_{Z_1}&\leq \left\{\begin{array}{ll}
\max([n_{1,3}-n_{1,4}]_+,n_{1,3}-(n_{1,4}-n_{2,4}),n_{2,3}-n_{2,4}),
&\text{ if } n_{1,3}+n_{2,4}\neq n_{1,4}+n_{2,3},\\
\max([n_{1,3}-n_{1,4}]_+,n_{2,3}-n_{2,4}),
&\text{ otherwise},
\end{array}\right.
\end{align*}
\begin{align*}
&r_{U_2}+r_{S_1}+r_{Z_1}\\&\leq
\left\{\begin{array}{ll}
\max([n_{1,3}-n_{1,4}]_+,n_{2,3}-n'_C,n_{1,3}-(n_{1,4}-n_{2,4}),n_{2,3}-n_{2,4}),
&\text{ if } n_{1,3}+n_{2,4}\neq n_{1,4}+n_{2,3},\\
\max([n_{1,3}-n_{1,4}]_+,n_{2,3}-n'_C,n_{2,3}-n_{2,4}),
&\text{ otherwise}.
\end{array}\right.
\end{align*}
By Fourier-Motzkin elimination, we may conclude that the achievable
$R_1+R_2$ is given by the smaller of $u_2(n_C),u_3(n_C),u_4(n_C),u_5$ and
\[ \frac{[n_C-\max(n_{1,3},n_{1,4})]_+ + \max(n_{2,4},n_{1,4}) +
\max([n_{1,3}-n_{1,4}]_+,n_{2,3}-n'_C) + n_C + n'_C +
[n_{2,4}-n_{2,3}]_+}{2}.\]
The above term can be shown to be not smaller than the minimum of
$u_2(n_C),u_3(n_C),u_4(n_C)$ and $u_5$ if $n'_C$ is chosen as was done in
case~(1) above.

We can represent the two cases above together using the following notation:
\begin{align*}
X_1&=V_1+Z_1+\bar{S}_{1,3}+S'_1,\\
X_2&=V_2+U_2+Z_2+\bar{S}_{2,3},
\end{align*}
where
\begin{align*}
\bar{S}_{1,3}&=\left\{\begin{array}{ll}
\wtilde{S}_{1,3},\qquad\qquad  &\text{ if } n_{2,4}\geq n_{1,4},\\
{\bf S}^{n_{1,4}-n_{2,4}}\wtilde{S}_{2,3},\qquad\qquad
 &\text{ otherwise},
\end{array}\right.\\
\bar{S}_{2,3}&=\left\{\begin{array}{ll}
-{\bf S}^{n_{2,4}-n_{1,4}}\wtilde{S}_{1,3}+S_{2,3}^\perp,
&\text{ if } n_{2,4}\geq n_{1,4},\\
-\wtilde{S}_{2,3}+S_{2,3}^\perp,
&\text{ otherwise},
\end{array}\right.
\end{align*}
with the result that
\begin{align*}
Y_3&={\bf S}^{n-n_{1,3}}(V_1+Z_1) + {\bf S}^{n-n_{2,3}}(V_2+U_2) + S_1,\\
Y_4&={\bf S}^{n-n_{2,4}}(V_2+U_2+Z_2) + {\bf S}^{n-n_{1,4}}(V_1),
\end{align*}
where
\[ S_1 \defineqq \left\{\begin{array}{ll}
({\bf S}^{n-n_{1,3}}-{\bf S}^{n-n_{2,3}+n_{2,4}-n_{1,4}})\wtilde{S}_{1,3}
+{\bf S}^{n-n_{2,3}}S_{2,3}^\perp,&\text{ if }n_{2,4}\geq n_{1,4},\\
({\bf S}^{n-n_{1,3}+n_{1,4}-n_{2,4}} - {\bf S}^{n-n_{2,3}})\wtilde{S}_{2,3}
+{\bf S}^{n-n_{2,3}}S_{2,3}^\perp, &\text{ otherwise}.
\end{array}\right.
\]
We will use a symmetric form of this notation below for regime~(iv).

\noindent{\em Regime (iv):} Application of
Theorem~\ref{thm:sourcecoopgenericschemes}(b) with the following auxiliary
random variables covers this regime: $U_1,U_2$ are constants.
$V_1,V_2,Z_1,Z_2,\wtilde{S}_{1,3},\wtilde{S}_{2,3},\wtilde{S}_{1,4},
\wtilde{S}_{2,4},S_{2,3}^\perp, S_{1,4}^\perp, S'_1, S'_2$ are
chosen to be independent and uniformly distributed over their alphabets. Also,
we choose the alphabets to be $V_1,V_2,\wtilde{S}_{1,3},\wtilde{S}_{2,3},\wtilde{S}_{1,4},
\wtilde{S}_{2,4}\in
{\mathbb F}^n$, $Z_1\in {\mathcal F}_{n_{1,4}}$, $Z_2\in {\mathcal
F}_{n_{2,3}}$, $S_{2,3}^\perp \in {\mathcal F}_{n_{2,4}}$,
$S_{1,4}^\perp \in {\mathcal F}_{n_{1,3}}$,  $S'_1 \in {\mathcal
F}_{\max(n_{1,3},n_{1,4})}$, and $S'_2 \in {\mathcal F}_{\max(n_{2,4},n_{2,3})}$. $W$ is independent of all these and has the same cardinality as
$(V_1,V_2)$. Further, we define $X_1$, and $X_2$ as follows
\begin{align*}
X_1&=V_1+U_1+Z_1+\bar{S}_{1,3} + \bar{S}_{1,4} + S_1'\\
X_2&=V_2+U_2+Z_2+\bar{S}_{2,3} + \bar{S}_{2,4} + S_2',
\end{align*}
where
\begin{align*}
\bar{S}_{1,3}&=\left\{\begin{array}{ll}
\wtilde{S}_{1,3},\qquad\qquad  &\text{ if } n_{2,4}\geq n_{1,4},\\
{\bf S}^{n_{1,4}-n_{2,4}}\wtilde{S}_{2,3},\qquad\qquad
 &\text{ otherwise},
\end{array}\right.\\
\bar{S}_{2,3}&=\left\{\begin{array}{ll}
-{\bf S}^{n_{2,4}-n_{1,4}}\wtilde{S}_{1,3}+S_{2,3}^\perp,
&\text{ if } n_{2,4}\geq n_{1,4},\\
-\wtilde{S}_{2,3}+S_{2,3}^\perp,
&\text{ otherwise},
\end{array}\right.\\
\bar{S}_{2,4}&=\left\{\begin{array}{ll}
\wtilde{S}_{2,4},\qquad\qquad  &\text{ if } n_{1,3}\geq n_{2,3},\\
{\bf S}^{n_{2,3}-n_{1,3}}\wtilde{S}_{1,4},\qquad\qquad
 &\text{ otherwise},
\end{array}\right.\\
\bar{S}_{1,4}&=\left\{\begin{array}{ll}
-{\bf S}^{n_{1,3}-n_{2,3}}\wtilde{S}_{2,4}+S_{1,4}^\perp,
&\text{ if } n_{1,3}\geq n_{2,3},\\
-\wtilde{S}_{1,4}+S_{1,4}^\perp,
&\text{ otherwise}.
\end{array}\right.
\end{align*}
The upshot of this is that
\begin{align*}
Y_3&={\bf S}^{n-n_{1,3}}(V_1+Z_1) + {\bf S}^{n-n_{2,3}}V_2 + S_1,\text{
and}\\
Y_4&={\bf S}^{n-n_{2,4}}(V_2+Z_2) + {\bf S}^{n-n_{2,3}}V_1 + S_2,
\end{align*}
where
\begin{align*}
S_1 &\defineqq \left\{\begin{array}{ll}
({\bf S}^{n-n_{1,3}}-{\bf S}^{n-n_{2,3}+n_{2,4}-n_{1,4}})\wtilde{S}_{1,3}
+{\bf S}^{n-n_{2,3}}S_{2,3}^\perp,&\text{ if }n_{2,4}\geq n_{1,4},\\
({\bf S}^{n-n_{1,3}+n_{1,4}-n_{2,4}} - {\bf S}^{n-n_{2,3}})\wtilde{S}_{2,3}
+{\bf S}^{n-n_{2,3}}S_{2,3}^\perp, &\text{ otherwise},
\end{array}\right.\\
S_2 &\defineqq \left\{\begin{array}{ll}
({\bf S}^{n-n_{2,4}}-{\bf S}^{n-n_{1,4}+n_{1,3}-n_{2,3}})\wtilde{S}_{2,4}
+{\bf S}^{n-n_{1,4}}S_{1,4}^\perp,&\text{ if }n_{1,3}\geq n_{2,3},\\
({\bf S}^{n-n_{2,4}+n_{2,3}-n_{1,3}} - {\bf S}^{n-n_{1,4}})\wtilde{S}_{1,4}
+{\bf S}^{n-n_{1,4}}S_{1,4}^\perp, &\text{ otherwise}.
\end{array}\right.
\end{align*}
With these choices, the conditions on the non-negative rates
$r_{V_1},r_{V_2},r_{S_1},r_{S_2},r_{Z_1},r_{Z_2}$ after removing redundant
conditions are
\begin{align*}
r_{S_1}&\leq [n_C-\max(n_{1,3},n_{1,4})]_+,\\
r_{V_1}+r_{Z_1}+r_{S_1}&\leq n_C,\\
\\
r_{Z_1}&\leq [n_{1,3}-n_{1,4}]_+,\\
r_{S_1}+r_{Z_1}&\leq \max([n_{1,3}-n_{1,4}]_+,n_{S_1}),\\
(r_{V_1}+r_{V_2})+r_{S_1}+r_{Z_1}
&\leq \max(n_{1,3},n_{2,3}),
\end{align*}
and the corresponding inequalities with subscripts 1 and 2 exchanged,
and 3 replaced by 4, where
\begin{align*} n_{S_1}=\left\{\begin{array}{ll}
\max(n_{1,3},n_{2,3}-(n_{2,4}-n_{1,4})),
&\text{ if } n_{2,4} \geq n_{1,4} \text{ and }
n_{1,3}+n_{2,4}\neq n_{1,4}+n_{2,3},\\
\max(n_{2,3},n_{1,3}-(n_{1,4}-n_{2,4})),
&\text{ if } n_{2,4} < n_{1,4} \text{ and }
n_{1,3}+n_{2,4}\neq n_{1,4}+n_{2,3},\\
 (n_{2,3}-n_{2,4})_{+},
&\text{ if }
n_{1,3}+n_{2,4} = n_{1,4}+n_{2,3}.
\end{array}\right.
\end{align*}
We may apply Fourier-Motzkin elimination to obtain the sum-rate supported
by this scheme. We get a sum-rate which is the minimum of
$u_2(n_C),u_3(n_C),u_4(n_C),u_5,$. This completes the achievability proof.

\section{Proof of achievability of Theorem~\ref{thm:sourcecoopG}}
\label{app:Gachieve}

We prove Theorem~\ref{thm:sourcecoopG} using
Theorem~\ref{thm:sourcecoopgenericschemes}. Note that we proved the latter
for discrete alphabets, but the extension to the continuous alphabet case
is standard and we will assume that version in this section.  This proof
will follow the proof of Theorem~\ref{thm:sourcecoopLD} closely. We first
make the following definitions:
\begin{align*}
n_{k_1,k_2}&\defineqq [\log|h_{k_1,k_2}|^2]_+,\;k_1\in\{ 1,2\},\,
\,k_2\in\{3,4\},\text{ and}\\
n_C &\defineqq [\log|h_C|^2]_+.
\end{align*}
First, we observe that the following four terms $u'_1, u'_2, u'_3$, and $u'_4$
are within a constant (7 bits) of the corresponding unprimed terms,
$u_1,u_2,u_3$, and $u_4$, respectively
\begin{align}
u'_1&=\max(n_{1,3}-n_{1,4}+n_C,n_{2,3},n_C) +
\max(n_{2,4}-n_{2,3}+n_C,n_{1,4},n_C),\label{eq:u'1}\\
u'_2&=
\max(n_{1,3},n_{2,3}) + \left(\max(n_{2,4},n_{2,3},n_C)-n_{2,3}\right),
\label{eq:u'2}\\
u'_3&=
\max(n_{2,4},n_{1,4}) + \left(\max(n_{1,3},n_{1,4},n_C)-n_{1,4}\right),
\label{eq:u'3}\\
u'_4&=\max(n_{1,3},n_C)+\max(n_{2,4},n_C).\label{eq:u'4}
\end{align}
Hence, it is enough to show that the minimum of the four terms above and
\begin{align}
u'_5 = \log\bigg( 1 &+
 \left(|h_{1,3}|^2+|h_{2,4}|^2+|h_{1,4}|^2+|h_{2,3}|^2\right)\notag\\
  &+ \left( |h_{1,3}h_{2,4}|^2 + |h_{1,4}h_{2,3}|^2 -
2|h_{1,3}h_{2,4}h_{1,4}h_{2,3}|\cos\theta\right) \bigg), \label{eq:u'5}
\end{align}
which is within a constant (2 bits) of $u_5$, is achievable. We again
consider the same four regimes as in the proof of
Theorem~\ref{thm:sourcecoopLD}:

\noindent{\em Regime (i):} $n_C\leq n_{\text{min}} \defineqq
\min(n_{1,3},n_{1,4},n_{2,3},n_{2,4}).$ The discussion for regime~(i) in
the linear deterministic case continues to hold here as well. Note that
$u_5$ is such that
\[ u_5 - 5 \leq u''_5 \defineqq \max(n_{1,3}+n_{2,4},n_{1,4}+n_{2,3}).\]
Thus, when condition \eqref{eq:LDu1cond} does not hold, the achievability
(within a gap of 9 bits from the upperbound) is implied by the results of
Etkin-Tse-Wang~\cite{EtkinTseWang08} (where 2-bit gap comes
from~\cite{EtkinTseWang08} and an additional 7 bits were incurred above).
And, when condition \eqref{eq:LDu1cond} holds, we need only show
achievability in the restricted regime of $n_C$ where
\[ u'_1(n_C) \leq \min(u'_2(n_C),u'_3(n_C),u'_4(n_C),u''_5).\]
We employ Theorem~\ref{thm:sourcecoopgenericschemes}(a) using the following
auxiliary random variables $W,V_1,U_1,Z_1,V_2,U_2,Z_2$ are zero-mean
Gaussian random variables and independent of each other with the following
variances:
\begin{align*}
\sigma_{V_1}^2=\sigma_{V_2}^2&=1/K,\\
\sigma_{U_1}^2=\sigma_{U_2}^2&=\frac{1/K}{\max(1,|h_C|)},\\
\sigma_{Z_1}^2&=\frac{1/K}{\max(1,|h_{1,4}|^2)},\text{ and}\\
\sigma_{Z_2}^2&=\frac{1/K}{\max(1,|h_{2,3}|^2)},
\end{align*}
where $K$ is a constant which will be specified soon. $W$ is independent of
all these and has the same distribution as $(V_1,V_2)$. $X_1$ and $X_2$ are
defined as
\begin{align*}
X_1&=V_1+U_1+Z_1,\\
X_2&=V_2+U_2+Z_2.
\end{align*}
In order for the power constraint to be satisfied, it is enough to have
$K<3$. This defines $p_Wp_{V_1,U_1,X_1|W}p_{V_2,U_2,X_2|W}$. These choices
are such that the ``private'' signal $Z_1$ appears at destination~4 with
less power than the noise, and, similarly, $Z_2$ appears at destination~3
with less power than the noise. With these choices, the conditions on the
non-negative rates $r_{V_1},r_{V_2},r_{U_1},r_{U_2},r_{Z_1},r_{Z_2}$ are
\begin{align*}
r_{V_1} &\leq \log\left(1+\frac{|h_C|^2/K}{2/K+1}\right)\\
r_{Z_1} &\leq \log\left(1+\frac{|h_{1,3}|^2/(\max(1,|h_{1,4}|^2)K)}{1/K+1}\right),\\
r_{U_1}+r_{Z_1} &\leq \log\left(1+\frac{|h_{1,3}|^2/(\max(1,|h_C|^2)K) +
|h_{1,3}|^2/(\max(1,|h_{1,4}|^2)K)}{1/K+1}\right),\\
r_{U_2}+r_{Z_1} &\leq \log\left(1+\frac{|h_{1,3}|^2/(\max(1,|h_{1,4}|^2)K) +
|h_{2,3}|^2/(\max(1,|h_C|^2)K)}{1/K+1}\right),\\
r_{U_1}+r_{U_2}+r_{Z_1} &\leq \log\left(1 +
\frac{\frac{|h_{1,3}|^2}{\max(1,|h_C|^2)K} +
 \frac{|h_{1,3}|^2}{\max(1,|h_{1,4}|^2)K} +
 \frac{|h_{2,3}|^2}{\max(1,|h_C|^2)K}}{1/K + 1}\right),\\
r_{V_1}+r_{V_2}+r_{U_1}+r_{U_2}+r_{Z_1} &\leq \log\left(1 + \frac{
\frac{|h_{1,3}|^2}{K} + \frac{|h_{1,3}|^2}{\max(1,|h_C|^2)K} +
\frac{|h_{1,3}|^2}{\max(1,|h_{1,4}|^2)K} +
\frac{|h_{2,3}|^2}{K} + \frac{|h_{2,3}|^2}{\max(1,|h_C|^2)K}}{1/K + 1}\right),
\end{align*}
Simplifying, we can show that these conditions imply that non-negative
rates which satisfy the same conditions as in the linear deterministic case
(up to a constant) are achievable.
\begin{align*}
r_{V_1} &\leq n_C - \log 5\\
r_{Z_1} &\leq [n_{1,3}-n_{1,4}]_+ - \log 4,\\
r_{U_1}+r_{Z_1} &\leq n_{1,3}-n_C - \log 4,\\
r_{U_2}+r_{Z_1} &\leq \max(n_{1,3}-n_{1,4},n_{2,3}-n_C) - \log 4,\\
r_{U_1}+r_{U_2}+r_{Z_1} &\leq \max(n_{1,3}-n_C,n_{2,3}-n_C) - \log 4,\\
r_{V_1}+r_{V_2}+r_{U_1}+r_{U_2}+r_{Z_1} &\leq \max(n_{1,3},n_{2,3}) - \log 4,
\end{align*}
and the corresponding inequalities with subscripts 1 and 2 exchanged, and 3
and 4 exchanged. Note that the right hand sides above should be
interpreted as zero if they evaluate to less than zero. We will tacitly
assume this for similar conditions in the sequel. Further, we make
the following choices for the rates.
\begin{align*}
r_{Z_1}&=[n_{1,3}-n_{1,4}]_+ - \log 4,\\
r_{Z_2}&=[n_{2,4}-n_{2,3}]_+ - \log 4,\\
r_{V_1}=r_{V_2}&=n_C -\log 5,\\
r_{U_1}&=\max(n_{2,4}-n_{2,3},n_{1,4}-n_C)-[n_{2,4}-n_{2,3}]_+ - \log 4,
\text{ and}\\
r_{U_2}&=\max(n_{1,3}-n_{1,4},n_{2,3}-n_C)-[n_{1,3}-n_{1,4}]_+ - \log 4,
\end{align*}
where we interpret the rates as zero if their values work out to less than
zero. It can be shown that under the restricted regime of $n_C$, these
choices satisfy all the conditions above. The resulting sum-rate is
$u'_1(n_C)$ within a constant gap (of at most 13 bits) as required.

When condition \eqref{eq:LDu1cond} does not hold, as we mentioned earlier,
it is enough to prove that the sum-rate at $n_C=0$ is achievable. We may
invoke the achievability proof of Etkin-Tse-Wang~\cite{EtkinTseWang08} to
conclude that a sum-rate which matches our upperbounds (up to a constant)
is achievable. Thus, overall, in regime~(i), we may conclude that the
upperbound is achievable within a constant gap of 20 bits.

\noindent{\em Regime (ii):} $n_{\text{min}} < n_C \leq
\min(n_{1,3},n_{2,4}).$ As in the linear deterministic case, the
achievability in this regime is implied by the achievability in regime~(i).

\noindent{\em Regime (iii):} $\min(n_{1,3},n_{2,4}) < n_C \leq
\max(n_{1,3},n_{2,4}).$ Without loss of generality, let us assume that
$n_{1,3}\leq n_C \leq n_{2,4}$. We will apply
Theorem~\ref{thm:sourcecoopgenericschemes}(c) to establish the
achievability. We consider two separate possibilities: (1)~$|h_{2,4}| \geq
|h_{1,4}|$ and (2)~$|h_{2,4}| < |h_{1,4}|$.

\noindent(1) When $|h_{2,4}| \geq |h_{1,4}|$ (which implies that $n_{2,4}
\geq n_{1,4}$), the auxiliary random variables are as follows: $U_1$ is set
to a constant. $U_2,V_1,V_2,Z_1,Z_2, \wtilde{S}_{1,3}, S_{2,3}^\perp$ and
$S'_1$ are independent zero-mean Gaussian random variables. Their variances
are as follows
\begin{align*}
\sigma_{V_1}^2=\sigma_{V_2}^2&=1/K,\\
\sigma_{U_2}^2&=\frac{1/K}{\max(1,|h'_C|^2)},\\
\sigma_{Z_1}^2&=\frac{1/K}{\max(1,|h_{1,4}|^2)},\\
\sigma_{Z_2}^2&=\frac{1/K}{\max(1,|h_{2,3}|^2)},\\
\sigma_{\wtilde{S}_{1,3}}^2&=1/K,\\
\sigma_{S'_1}^2 &= \frac{1/K}{\max(1,|h_{1,3}|^2,|h_{1,4}|^2)},\\
\sigma_{S_{2,3}^\perp}^2&=\frac{1/K}{\max(1,|h_{2,4}|^2)},
\end{align*}
where $K$ and $h'_C$ $(0 < h'_C)$ will be specified. Let us define
$n'_C\defineqq [\log {h'}_C^2]_+$. We will pick a $h'_C$ such that $n'_C
\leq \min(n_C,n_{2,3})$. We set $W$ to be independent of
all these and have the same distribution as $(V_1,V_2)$.
We define $X_1$, and $X_2$ as follows
\begin{align*}
X_1&=V_1+Z_1+\wtilde{S}_{1,3}+S'_1,\\
X_2&=V_2+U_2+Z_2-\frac{|h_{1,4}|e^{j\theta/2}}{|h_{2,4}|}\wtilde{S}_{1,3}+S_{2,3}^\perp.
\end{align*}
This satisfies the power constraint if $K<5$ (where we used the fact that
$|h_{2,4}| \geq |h_{1,4}|$). Let $S_1=(\wtilde{S}_{1,3},S_{2,3}^\perp)$ and
\[\bar{S}_1=\left(|h_{1,3}|
   - \frac{|h_{1,4}||h_{2,3}|e^{j\theta}}{|h_{2,4}|}\right)\wtilde{S}_{1,3}
 + |h_{2,3}|e^{j\theta/2}S_{2,3}^\perp.\]
Then
\begin{align*}
Y_3&=|h_{1,3}|(V_1+Z_1+S'_1) + \bar{S}_1 + N_3,
\\
Y_4&=|h_{2,4}|(V_2+U_2+Z_2) + |h_{1,4}|e^{j\theta/2}(V_1+Z_1+S'_1) +
|h_{2,4}|S_{2,3}^\perp +N_4.
\end{align*}
The conditions on the rates are
\begin{align*}
r_{S_1}&\leq \log\left( 1 + {|h_C|^2/(\max(1,|h_{1,3}|^2,|h_{1,4}|^2)K)}\right),\\
r_{Z_1}+r_{S_1} &\leq \log\left( 1 +
\frac{|h_C|^2}{\max(1,|h_{1,3}|^2,|h_{1,4}|^2)K} +
\frac{|h_C|^2}{\max(1,|h_{1,4}|^2)K}\right),\\
r_{V_1}+r_{Z_1}+r_{S_1}&\leq \log\left( 1 + \frac{|h_C|^2}{K} +
\frac{|h_C|^2}{\max(1,|h_{1,3}|^2,|h_{1,4}|^2)K} +
\frac{|h_C|^2}{\max(1,|h_{1,4}|^2)K} \right),\\
\\
r_{V_2}&\leq \log\left( 1 + \frac{|h_C|^2/K}{|h_C|^2/(\max(1,|{h'_C}|^2)K) +
|h_C|^2/(\max(1,|h_{2,3}|^2)K) + 1}\right),\\
\\
r_{Z_1}&\leq \log\left( 1 + \frac{|h_{1,3}|^2/(\max(1,|h_{1,4}|^2)K)}{2/K + 1}\right),\\
r_{U_2}+r_{Z_1}&\leq \log\left( 1 + \frac{|h_{1,3}|^2/(\max(1,|h_{1,4}|^2)K) +
|h_{2,3}|^2/(\max(1,|{h'_C}|^2)K)}{2/K + 1}\right),\\
r_{S_1}+r_{Z_1}&\leq \log\left( 1 +
\frac{\frac{|h_{1,3}|^2}{\max(1,{|h_{1,4}|^2})K} +
 \frac{\left| |h_{1,3}||h_{2,4}| - |h_{1,4}||h_{2,3}|e^{j\theta}\right|^2}
   {{|h_{2,4}|^2}K} +
 \frac{|h_{2,3}|^2}{\max(1,{|h_{2,4}|^2})K}}{2/K+1}\right),
\end{align*}
\begin{align*}
&r_{U_2}+r_{S_1}+r_{Z_1}\\&\leq
\log\left( 1 + \frac{\frac{|h_{1,3}|^2}{\max(1,{|h_{1,4}|^2})K}
+ \frac{\left| |h_{1,3}||h_{2,4}| -
|h_{1,4}||h_{2,3}|e^{j\theta}\right|^2}{{|h_{2,4}|^2}K}
+ \frac{|h_{2,3}|^2}{\max(1,{|h_{2,4}|^2})K}
+ \frac{|h_{2,3}|^2}{\max(1,|{h'_C}|^2)K}}{2/K+1} \right),\\
&(r_{V_1}+r_{V_2})+r_{U_2}+r_{S_1}+r_{Z_1}\\&\leq
\log\left( 1 + \frac{\frac{|h_{1,3}|^2}{K} + \frac{|h_{1,3}|^2}{\max(1,{|h_{1,4}|^2})K}
+ \frac{\left| |h_{1,3}||h_{2,4}| -
|h_{1,4}||h_{2,3}|e^{j\theta}\right|^2}{{|h_{2,4}|^2}K}
+ \frac{|h_{2,3}|^2}{K} + \frac{|h_{2,3}|^2}{\max(1,{|h_{2,4}|^2})K}
+ \frac{|h_{2,3}|^2}{\max(1,|{h'_C}|^2)K}}{2/K+1} \right),
\end{align*}
\begin{align*}
r_{Z_2}&\leq \log\left( 1 +
 \frac{|h_{2,4}|^2/(\max(1,|h_{2,3}|^2)K}{3/K + 1}\right),\\
r_{U_2}+r_{Z_2}&\leq \log\left( 1 +
 \frac{|h_{2,4}|^2/(\max(1,|{h'_C}|^2)K) + |h_{2,4}|^2/(\max(1,|h_{2,3}|^2)K}{3/K + 1}\right),\\
(r_{V_1}+r_{V_2})+r_{U_2}+r_{Z_2} &\leq \log\left( 1 +
 \frac{\frac{|h_{2,4}|^2}{K} + \frac{|h_{2,4}|^2}{\max(1,|{h'_C}|^2)K} +
\frac{|h_{2,4}|^2}{\max(1,|h_{2,3}|^2)K} + \frac{|h_{1,4}|^2}{K}}{3/K + 1}\right).
\end{align*}
Upon simplification, the above conditions imply that non-negative rates
which satisfy the conditions below are achievable.
\begin{align*}
r_{S_1}&\leq [n_C-\max(n_{1,3},n_{1,4})]_+ - \log 5,\\
r_{V_1}+r_{Z_1}+r_{S_1}&\leq n_C - \log 5,\\
\\
r_{V_2}&\leq n'_C -\log 7,\\
\\
r_{Z_1}&\leq [n_{1,3}-n_{1,4}]_+ - \log 7,\\
r_{U_2}+r_{Z_1}&\leq \max([n_{1,3}-n_{1,4}]_+,n_{2,3}-n'_C) - \log 7,
\end{align*}
\begin{align*}
&r_{S_1}+r_{Z_1}\\&\leq \log\left( 1 +
 \left|\frac{h_{1,3}}{\max(1,|h_{1,4}|)}\right|^2 +
 \left| \frac{|h_{1,3}||h_{2,4}| - |h_{1,4}||h_{2,3}|e^{j\theta}}
    {h_{2,4}}\right|^2 +
\left|\frac{h_{2,3}}{\max(1,|h_{2,4}|)}\right|^2\right) - \log 7,\\
&r_{U_2}+r_{S_1}+r_{Z_1}\\&\leq
\log\left( 1 + \left|\frac{h_{2,3}}{\max(1,|h_C'|)}\right|^2 +
 \left|\frac{h_{1,3}}{h_{1,4}}\right|^2 +
 \left| \frac{|h_{1,3}||h_{2,4}| - |h_{1,4}||h_{2,3}|e^{j\theta}}{h_{2,4}}\right|^2 +
 \left|\frac{h_{2,3}}{\max(1,|h_{2,4}|)}\right|^2\right) - \log 7,
\end{align*}
\begin{align*}
(r_{V_1}+r_{V_2})+r_{U_2}+r_{S_1}+r_{Z_1}
&\leq \max(n_{1,3},n_{2,3}) - \log 7,\\
\\
r_{Z_2}&\leq [n_{2,4}-n_{2,3}]_+ - \log 8,\\
r_{U_2}+r_{Z_2}&\leq \max([n_{2,4}-n_{2,3}]_+,n_{2,4}-n'_C) - \log 8,\\
(r_{V_1}+r_{V_2})+r_{U_2}+r_{Z_2} &\leq \max(n_{2,4},n_{1,4}) -\log 8.
\end{align*}
Note that the conditions on the rates are exactly as in the linear
deterministic case up to a constant except for the constraints on
$r_{S_1}+r_{Z_1}$ and $r_{U_2}+r_{S_1}+r_{Z_1}$. By Fourier-Motzkin
elimination, we may conclude that a sum-rate $R_1+R_2$ within a constant
(9 bits) of the smaller of $u'_2(n_C),u'_3(n_C), u'_4(n_C), u'_5$ and
\[ \frac{[n_C-\max(n_{1,3},n_{1,4})]_+ + \max(n_{2,4},n_{1,4}) +
\max([n_{1,3}-n_{1,4}]_+,n_{2,3}-n'_C) + n_C + n'_C +
[n_{2,4}-n_{2,3}]_+}{2}\]
is achievable.
The last term can be shown to be not smaller than the minimum of
$u'_2(n_C),u'_3(n_C),u'_4(n_C)$ and $u'_5$, if $n'_C$ is chosen to be such
that $u_1(n'_C) = \min(u_2(n_C),u_3(n_C),u_4(n_C),u_5)$ when
$u_1(0)<\min(u_2(0),u_3(0),u_4(0),u_5)$, and $n'_C=0$,
otherwise. Note that from earlier discussion, we know that this choice of
$n'_C$ must be less than or equal to $n_C$ and all $n_{i,j}$,
$i\in\{1,2\}$, $j\in\{3,4\}$, and in particular $n_{2,3}$.

\noindent(2) When $|h_{2,4}| < |h_{1,4}|$ (which implies that $n_{2,4} \leq
n_{1,4}$),
we apply Theorem~\ref{thm:sourcecoopgenericschemes}(c) as in case~(1) above
with the same choices for the auxiliary random variables
$W,V_1,V_2,U_1,U_2,Z_1,Z_2,S_{2,3}^\perp,S'_1$. But, instead of
$\wtilde{S}_{1,3}$ we now define an independent, zero-mean Gaussian random
variable $\wtilde{S}_{2,3}$ with variance $\sigma_{\wtilde{S}_{2,3}}^2 =
1/K$. The conditional distributions of $X_1$ and $X_2$ are defined through
\begin{align*}
X_1&=V_1+Z_1 - \frac{|h_{2,4}|}{|h_{1,4}|e^{j\theta/2}}\wtilde{S}_{2,3}+S'_1,\\
X_2&=V_2+U_2+Z_2 + \wtilde{S}_{2,3}+S_{2,3}^\perp,
\end{align*}
which satisfies the power constraint if we set $K<4$ (since $|h_{2,4}| <
|h_{1,4}|$). We define $S_1=(\wtilde{S}_{2,3},S_{2,3}^\perp)$ and
\[ \bar{S}_1=\left(|h_{2,3}|e^{j\theta/2} -
\frac{|h_{1,3}||h_{2,4}|e^{-j\theta/2}}{|h_{1,4}|}\right)\wtilde{S}_{2,3} +
|h_{2,3}|e^{j\theta/2}S_{2,3}^\perp.\]
The joint distribution of the signals received at the destinations
is given by
\begin{align*}
Y_3&=|h_{1,3}|(V_1+Z_1+S'_1) + |h_{2,3}|e^{j\theta}(V_2+U_2+Z_2) +
\bar{S}_1 + N_3,
\\
Y_4&=|h_{2,4}|(V_2+U_2+Z_2) + |h_{1,4}|e^{j\theta}(V_1+Z_1+S'_1) +
|h_{2,4}|S_{2,3}^\perp +N_4.
\end{align*}
The simplified conditions on the rates are identical to those in case~(1),
except for the following two
\begin{align*}
r_{S_1}+r_{Z_1}&\leq \log\left( 1 + \left|\frac{h_{1,3}}{h_{1,4}}\right|^2
+ \left| \frac{|h_{1,3}||h_{2,4}| -
|h_{1,4}||h_{2,3}|e^{j\theta}}{h_{1,4}}\right|^2 +
\left|\frac{h_{2,3}}{h_{2,4}}\right|^2\right) - \log 6,\\
r_{U_2}+r_{S_1}+r_{Z_1}&\leq
\log\left( 1 + \left|\frac{h_{2,4}}{h_C'}\right|^2 +
 \left|\frac{h_{1,3}}{h_{1,4}}\right|^2 +
 \left| \frac{|h_{1,3}||h_{2,4}| - |h_{1,4}||h_{2,3}|e^{j\theta}}{h_{1,4}}\right|^2 +
 \left|\frac{h_{2,3}}{h_{2,4}}\right|^2\right) - \log 6.
\end{align*}
Applying Fourier-Motzkin elimination and choosing $n'_C$ as in case~(1)
completes the achievability proof (to within 9 bits of the of the smaller
of $u'_2(n_C),u'_3(n_C), u'_4(n_C), u'_5$).

Thus, we may conclude that the upperbound is achievable with a gap of
at most 16 bits in regime~(iii). Note that we may represent the two cases
together as follows:
\begin{align*}
X_1&=V_1+Z_1+\bar{S}_{1,3}+S'_1,\\
X_2&=V_2+U_2+Z_2+\bar{S}_{2,4},
\end{align*}
where
\begin{align*}
\bar{S}_{1,3}&=\left\{\begin{array}{ll}
\wtilde{S}_{1,3},&\text{ if }|h_{2,4}|\geq |h_{1,4}|,\\
-\frac{|h_{2,4}|}{|h_{1,4}|e^{j\theta/2}}\wtilde{S}_{2,3},
 &\text{ otherwise}.
\end{array}\right.\\
\bar{S}_{2,3}&=\left\{\begin{array}{ll}
-\frac{|h_{1,4}|e^{j\theta/2}}{|h_{2,4}|}\wtilde{S}_{1,3} + S_{2,3}^\perp,&
\text{ if }|h_{2,4}|\geq |h_{1,4}|,\\
\wtilde{S}_{2,3} + S_{2,3}^\perp,& \text{ otherwise}.
\end{array}\right.
\end{align*}
This gives
\begin{align*}
Y_3&=|h_{1,3}|(V_1+Z_1+S'_1) + |h_{2,3}|e^{j\theta/2}(V_2+U_2+Z_2) +
\bar{S}_1+N_3,
\\
Y_4&=|h_{2,4}|(V_2+U_2+Z_2) + |h_{2,3}|e^{j\theta/2}(V_1+Z_1+S'_1) +
|h_{2,4}|S_{2,3}^\perp+N_4,
\end{align*}
where, we define $S_1=(\bar{S}_{2,3},\bar{S}_{1,3},S_{2,3}^\perp)$ and
\[ \bar{S}_1=|h_{2,3}|e^{j\theta/2}\bar{S}_{2,3} + |h_{1,3}|\bar{S}_{1,3} +
|h_{2,3}|e^{j\theta/2}S_{2,3}^\perp.\]
The distribution of $S_1,S_2$ we will employ in regime (iv) below is a
symmetric generalization of this.

\noindent{\em Regime (iv):} $\max(n_{1,3},n_{2,4}) < n_C$. In this regime,
we employ Theorem~\ref{thm:sourcecoopgenericschemes}(b) as we did for the
linear deterministic case. $U_1,U_2$ are constants.
$V_1,V_2,Z_1,Z_2, \wtilde{S}_{1,3},
\wtilde{S}_{2,3},\wtilde{S}_{1,4},\wtilde{S}_{2,4}, S_{2,3}^\perp,
S_{1,4}^\perp, S'_1, S'_2$ are independent zero-mean
Gaussian random variables. Their variances are as follows
\begin{align*}
\sigma_{V_1}^2=\sigma_{V_2}^2&=1/K,\\
\sigma_{Z_1}^2&=\frac{1/K}{\max(1,|h_{1,4}|^2)},\\
\sigma_{Z_2}^2&=\frac{1/K}{\max(1,|h_{2,3}|^2)},\\
\sigma_{\wtilde{S}_{i,j}}^2&=1/K,\;\; i\in \{1,2\},\;j\in \{3,4\},\\
\sigma_{S_{1,4}^\perp}^2&=\frac{1/K}{\max(1,|h_{1,3}|^2)},\\
\sigma_{S_{2,3}^\perp}^2&=\frac{1/K}{\max(1,|h_{2,4}|^2)},\\
\sigma_{S'_1}^2 &= \frac{1/K}{\max(1,|h_{1,3}|^2,|h_{1,4}|^2)},\\
\sigma_{S'_2}^2 &= \frac{1/K}{\max(1,|h_{2,4}|^2,|h_{2,3}|^2)}.
\end{align*}
where $K$ is to be specified. We set $W$ to be independent of
all these and have the same distribution as $(V_1,V_2)$.
We define $X_1$, and $X_2$ as follows
\begin{align*}
X_1&=V_1+Z_1+\bar{S}_{1,3}+\bar{S}_{1,4}+S'_1,\\
X_2&=V_2+Z_2+\bar{S}_{2,4}+\bar{S}_{1,3}+S'_2,
\end{align*}
where
\begin{align*}
\bar{S}_{1,3}&=\left\{\begin{array}{ll}
\wtilde{S}_{1,3},&\text{ if }|h_{2,4}|\geq |h_{1,4}|,\\
-\frac{|h_{2,4}|}{|h_{1,4}|e^{j\theta/2}}\wtilde{S}_{2,3},\quad\qquad&\text{
otherwise}.
\end{array}\right.\\
\bar{S}_{2,3}&=\left\{\begin{array}{ll}
-\frac{|h_{1,4}|e^{j\theta/2}}{|h_{2,4}|}\wtilde{S}_{1,3} + S_{2,3}^\perp,&
\text{ if }|h_{2,4}|\geq |h_{1,4}|,\\
\wtilde{S}_{2,3} + S_{2,3}^\perp,& \text{ otherwise}.
\end{array}\right.\\
\bar{S}_{2,4}&=\left\{\begin{array}{ll}
\wtilde{S}_{2,4},&\text{ if }|h_{1,3}|\geq |h_{2,3}|,\\
-\frac{|h_{1,3}|}{|h_{2,3}|e^{j\theta/2}}\wtilde{S}_{1,4},\quad\qquad&\text{
otherwise}.
\end{array}\right.\\
\bar{S}_{1,4}&=\left\{\begin{array}{ll}
-\frac{|h_{2,3}|e^{j\theta/2}}{|h_{1,3}|}\wtilde{S}_{2,4} + S_{1,4}^\perp,&\text{ if }
|h_{1,3}|\geq |h_{2,3}|,\\
\wtilde{S}_{1,4} + S_{1,4}^\perp,& \text{ otherwise}.
\end{array}\right.
\end{align*}
To satisfy the power constraint, it is enough to choose $K<7$. Also, we
define $S_1=(\bar{S}_{2,3},\bar{S}_{1,3},S_{2,3}^\perp)$,
$S_2=(\bar{S}_{1,4},\bar{S}_{2,4},S_{1,4}^\perp)$, and
\begin{align*}
\bar{S}_1=|h_{2,3}|e^{j\theta/2}\bar{S}_{2,3} + |h_{1,3}|\bar{S}_{1,3} +
|h_{2,3}|e^{j\theta/2}S_{2,3}^\perp,\\
\bar{S}_2=|h_{1,4}|e^{j\theta/2}\bar{S}_{1,4} + |h_{2,4}|\bar{S}_{2,4} +
|h_{1,4}|e^{j\theta/2}S_{1,4}^\perp.
\end{align*}
Thus, $S_1$ and $S_2$ are independent of each other. Note that we defined
$\bar{S}_{1,3}$ and $\bar{S}_{2,3}$ in the same way as we did in
regime~(iii). The destinations receive
\begin{align*}
Y_3&=|h_{1,3}|(V_1+Z_1+S'_1) + |h_{2,3}|e^{j\theta/2}(V_2+Z_2+S'_2) +
\bar{S}_1 + |h_{1,3}|S_{1,4}^\perp + N_3,\\
Y_4&=|h_{2,4}|(V_2+Z_2+S'_2) + |h_{1,4}|e^{j\theta/2}(V_1+Z_1+S'_1) +
\bar{S}_2 + |h_{2,4}|S_{2,3}^\perp + N_4.
\end{align*}
It must be noted that $|h_{1,3}|S_{1,4}^\perp$ and $|h_{2,4}|S_{2,3}^\perp$
have variances at most unity (which is the variance of the noise).
Since $U_1$ and $U_2$ are constants, we must set $r_{U_1}=r_{U_2}=0$. The
conditions on the non-negative rates are as follows.
\begin{align*}
r_{S_1}&\leq \log\left( 1 + |h_C|^2/(\max(1,|h_{1,3}|^2,|h_{1,4}|^2)K) \right),\\
r_{Z_1}+r_{S_1}&\leq \log\left( 1 + |h_C|^2/(\max(1,|h_{1,4}|^2)K) +
|h_C|^2/(\max(1,|h_{1,3}|^2,|h_{1,4}|^2)K) \right)\\
r_{V_1}+r_{Z_1}+r_{S_1}&\leq \log\left( 1 + \frac{|h_C|^2}{K} +
\frac{|h_C|^2}{\max(1,|h_{1,4}|^2)K} +
\frac{|h_C|^2}{\max(1,|h_{1,3}|^2,|h_{1,4}|^2)K}\right),\\
\\
r_{Z_1}&\leq \log\left( 1 +
   \frac{|h_{1,3}|^2/(\max(1,|h_{1,4}|^2)K)}{4/K + 1}\right),\\
r_{S_1}+r_{Z_1} &\leq
\left\{\begin{array}{ll}
\log\left( 1 + \frac{\left(\left|\frac{|h_{1,3}|}{|h_{1,4}|}\right|^2 +
 \left| \frac{|h_{1,3}||h_{2,4}| -
|h_{1,4}||h_{2,3}|e^{j\theta}}{|h_{2,4}|}\right|^2 +
 \left|\frac{|h_{2,3}|}{|h_{2,4}|}\right|^2\right)/K}{4/K+1}\right),&\text{ if }|h_{2,4}| \geq |h_{1,4}|\\
\log\left( 1 + \frac{\left(\left|\frac{|h_{1,3}|}{|h_{1,4}|}\right|^2
+ \left| \frac{|h_{1,3}||h_{2,4}| -
|h_{1,4}||h_{2,3}|e^{j\theta}}{|h_{1,4}|}\right|^2 +
\left|\frac{|h_{2,3}|}{|h_{2,4}|}\right|^2\right)/K}{4/K+1}\right),& \text{
otherwise}
\end{array}\right.
\end{align*}
\begin{align*}
&(r_{V_1}+r_{V_2})+r_{S_1}+r_{Z_1}\\
&\leq \left\{\begin{array}{ll}
\log\left( 1 + \frac{\left( |h_{1,3}|^2 +
 \left|\frac{|h_{1,3}|}{|h_{1,4}|}\right|^2 + |h_{2,3}|^2 +
 \left| \frac{|h_{1,3}||h_{2,4}| - |h_{1,4}||h_{2,3}|e^{j\theta}}{|h_{2,4}|}\right|^2 +
 \left|\frac{|h_{2,3}|}{|h_{2,4}|}\right|^2\right)/K}{4/K+1}\right),&\text{ if }|h_{2,4}| \geq |h_{1,4}|\\
\log\left( 1 + \frac{\left( |h_{1,3}|^2 +
 \left|\frac{|h_{1,3}|}{|h_{1,4}|}\right|^2 + |h_{2,3}|^2 +
 \left| \frac{|h_{1,3}||h_{2,4}| - |h_{1,4}||h_{2,3}|e^{j\theta}}{|h_{1,4}|}\right|^2 +
\left|\frac{|h_{2,3}|}{|h_{2,4}|}\right|^2\right)/K}{4/K+1}\right),& \text{
otherwise}
\end{array}\right.
\end{align*}
and the corresponding conditions with 1 and 2 interchanged and 3 and 4
interchanged. On simplification, this implies that non-negative rates which
satisfy the following conditions are achievable.
\begin{align*}
r_{S_1}&\leq [n_C-\max(n_{1,3},n_{1,4})]_+ - \log 7,\\
r_{V_1}+r_{Z_1}+r_{S_1}&\leq n_C - \log 7,\\
\\
r_{Z_1}&\leq [n_{1,3}-n_{1,4}]_+ - \log 11,
\end{align*}
\begin{align*}
r_{S_1}+r_{Z_1} &\leq
\left\{\begin{array}{ll}
\log\left( 1 + \left|\frac{|h_{1,3}|}{|h_{1,4}|}\right|^2 +
 \left| \frac{|h_{1,3}||h_{2,4}| - |h_{1,4}||h_{2,3}|e^{j\theta}}{|h_{2,4}|}\right|^2 +
 \left|\frac{|h_{2,3}|}{|h_{2,4}|}\right|^2\right) - \log 11 ,&\text{ if }|h_{2,4}| \geq |h_{1,4}|\\
\log\left( 1 + \left|\frac{|h_{1,3}|}{|h_{1,4}|}\right|^2
+ \left| \frac{|h_{1,3}||h_{2,4}| -
|h_{1,4}||h_{2,3}|e^{j\theta}}{|h_{1,4}|}\right|^2 +
\left|\frac{|h_{2,3}|}{|h_{2,4}|}\right|^2\right) - \log 11,& \text{
otherwise},
\end{array}\right.
\end{align*}
\begin{align*}
(r_{V_1}+r_{V_2})+r_{S_1}+r_{Z_1}
&\leq \max(n_{1,3},n_{2,3}) - \log 11,
\end{align*}
and the corresponding conditions with 1 and 2 interchanged and 3 and 4
interchanged. Note that these conditions are identical to the ones for the
linear deterministic case (up to a constant) except for the ones on
$r_{S_1}+r_{Z_1}$ and $r_{S_2}+r_{Z_2}$.
Fourier-Motzkin elimination shows that a sum-rate given by the minimum of
$u'_2(n_C),u'_3(n_C),u'_4(n_C),$ and $u'_5$ is achievable in regime~(iv) up
to a constant of 7 bits.

Thus, by combining the results for all regimes, we may conclude that the
upperbound is achievable within a constant gap (of 20 bits).

\section{Source cooperation: upperbounds}\label{app:sourcecoopupperbounds}

We prove four upperbounds to the sum-rate which will together imply the
upperbounds in Theorems~\ref{thm:sourcecoopLD} and~\ref{thm:sourcecoopG}.

\noindent{\em Upperbound 1:}
From Fano's inequality,
\begin{align*}
T(R_1+R_2-o(\epsilon))&\leq I(M_1;Y_3^T)+I(M_2;Y_4^T)\\
&\leq
I(M_1;Y_3^T,h_{1,4}^\ast(X_1^T),h_{1,2}^\ast(X_1^T),h_{2,1}^\ast(X_2^T))\\
 &\qquad+ I(M_2;Y_4^T,h_{2,3}^\ast(X_2^T),h_{1,2}^\ast(X_1^T),h_{2,1}^\ast(X_2^T)).
\end{align*}
Note that we have provided additional signals to both the destinations --
for instance, destination~4 now has access to $h_{1,4}^\ast(X_1^T)$,
$h_{1,2}^\ast(X_1^T)$, and $h_{2,1}^\ast(X_2^T))$ in addition to its
channel output $Y_3^T$. We will now upperbound the two symmetric mutual
information terms above; only the first term is shown below.
\begin{align*}
I(M_1;&Y_3^T,h_{1,4}^\ast(X_1^T),h_{1,2}^\ast(X_1^T),h_{2,1}^\ast(X_2^T))\\
&= H(h_{1,4}^\ast(X_1^T),h_{1,2}^\ast(X_1^T),h_{2,1}^\ast(X_2^T))
 + H(Y_3^T|h_{1,4}^\ast(X_1^T),h_{1,2}^\ast(X_1^T),h_{2,1}^\ast(X_2^T))\\
 &\quad- H(Y_3^T,h_{1,4}^\ast(X_1^T),h_{1,2}^\ast(X_1^T),h_{2,1}^\ast(X_2^T)|M_1)\\
&\leq H(h_{1,4}^\ast(X_1^T),h_{1,2}^\ast(X_1^T),h_{2,1}^\ast(X_2^T))
 + H(Y_3^T|h_{1,4}^\ast(X_1^T),h_{2,1}^\ast(X_2^T))  \\
 &\quad- H(Y_3^T,h_{1,4}^\ast(X_1^T),h_{1,2}^\ast(X_1^T),h_{2,1}^\ast(X_2^T)|M_1).
\end{align*}
We now derive an upperbound for the third term.
\begin{align*}
&H(Y_3^T,h_{1,4}^\ast(X_1^T),h_{1,2}^\ast(X_1^T),h_{2,1}^\ast(X_2^T)|M_1)\\
&= \sum_{t=1}^T H(Y_3(t),h_{1,4}^\ast(X_1(t)),h_{1,2}^\ast(X_1(t)),h_{2,1}^\ast(X_2(t))
   |Y_3^{t-1},h_{1,4}^\ast(X_1^{t-1}),h_{1,2}^\ast(X_1^{t-1}),h_{2,1}^\ast(X_2^{t-1}),M_1)\\
&\stackrel{\text{(a)}}{=} \sum_{t=1}^T
H\bigg(h_{2,3}^\ast(X_2(t)),h_{1,4}^\ast(X_1(t)),
 h_{1,2}^\ast(X_1(t)),h_{2,1}^\ast(X_2(t))\bigg|X_1^t,h_{2,3}^\ast(X_2^{t-1}),h_{1,4}^\ast(X_1^{t-1}),\\&\qquad\qquad\qquad\qquad\qquad\qquad\qquad\qquad\qquad\qquad\qquad\qquad\qquad\qquad
h_{1,2}^\ast(X_1^{t-1}),h_{2,1}^\ast(X_2^{t-1}),M_1\bigg)\\
&\stackrel{\text{(b)}}{=} \sum_{t=1}^T H(h_{1,4}^\ast(X_1(t))|X_1(t)) + H(h_{1,2}^\ast(X_1(t))|X_1(t))\\
 &\qquad\quad + H(h_{2,3}^\ast(X_2(t)),h_{2,1}^\ast(X_2(t))|X_1^t,h_{2,3}^\ast(X_2^{t-1}),h_{1,4}^\ast(X_1^{t-1}),h_{1,2}^\ast(X_1^{t-1}),h_{2,1}^\ast(X_2^{t-1}),M_1)\\
&\stackrel{\text{(c)}}{=} \sum_{t=1}^T H(h_{1,4}^\ast(X_1(t))|X_1(t))
 + H(h_{1,2}^\ast(X_1(t))|X_1(t))\\
 &\qquad\quad+ H(h_{2,3}^\ast(X_2(t)),h_{2,1}^\ast(X_2(t))|h_{2,3}^\ast(X_2^{t-1}),h_{1,2}^\ast(X_1^{t-1}),
     h_{2,1}^\ast(X_2^{t-1}))\\
&\stackrel{\text{(d)}}{=} \sum_{t=1}^T H(h_{1,4}^\ast(X_1(t))|X_1(t))
 + H(h_{1,2}^\ast(X_1(t))|X_1(t))\\
 &\qquad\quad + H(h_{2,3}^\ast(X_2(t)),h_{1,2}^\ast(X_1(t)),h_{2,1}^\ast(X_2(t))|
     h_{2,3}^\ast(X_2^{t-1}),h_{1,2}^\ast(X_1^{t-1}),h_{2,1}^\ast(X_2^{t-1})) \\
 &\qquad\quad
 - H(h_{1,2}^\ast(X_1(t))|h_{2,3}^\ast(X_2^t),h_{1,2}^\ast(X_1^{t-1}),h_{2,1}^\ast(X_2^{t}))\\
&\stackrel{\text{(e)}}{\geq} \sum_{t=1}^T H(h_{1,4}^\ast(X_1(t))|X_1(t))
 + H(h_{1,2}^\ast(X_1(t))|X_1(t))\\
 &\qquad\quad + H(h_{2,3}^\ast(X_2(t)),h_{1,2}^\ast(X_1(t)),h_{2,1}^\ast(X_2(t))|
     h_{2,3}^\ast(X_2^{t-1}),h_{1,2}^\ast(X_1^{t-1}),h_{2,1}^\ast(X_2^{t-1})) \\
&\qquad\quad
 - H(h_{1,2}^\ast(X_1(t))|h_{1,2}^\ast(X_1^{t-1}))\\
&= H(h_{2,3}^\ast(X_2^T),h_{1,2}^\ast(X_1^T),h_{2,1}^\ast(X_2^T)) -
H(h_{1,2}^\ast(X_1^T))\\ &\qquad\qquad +
\left(\sum_{t=1}^T H(h_{1,4}^\ast(X_1(t))|X_1(t)) +H(h_{1,2}^\ast(X_1(t))|X_1(t))\right)
\end{align*}
where (a) follows from the following facts: (1) $Y_3(s)=h_{1,3}(X_1(s))+h_{2,3}^\ast(X_2(s))$, (2) $h_{1,3}$ is a deterministic function, and
(3) $X_1(s)$ is a deterministic function of $M_1,h_{2,1}^\ast(X_2^{t-1})$,
for all $s\leq t$. Equality (b) can be seen from the channel model by which,
conditioned on $X_1(t)$, the following three sets of random variables are
independent: (1) $h_{1,4}^\ast(X_1(t))$, (2) $h_{1,2}^\ast(X_1(t))$, and
(3) $M_1$ and all other quantities with indices up to and including $t$. The
next equality (c) is a consequence of the fact that the following is a Markov
chain
\begin{align*}
(M_2,X_2^t,h_{2,3}^\ast(X_2^t)) - (h_{1,2}^\ast(X_1^{t-1}),h_{2,1}^\ast(X_2^{t-1})) - (M_1,X_1^t).
\end{align*}
This follows from (1) the channel $p_{h_{1,2}^\ast(X_1),h_{2,1}^\ast(X_2),
h_{2,3}^\ast(X_2)|X_1,X_2}=p_{h_{1,2}^\ast(X_1)|X_1}p_{h_{2,1}^\ast(X_2)|X_1}
p_{h_{2,3}^\ast(X_2)|X_2}$, (2) the independence of $M_1$ and $M_2$, and
(3) the fact that the channel inputs depend deterministically on the
messages at the respective sources and what these sources have received.
Equality (d) is just the chain rule of entropy, and the inequality (e)
follows from the non-negativity of mutual information.

\noindent Combining everything, we have
\begin{align*}
T(R_1+R_2-o(\epsilon))\leq&
\bigg(
 \left\{H(Y_3^T|h_{1,4}^\ast(X_1^T),h_{2,1}^\ast(X_2^T))-\sum_{t=1}^TH(h_{2,3}^\ast(X_2(t))|X_2(t))\right\}\\
&\qquad\qquad\qquad\qquad+\left\{H(h_{1,2}^\ast(X_1^T))-\sum_{t=1}^TH(h_{1,2}^\ast(X_1(t))|X_1(t))\right\}\bigg)\\
&+\bigg(
 \left\{H(Y_4^T|h_{2,3}^\ast(X_2^T),h_{1,2}^\ast(X_1^T))-\sum_{t=1}^TH(h_{1,4}^\ast(X_1(t))|X_1(t))\right\}\\
&\qquad\qquad\qquad\qquad+\left\{H(h_{2,1}^\ast(X_2^T))-\sum_{t=1}^TH(h_{2,1}^\ast(X_2(t))|X_2(t))\right\}
 \bigg).
\end{align*}

\noindent{\em Linear deterministic model:} Evaluating the bound directly
gives us
\begin{align*}
R_1+R_2\leq\max(n_{1,3}-n_{1,4},n_{2,3}-n_C,0) + n_C +
\max(n_{2,4}-n_{2,3},n_{1,4}-n_C,0) + n_C.
\end{align*}

\noindent{\em Gaussian model:}
Consider the first bracketed term.
\begin{align*}
\frac{1}{T}&\left\{H(Y_3^T|h_{1,4}^\ast(X_1^T),h_{2,1}^\ast(X_2^T))-\sum_{t=1}^TH(h_{2,3}^\ast(X_2(t))|X_2(t))\right\}\\
&\qquad\qquad\leq \frac{1}{T}H(Y_3^T-h_{1,3}h_{1,4}^{-1}h_{1,4}^\ast(X_1^T)-h_{2,3}h_{2,1}^{-1}h_{2,3}^\ast(X_2^T)) - H(N_3)\\
&\qquad\qquad= H(N_3-h_{1,3}h_{1,4}^{-1}N_4-h_{2,3}h_{2,1}^{-1}N_1) - H(N_3)\\
&\qquad\qquad\leq \log\left(1+\left|\frac{h_{1,3}}{h_{1,4}}\right|^2+\left|\frac{h_{2,3}}{h_{2,1}}\right|^2\right)
.
\\
\frac{1}{T}&
\left\{H(h_{1,2}^\ast(X_1^T))-\sum_{t=1}^TH(h_{1,2}^\ast(X_1(t))|X_1(t))\right\}
\\
&\qquad\qquad= \frac{1}{T}H(h_{1,2}X_1^T+N_2^T) - H(N_2)\\
&\qquad\qquad\leq \log(1+|h_{1,2}|^2)
.
\end{align*}

If $|h_C|<1$, we do not subtract the term $h_{2,3}h_{2,1}^{-1}h_{2,3}^\ast(X_2^T)$ while upperbounding the first term. This gives
\begin{align*}
\frac{1}{T}&\left\{H(Y_3^T|h_{1,4}^\ast(X_1^T),h_{2,1}^\ast(X_2^T))-\sum_{t=1}^TH(h_{2,3}^\ast(X_2(t))|X_2(t))\right\}\\
&\qquad\qquad\leq \frac{1}{T}H(Y_3^T-h_{1,3}h_{1,4}^{-1}h_{1,4}^\ast(X_1^T))
- H(N_3)\\
&\qquad\qquad= H(N_3^T-h_{1,3}h_{1,4}^{-1}N_4^T+h_{2,3}X_2^T) - H(N_3)\\
&\qquad\qquad\leq \log\left(1+\left|\frac{h_{1,3}}{h_{1,4}}\right|^2+\left|h_{2,3}\right|^2\right)
.
\end{align*}
And similarly, we do not subtract the term
$h_{1,3}h_{1,4}^{-1}h_{1,4}^\ast(X_1^T)$ if $|h_{1,4}|<1$ which gives the following upperbound for the first term.
\begin{align*}
\frac{1}{T}&\left\{H(Y_3^T|h_{1,4}^\ast(X_1^T),h_{2,1}^\ast(X_2^T))-\sum_{t=1}^TH(h_{2,3}^\ast(X_2(t))|X_2(t))\right\}\\
&\qquad\qquad\leq \log\left(1+\left|{h_{1,3}}\right|^2+\left|\frac{h_{2,3}}{h_{2,1}}\right|^2\right)
.
\end{align*}
If both $|h_{1,4}|<1$ and $|h_C|<1$, we do not subtract either terms in which case
we get the upperbound
\begin{align*}
\frac{1}{T}&\left\{H(Y_3^T|h_{1,4}^\ast(X_1^T),h_{2,1}^\ast(X_2^T))-\sum_{t=1}^TH(h_{2,3}^\ast(X_2(t))|X_2(t))\right\}\\
&\qquad\qquad\leq \log\left(1+\left(\left|{h_{1,3}}\right|+\left|{h_{2,3}}\right|\right)^2\right)
.
\end{align*}
Similarly, upperbounding the second bracketed term and combining, we have
\begin{align*}
R_1+R_2&\leq
\log\left(1+\left(\frac{\left|h_{1,3}\right|}{\max\left(1,\left|h_{1,4}\right|\right)}+\frac{\left|h_{2,3}\right|}{\max\left(1,\left|h_C\right|\right)}\right)^2\right)(1+|h_C|^2)\\
&\quad+
\log\left(1+\left(\frac{\left|h_{2,4}\right|}{\max\left(1,\left|h_{2,3}\right|\right)}+\frac{\left|h_{1,4}\right|}{\max\left(1,\left|h_C\right|\right)}\right)^2\right)(1+|h_C|^2).
\end{align*}

\noindent{\em Upperbounds 2 and 3:}
We prove upperbound~2 below, the third one follows from a similar argument.
From Fano's inequality,
\begin{align*}
&T(R_1+R_2-o(\epsilon))\\&\leq I(M_1;Y_3^T)+I(M_2;Y_4^T)\\
&\leq I(M_1;Y_3^T) + I(M_2;Y_4^T,h_{2,3}^\ast(X_2^T),h_{1,2}^\ast(X_1^T),h_{2,1}^\ast(X_2^T),M_1)\\
&\leq I(M_1;Y_3^T) + I(M_2;Y_4^T,h_{2,3}^\ast(X_2^T),h_{1,2}^\ast(X_1^T),h_{2,1}^\ast(X_2^T)|M_1)\\
&\leq \sum_{t=1}^T I(M_1;Y_3(t)|Y_3^{t-1}) \\ &\qquad\quad+
            I(M_2;Y_4(t),h_{2,3}^\ast(X_2(t)),h_{1,2}^\ast(X_1(t)),h_{2,1}^\ast(X_2(t))|Y_4^{t-1},h_{2,3}^\ast(X_2^{t-1}),h_{1,2}^\ast(X_1^{t-1}),h_{2,1}^\ast(X_2^{t-1}),M_1).
\end{align*}
Below, we upperbound these terms separately.
\begin{align*}
I(M_1;Y_3(t)|Y_3^{t-1})&=H(Y_3(t)|Y_3^{t-1}) - H(Y_3(t)|Y_3^{t-1},M_1)\\
&\leq H(Y_3(t)) - H(Y_3(t)|Y_3^{t-1},Y_4^{t-1},h_{1,2}^\ast(X_1^{t-1}),h_{2,1}^\ast(X_2^{t-1}),M_1)\\
&\stackrel{\text{(a)}}{\leq} H(Y_3(t)) - H(h_{2,3}^\ast(X_2(t))|Y_4^{t-1},h_{2,3}^\ast(X_2^{t-1}),h_{1,2}^\ast(X_1^{t-1}),h_{2,1}^\ast(X_2^{t-1}),M_1),
\end{align*}
where (a) follows from the fact that $Y_3(t)=h_{1,3}(X_1(t))+h_{2,3}^\ast(X_2(t))$,
and, $h_{1,3}$ is a deterministic function and $X_1(t)$ is a deterministic
function $f_{1,t}$ of $(M_1,h_{2,1}^\ast(X_2^{t-1}))$.
\begin{align*}
I(M_2;&Y_4(t),h_{2,3}^\ast(X_2(t)),h_{1,2}^\ast(X_1(t)),h_{2,1}^\ast(X_2(t))|Y_4^{t-1},h_{2,3}^\ast(X_2^{t-1}),h_{1,2}^\ast(X_1^{t-1}),h_{2,1}^\ast(X_2^{t-1}),M_1)\\
&=I(M_2;h_{2,3}^\ast(X_2(t))|Y_4^{t-1},h_{2,3}^\ast(X_2^{t-1}),h_{1,2}^\ast(X_1^{t-1}),h_{2,1}^\ast(X_2^{t-1}),M_1)\\
&\quad+I(M_2;Y_4(t)|Y_4^{t-1},h_{2,3}^\ast(X_2^t),h_{1,2}^\ast(X_1^{t-1}),h_{2,1}^\ast(X_2^{t-1}),M_1)\\
&\quad+I(M_2;h_{1,2}^\ast(X_1(t)),h_{2,1}^\ast(X_2(t))|Y_4^{t},h_{2,3}^\ast(X_2^t),h_{1,2}^\ast(X_1^{t-1}),h_{2,1}^\ast(X_2^{t-1}),M_1)
\end{align*}
We upperbound these three terms separately now.
\begin{align*}
I(M_2;h_{2,3}^\ast(X_2(t))|&Y_4^{t-1},h_{2,3}^\ast(X_2^{t-1}),h_{1,2}^\ast(X_1^{t-1}),h_{2,1}^\ast(X_2^{t-1}),M_1)\\
&=H(h_{2,3}^\ast(X_2(t))|Y_4^{t-1},h_{2,3}^\ast(X_2^{t-1}),h_{1,2}^\ast(X_1^{t-1}),h_{2,1}^\ast(X_2^{t-1}),M_1)\\
&\quad-H(h_{2,3}^\ast(X_2(t))|Y_4^{t-1},h_{2,3}^\ast(X_2^{t-1}),h_{1,2}^\ast(X_1^{t-1}),h_{2,1}^\ast(X_2^{t-1}),M_1,M_2).\\
I(M_2;Y_4(t)|Y_4^{t-1},&h_{2,3}^\ast(X_2^t),h_{1,2}^\ast(X_1^{t-1}),h_{2,1}^\ast(X_2^{t-1}),M_1)\\
&=H(Y_4(t)|Y_4^{t-1},h_{2,3}^\ast(X_2^t),h_{1,2}^\ast(X_1^{t-1}),h_{2,1}^\ast(X_2^{t-1}),M_1)\\
 &\quad- H(Y_4(t)|Y_4^{t-1},h_{2,3}^\ast(X_2^t),h_{1,2}^\ast(X_1^{t-1}),h_{2,1}^\ast(X_2^{t-1}),M_1,M_2)\\
&\stackrel{\text{(a)}}{\leq} H(Y_4(t)|X_1(t),h_{2,3}^\ast(X_2(t))) - H(Y_4(t)|X_1(t),X_2(t)),
\end{align*}
where (a) follows from the channel model (memorylessness and independence of
the noise processes at the different nodes) and the fact that
$X_1(t)$ ($X_2(t)$, resp.) is a deterministic function $f_{1,t}$ ($f_{2,t}$,
resp.) of $(M_1,h_{2,1}^\ast(X_2^{t-1}))$ ($(M_2,h_{1,2}^\ast(X_1^{t-1}))$, resp.).
\begin{align*}
I(M_2&;h_{1,2}^\ast(X_1(t)),h_{2,1}^\ast(X_2(t))|Y_4^{t},h_{2,3}^\ast(X_2^t),h_{1,2}^\ast(X_1^{t-1}),h_{2,1}^\ast(X_2^{t-1}),M_1)\\
&\stackrel{\text{(a)}}{=}I(M_2;h_{2,1}^\ast(X_2(t))|Y_4^{t},h_{2,3}^\ast(X_2^t),h_{1,2}^\ast(X_1^{t-1}),h_{2,1}^\ast(X_2^{t-1}),M_1)\\
&= H(h_{2,1}^\ast(X_2(t))|Y_4^{t},h_{2,3}^\ast(X_2^t),h_{1,2}^\ast(X_1^{t-1}),h_{2,1}^\ast(X_2^{t-1}),M_1)\\
&\quad - H(h_{2,1}^\ast(X_2(t))|Y_4^{t},h_{2,3}^\ast(X_2^t),h_{1,2}^\ast(X_1^{t-1}),h_{2,1}^\ast(X_2^{t-1}),M_1,M_2)\\
&\stackrel{\text{(b)}}{\leq} H(h_{2,1}^\ast(X_2(t))|Y_4(t),h_{2,3}^\ast(X_2(t)),X_1(t)) - H(h_{2,1}^\ast(X_2(t))|X_2(t)),
\end{align*}
where (a) follows from the channel model (memorylessness and the independence
of the noise processes at the different nodes) and the fact that $X_1(t)$ is a
deterministic function $f_{1,t}$ of $(M_1,h_{2,1}^\ast(X_2^{t-1}))$, and (b)
follows from the facts above, its analogue for $X_2(t)$ and the channel model.

\noindent Combining everything, we have
\begin{align*}
T(R_1+R_2-o(\epsilon)) &\leq
\sum_{t=1}^T \bigg\{ H(Y_3(t)|Y_3^{t-1})\\
&\qquad\qquad-H(h_{2,3}^\ast(X_2(t))|Y_4^{t-1},h_{2,3}^\ast(X_2^{t-1}),h_{1,2}^\ast(X_1^{t-1}),h_{2,1}^\ast(X_2^{t-1}),M_1,M_2)\bigg\}\\
&\quad\qquad +\left\{ H(Y_4(t)|X_1(t),h_{2,3}^\ast(X_2(t))) - H(Y_4(t)|X_1(t),X_2(t)) \right\}\\
&\quad\qquad +\left\{ H(h_{2,1}^\ast(X_2(t))|Y_4(t),h_{2,3}^\ast(X_2(t)),X_1(t)) - H(h_{2,1}^\ast(X_2(t))|X_2(t))\right\}.
\end{align*}

\noindent{\em Linear deterministic model:}
Suppose $n_{C}\leq\max(n_{2,4},n_{2,3})$, then evaluating the upperbound,
\begin{align*}
R_1+R_2 &\leq \{\max(n_{1,3},n_{2,3})-0\} + \{[n_{2,4}-n_{2,3}]_+-0\} + \{0-0\}.
\end{align*}
Otherwise, if $n_{C}>\max(n_{2,4},n_{2,3})$ we have either: (a)
$n_{C}>n_{2,4}\geq n_{2,3}$ in which case, we lower the noise level at
node~4 by $n_{C}-n_{2,4}$ level by defining $n_{2,4}'=n_{2,4}+(n_{C}-n_{2,4})$
and $n_{1,4}'=n_{1,4}+(n_{C}-n_{2,4})$. Now the upperbound evaluates to
\begin{align*}
R_1+R_2&\leq \max(n_{1,3},n_{2,3}) + n_{C}-n_{2,3}
 = n_{C}+[n_{1,3}-n_{2,3}]_+,
\end{align*}
or (b) $n_{C}>n_{2,3}>n_{2,4}$ in which case, we lower the noise level at
node~3 by $n_{C}-n_{2,3}$ by defining $n_{1,3}'=n_{1,3}+(n_{C}-n_{2,3})$ and
$n_{2,3}'=n_{2,3}+(n_{C}-n_{2,3})$. Now the upperbound becomes
\begin{align*}
R_1+R_2 &\leq \max(n_{1,3},n_{2,3}) + n_{C}-n_{2,3}
 = n_{C}+[n_{1,3}-n_{2,3}]_+.
\end{align*}
Hence, without any conditions on $n_{C}, n_{2,4}, n_{2,3}$, we have
\begin{align*}
R_1+R_2 \leq \max(n_{2,4},n_{2,3},n_{C}) + [n_{1,3}-n_{2,3}]_+.
\end{align*}
By symmetry, we also have
\begin{align*}
R_1+R_2 \leq \max(n_{1,3},n_{1,4},n_{C}) + [n_{2,4}-n_{1,4}]_+.
\end{align*}

\noindent{\em Gaussian model:} When $|h_C|\leq \max(|h_{2,3}|,|h_{2,4}|)$,
\begin{align*}
\frac{1}{T}&\sum_{t=1}^T \left\{ H(Y_3(t)|Y_3^{t-1})
-H(h_{2,3}^\ast(X_2(t))|Y_4^{t-1},h_{2,3}^\ast(X_2^{t-1}),h_{1,2}^\ast(X_1^{t-1}),h_{2,1}^\ast(X_2^{t-1}),M_1,M_2)\right\}\\
&\qquad\qquad\leq \frac{1}{T}\sum_{t=1}^T H(Y_3(t)) - H(N_3(t))\\
&\qquad\qquad\leq \log(1+(|h_{1,3}|+|h_{2,3}|)^2)
.
\\
\frac{1}{T}&\sum_{t=1}^T\left\{ H(Y_4(t)|X_1(t),h_{2,3}^\ast(X_2(t)))
  - H(Y_4(t)|X_1(t),X_2(t)) \right\}\\
&\qquad\qquad\leq \frac{1}{T}\sum_{t=1}^T H(Y_4(t)-h_{1,4}X_1(t)
 -h_{2,4}h_{2,3}^{-1}h_{2,3}^\ast(X_2(t))) - H(N_4(t))\\
&\qquad\qquad= \frac{1}{T}\sum_{t=1}^T H(N_4(t)-h_{2,4}h_{2,3}^{-1}N_3(t))
  - H(N_4(t))\\
&\qquad\qquad\leq \log\left(1+\left|\frac{h_{2,4}}{h_{2,3}}\right|^2\right)
.
\\
\frac{1}{T}&\sum_{t=1}^T\left\{ H(h_{2,1}^\ast(X_2(t))|Y_4(t),
h_{2,3}^\ast(X_2(t)),X_1(t)) - H(h_{2,1}^\ast(X_2(t))|X_2(t))\right\}\\
&\qquad\qquad\leq
\min\bigg(
\frac{1}{T}\sum_{t=1}^T
  H(h_{2,1}^\ast(X_2(t))-h_{2,1}h_{2,3}^{-1}h_{2,3}^\ast(X_2(t)))-
H(N_1(t)),\\
 &\qquad\qquad\qquad\qquad\qquad\qquad
\frac{1}{T}\sum_{t=1}^T
  H(h_{2,1}^\ast(X_2(t))-h_{2,1}h_{2,4}^{-1}(Y_4(t)-h_{1,4}X_1(t))) - H(N_1(t))
\bigg)\\
&\qquad\qquad=
\min\bigg(
 \frac{1}{T}\sum_{t=1}^T
 H(N_1(t)-h_{2,1}h_{2,3}^{-1}N_3(t))-H(N_1(t)),\\
 &\qquad\qquad\qquad\qquad\qquad\qquad
 \frac{1}{T}\sum_{t=1}^T H(N_1(t)-h_{2,1}h_{2,4}^{-1}N_4(t))-H(N_1(t))
 \bigg)\\
&\qquad\qquad\leq
\log\left(1+\min(|h_C|^2/|h_{2,3}|^2,|h_C|^2/|h_{2,4}|^2)\right)\\
&\qquad\qquad\leq 1.
\end{align*}
Also, (when $|h_{2,3}|<1$) we may upperbound the second term without
subtracting $h_{2,4}h_{2,3}^{-1}h_{2,3}^\ast(X_2(t))$
\begin{align*}
\frac{1}{T}\sum_{t=1}^T\left\{ H(Y_4(t)|X_1(t),h_{2,3}^\ast(X_2(t)))
  - H(Y_4(t)|X_1(t),X_2(t)) \right\}
&\leq \log\left(1+\left|{h_{2,4}}\right|^2\right)
.
\end{align*}
Thus, we have
\[ R_1+R_2\leq
\log2(1+(|h_{1,3}|+|h_{2,3}|)^2)
+\log\left(1+\frac{|h_{2,4}|^2}{\max\left(1,|h_{2,3}|^2\right)}\right).\]

If $|h_C| > |h_{2,4}| \geq |h_{2,3}|$, we consider the following
``enhanced'' channel with channel coefficients indicated by primed
quantities defined by
\begin{align*}
\frac{h_{2,4}^\prime}{h_{2,4}}&=\frac{|h_C|}{|h_{2,4}|},\\
\frac{h_{1,4}^\prime}{h_{1,4}}&=\frac{|h_C|}{|h_{2,4}|}.
\end{align*}
It is easy to see that this is equivalent to lowering the noise variance at
destination~4 from unity to $|h_{2,4}|^2/|h_C|^2$. Hence, an upperbound on
the ``enhanced'' channel is also an upperbound for the original. Thus, we
have
\begin{align*}
 R_1+R_2&\leq
\log2(1+(|h_{1,3}|+|h_{2,3}|)^2)
+\log\left(1+\frac{|h'_{2,4}|^2}{\max\left(1,|h_{2,3}|^2\right)}\right)\\
 &=
\log2(1+(|h_{1,3}|+|h_{2,3}|)^2)
+\log\left(1+\frac{|h_C|^2}{\max\left(1,|h_{2,3}|^2\right)}\right)
\end{align*}
Similarly, if $|h_C| > |h_{2,3}| \geq |h_{2,4}|$,  we consider the following
``enhanced'' channel
\begin{align*}
\frac{h_{1,3}^\prime}{h_{1,3}}&=\frac{|h_C|}{|h_{2,3}|},\\
\frac{h_{2,3}^\prime}{h_{2,3}}&=\frac{|h_C|}{|h_{2,3}|},
\end{align*}
which is equivalent to lowering the noise variance at destination~3 from
unity to $|h_{2,3}|^2/|h_C|^2$. Then, for $|h_{2,3}|\geq 1$,
\begin{align*}
 R_1+R_2&\leq
\log2(1+(|h'_{1,3}|+|h'_{2,3}|)^2)
+\log\left(1+\frac{|h_{2,4}|^2}{\max\left(1,|h'_{2,3}|^2\right)}\right)\\
 &\leq
\log2(1+(|h_{1,3}|+|h_{2,3}|)^2)+\log\left(1+\frac{|h_{2,4}|^2}{|h_{2,3}|^2}\right).
\end{align*}
For $|h_{2,3}|<1$, we upperbound the three terms directly
\begin{align*}
\frac{1}{T}&\sum_{t=1}^T \left\{ H(Y_3(t)|Y_3^{t-1})
-H(h_{2,3}^\ast(X_2(t))|Y_4^{t-1},h_{2,3}^\ast(X_2^{t-1}),h_{1,2}^\ast(X_1^{t-1}),h_{2,1}^\ast(X_2^{t-1}),M_1,M_2)\right\}\\
&\qquad\qquad\leq \frac{1}{T}\sum_{t=1}^T H(Y_3(t)) - H(N_3(t))\\
&\qquad\qquad\leq \log(1+(|h_{1,3}|+|h_{2,3}|)^2)
.
\\
\frac{1}{T}&\sum_{t=1}^T\left\{ H(Y_4(t)|X_1(t),h_{2,3}^\ast(X_2(t)))
  - H(Y_4(t)|X_1(t),X_2(t)) \right\}\\
&\qquad\qquad\leq \frac{1}{T}\sum_{t=1}^T H(Y_4(t)-h_{1,4}X_1(t)) - H(N_4(t))\\
&\qquad\qquad= \frac{1}{T}\sum_{t=1}^T H(N_4(t)+h_{2,4}X_2(t))
  - H(N_4(t))\\
&\qquad\qquad\leq \log\left(1+\left|{h_{2,4}}\right|^2\right)\\
&\qquad\qquad\leq \log2
.
\\
\frac{1}{T}&\sum_{t=1}^T\left\{ H(h_{2,1}^\ast(X_2(t))|Y_4(t),
h_{2,3}^\ast(X_2(t)),X_1(t)) - H(h_{2,1}^\ast(X_2(t))|X_2(t))\right\}\\
&\qquad\qquad\leq
 \frac{1}{T}\sum_{t=1}^T
  H(h_{2,1}^\ast(X_2(t)))- H(N_1(t))\\
&\qquad\qquad\leq
\log\left(1+|h_C|^2\right).
\end{align*}

Thus, in general, we have
\begin{align*}
R_1+R_2 &\leq
\log2(1+(|h_{1,3}|+|h_{2,3}|)^2)
+\log\left(1+\frac{\max(|h_{C}|^2,|h_{2,3}|^2,|h_{2,4}|^2)}{\max(1,|h_{2,3}|^2)}\right).
\end{align*}

\noindent{\em Upperbound 4:}

This is a simple cut-set upperbound~\cite{CoverThomas} with nodes 1 and 4
on one side of the cut and nodes 2 and 3 on the other. It is easy to verify
that
\begin{align*}
 R_1 &\leq \max_{p_{X_1}} I(X_1;Y_3,Y_2),\\
 R_2 &\leq \max_{p_{X_2}} I(X_2;Y_4,Y_1).
\end{align*}
Under the linear deterministic model, this translates to an upperbound on
the sum-rate of
\[ R_1 + R_2 \leq \max(n_{1,3},n_C) + \max(n_{2,4},n_C),\]
and for the Gaussian case, we get an upperbound of
\[ R_1 + R_2 \leq \log\left( 1 + |h_{1,3}|^2 + |h_C|^2\right) + \log\left(
1 + |h_{2,4}|^2 + |h_C|^2\right).\]

\noindent{\em Upperbound 5:}

This is also a simple cut-set upperbound. Nodes 1 and 2 are on one side of
the cut and nodes 3 and 4 are on the other. The resulting upperbound on the
sum-rate is
\begin{align*}
 R_1 + R_2 &\leq \max_{p_{X_1,X_2}} I(X_1;X_2; Y_3,Y_4).
\end{align*}
For the linear deterministic case, this gives
\[R_1 + R_2 \leq  \left\{\begin{array}{ll}
\max(n_{1,3}+n_{2,4},n_{1,4}+n_{2,3}),
 &\text{ if }n_{1,3}-n_{2,3}\neq n_{1,4}-n_{2,4},\\
\max(n_{1,3},n_{2,4},n_{1,4},n_{2,3}),
 &\text{ otherwise},
\end{array}\right.
\]
and for the Gaussian case, using the fact the eigenvalues of the input
($[X_1, X_2]$) covariance matrix cannot exceed 2, we may upperbound the
sum-rate by
\begin{align*}
 R_1 + R_2 \leq \log\bigg( 1 &+
2 \left(|h_{1,3}|^2+|h_{2,4}|^2+|h_{1,4}|^2+|h_{2,3}|^2\right)\\ &+
4 \left( |h_{1,3}h_{2,4}|^2 + |h_{1,4}h_{2,3}|^2 -
2|h_{1,3}h_{2,4}h_{1,4}h_{2,3}|\cos\theta\right) \bigg).
\end{align*}

Upperbounds~1, 2, and~3 can be further tightened to obtain a smaller
constant gap in Theorem~\ref{thm:sourcecoopG} by (1) considering the
correlation between the noise processes at the destinations, as well as (2)
modifying the correlation of the Gaussian noise processes in the additional
signals we provide to the destinations. Also, the correlation between the
input signals can be explicitly accounted for instead of assuming the
worst-case correlation at different stages as we do here. Upperbound~5 can
be easily improved by choosing the optimal input covariance matrix.
However, we will not pursue any of these directions in this paper.

\section{Upperbounds for the Gaussian interference channel with feedback}
\label{app:feedback}

Below, we will show that \eqref{eq:u2GaussFeedback} is an upperbound on the
sum-rate.  To see that this is the only biting upperbound in this regime,
let us consider the achievability proof. The achievability proof in
appendix~\ref{app:Gachieve} only depends on the marginals of the noises and
hence holds without change for the feedback case. Hence, the sum-rate
achieved in appendix~\ref{app:Gachieve} is achievable for the feedback
problem as well. Moreover, for the symmetric case, $u'_4$ of \eqref{eq:u'4}
in the achievability proof (appendix~\ref{app:Gachieve}) is strictly
subsumed by $u'_5$ of \eqref{eq:u'5}. Also, since for noiseless feedback,
$n_C$ and $n_\text{min}$ as defined in appendix~\ref{app:Gachieve} are
equal, $u'_1$ of \eqref{eq:u'1} is subsumed by $\min(u'_2,u'_3,u'_4,u'_5)$
as we argued in that appendix (where $u'_2$ through $u'_5$ are defined in
\eqref{eq:u'2}-\eqref{eq:u'5}. For the symmetric channel $u'_2=u'_3$.
Evaluating $u'_5$ and $u'_2$ for the symmetric channel with noiseless
feedback reveals that $u'_5$ is never smaller than $u'_2$ by more than 1
bit. Thus, the achievability proof is appendix~\ref{app:Gachieve} when
applied to the symmetric channel with noiseless feedback achieves a
sum-rate of $u'_2$ within a gap of at most 13 bits. Also, $u'_2$ is  within
a gap of at most 5 bits from \eqref{eq:u2GaussFeedback}. Thus, overall, the
achievability proof is appendix~\ref{app:Gachieve} achieves
\eqref{eq:u2GaussFeedback} with a gap of at most 19 bits.

It only remains to show that \eqref{eq:u2GaussFeedback} is an upperbound to
the sum-rate.The line of argument is similar to the one in the proof of
upperbound~2 for the cooperation case; the main difference is that
the genie does not provide $h_{1,2}^\ast(X_1^T)$ to destination~4. Let
us define $h_{2,3}^\ast(X_2) = h_{2,3}X_2 + N_1$ as before. Starting from
Fano's inequality, we write
\begin{align*}
T(R_1+R_2-o(\epsilon))&\leq I(M_1;Y_3^T)+I(M_2;Y_4^T)\\
&\leq I(M_1;Y_3^T) + I(M_2;Y_4^T,h_{2,3}^\ast(X_2^T),M_1)\\
&\leq I(M_1;Y_3^T) + I(M_2;Y_4^T,h_{2,3}^\ast(X_2^T)|M_1)\\
&\leq \sum_{t=1}^T I(M_1;Y_3(t)|Y_3^{t-1}) +
            I(M_2;Y_4(t),h_{2,3}^\ast(X_2(t))|Y_4^{t-1},h_{2,3}^\ast(X_2^{t-1}),M_1).
\end{align*}
Below, we upperbound these terms separately.
\begin{align*}
I(M_1;Y_3(t)|Y_3^{t-1})&=H(Y_3(t)|Y_3^{t-1}) - H(Y_3(t)|Y_3^{t-1},M_1)\\
&\leq H(Y_3(t)|Y_3^{t-1}) - H(Y_3(t)|Y_3^{t-1},Y_4^{t-1},M_1)\\
&\stackrel{\text{(a)}}{\leq} H(Y_3(t)|Y_3^{t-1}) - H(h_{2,3}^\ast(X_2(t))|Y_4^{t-1},h_{2,3}^\ast(X_2^{t-1}),M_1),
\end{align*}
where (a) follows from the fact that $Y_3(t)=h_{1,3}X_1(t)
+h_{2,3}^\ast(X_2(t))$, and $X_1(t)$ is a deterministic function $f_{1,t}$
of $(M_1)$.
\begin{align*}
I(M_2;Y_4(t),h_{2,3}^\ast(X_2(t))&|Y_4^{t-1},h_{2,3}^\ast(X_2^{t-1}),M_1)\\
&=I(M_2;h_{2,3}^\ast(X_2(t))|Y_4^{t-1},h_{2,3}^\ast(X_2^{t-1}),M_1)
 +I(M_2;Y_4(t)|Y_4^{t-1},h_{2,3}^\ast(X_2^t),M_1).
\end{align*}
We upperbound the above two terms separately now.
\begin{align*}
I(M_2;h_{2,3}^\ast(X_2(t))|&Y_4^{t-1},h_{2,3}^\ast(X_2^{t-1}),M_1)\\
&=H(h_{2,3}^\ast(X_2(t))|Y_4^{t-1},h_{2,3}^\ast(X_2^{t-1}),M_1)
 -H(h_{2,3}^\ast(X_2(t))|Y_4^{t-1},h_{2,3}^\ast(X_2^{t-1}),M_1,M_2)\\
&\stackrel{\text{(a)}}{=}H(h_{2,3}^\ast(X_2(t))|Y_4^{t-1},h_{2,3}^\ast(X_2^{t-1}),M_1)-H(N_3(t)),\\
I(M_2;Y_4(t)|Y_4^{t-1},&h_{2,3}^\ast(X_2^t),M_1)\\
&=H(Y_4(t)|Y_4^{t-1},h_{2,3}^\ast(X_2^t),M_1)
  - H(Y_4(t)|Y_4^{t-1},h_{2,3}^\ast(X_2^t),M_1,M_2)\\
&\stackrel{\text{(b)}}{\leq} H(Y_4(t)|X_1(t),h_{2,3}^\ast(X_2(t))) - H(N_4(t)),
\end{align*}
where (a) and (b) follow from (1) the channel model (memorylessness and
independence of the noise processes at the two destination nodes, and
$Y_4(t) = h_{2,4}X_2(t) + h^\ast_{1,4}(X_1(t)) = h_{2,4}X_2(t) +
h_{1,4}X_1(t) + N_4(t)$), and (2) the fact that
$X_1(t)$ and $X_2(t)$, resp., are deterministic functions of
$(M_1,h_{2,3}^\ast(X_2^{t-1}))$ and $(M_2,h_{1,4}^\ast(X_1^{t-1}))$, resp.
\noindent Combining everything, we have
\begin{align*}
T(R_1+R_2-o(\epsilon)) &\leq
\sum_{t=1}^T \left\{ H(Y_3(t)|Y_3^{t-1}) - H(N_3(t))\right\}
 +\left\{ H(Y_4(t)|X_1(t),h_{2,3}^\ast(X_2(t))) - H(N_4(t)) \right\}.
\end{align*}
The rest of the argument is exactly as in
Appendix~\ref{app:sourcecoopupperbounds}.

\section{A Gaussian example}\label{app:Example}

The upperbounds follow from Appendix~\ref{app:sourcecoopupperbounds}.
We can show that the sum-rate is upperbounded by all of the following for
the symmetric channel with $h_I=\sqrt{h_D}$.
\begin{align*}
U_1&=2\log\left(1+2h_D\right)\left(1+h_C^2\right),\\
u_2=u_3&=\log2\left(1+\left(h_D+\sqrt{h_D}\right)^2\right)
\left(1+\frac{\max(h_D^2,h_C^2)}{h_D}\right),\\
u_5&=\log\left( 1 + 4(h_D^2+h_D) + 4\left(h_D^2-h_D\right)^2\right).
\end{align*}
Note that $u_1$ above is slightly stronger than the one on
Theorem~\ref{thm:sourcecoopG}, but follows directly from the proof in
appendix~\ref{app:sourcecoopupperbounds} when specialized to the symmetric
channel with $h_I=\sqrt{h_D}$. Also, we have left out $u_4$ since this
upperbound is not important for this channel. The above upperbounds imply
that for any $\epsilon>0$, the sum-capacity is upperbounded by
$C+\epsilon$, for sufficiently large $h_D$.

The achievability again depends on different schemes depending on the
regime. For $h_C\leq \sqrt{h_D}$, we apply
Theorem~\ref{thm:sourcecoopgenericschemes}(a). We choose
$Z_1,Z_2,U_1,U_2,V_1,V_2$ as independent,zero-mean Gaussian auxiliary
random variables with the following variances.
\begin{align*}
\sigma_{Z_1}^2=\sigma_{Z_2}^2&=\frac{1}{h_D},\\
\sigma_{U_1}^2=\sigma_{U_2}^2&=\frac{1}{h_C^2},\text{and}\\
\sigma_{V_1}^2=\sigma_{V_2}^2&=1-\frac{1}{h_D2}-\frac{1}{h_C^2}.
\end{align*}
$W$ is independent of all these and has the same distribution as
$(V_1,V_2)$. $X_1$ and $X_2$ are as follows.
\begin{align*}
X_1&=V_1+U_1+Z_1,\\
X_2&=V_2+U_2+Z_2.
\end{align*}
Evaluating the expressions in Theorem~\ref{thm:sourcecoopgenericschemes}(a)
and simplifying using Fourier-Motzkin elimination, it can be shown that the
upperbound is achievable in the regime $h_C\leq \sqrt{h_D}$ within a gap of
6 bits for sufficiently large $h_D^2$.

In the range of $\sqrt{h_D}\leq h_C \leq h_D$, we find that, for
sufficiently large $h_D$, $u_2=$\\ $\log2\left(1+\left(h_D+\sqrt{h_D}\right)^2\right)\left(1+h_D\right)$,
which is independent of $h_C$ in this regime, dominates the other bounds.
Thus, the achievability in the regime $h_C\leq \sqrt{h_D}$ implies
achievability in this regime as well.

For $h_C> h_D$, we apply Theorem~\ref{thm:sourcecoopgenericschemes}(b)
with the following choices for the auxiliary random variables. $S'_1,S'_2,
Z_1,Z_2, S_1,S_2, V_1,V_2$ are independent, zero-mean Gaussian auxiliary
random variables with the following variances.
\begin{align*}
\sigma_{S'_1}^2=\sigma_{S'_2}^2&=\frac{1}{h_D^2},\\
\sigma_{Z_1}^2=\sigma_{Z_2}^2&=\frac{1/2}{h_D},\\
\sigma_{S_1}^2 = \sigma_{S_2}^2
  &= \frac{1}{2\left(1+\frac{1}{h_D}\right)}
     \left(1 - \frac{1}{h_D^2} - \frac{1/2}{h_D}\right),
   \text{and}\\
\sigma_{V_1}^2=\sigma_{V_2}^2
  &=\frac{1}{2}\left(1-\frac{1}{h_D^2} - \frac{1/2}{h_D}\right).
\end{align*}
$W$ is independent of all these and has the same distribution as
$(V_1,V_2)$, and $U_1,U_2$ are constants. $X_1$ and $X_2$ are as follows.
\begin{align*}
X_1&=V_1+S_1-\frac{h_X}{h_D}S_2+Z_1+S'_1,\\
X_2&=V_2+S_2-\frac{h_X}{h_D}S_1+Z_2+S'_2.
\end{align*}
\begin{align*}
Y_3&=h_D\left(V_1+Z_1+S'_1\right) + h_X\left(V_2+Z_2+S'_2\right)
     + \left(h_D - \frac{h_X^2}{h_D}\right)S_1 + N_3,\\
Y_4&=h_D\left(V_2+Z_2+S'_2\right) + h_X\left(V_1+Z_1+S'_1\right)
     + \left(h_D - \frac{h_X^2}{h_D}\right)S_2 + N_4,
\end{align*}
Evaluating the expressions in Theorem~\ref{thm:sourcecoopgenericschemes}(b)
and simplifying using Fourier-Motzkin elimination, it can be shown that the
upperbound is achievable in the regime $h_C>h_D$ within a gap of 5 bits for
sufficiently large $h_D$.

\section{Proof of Theorem~\ref{thm:reversibility}}\label{app:reversibility}

Consider the following theorem from~\cite{DestCoop} which is the
destination cooperation analog of Theorem~\ref{thm:sourcecoopG}.
\begin{thm}\cite{DestCoop}\label{thm:dstcoopG}
The sum-capacity of the Gaussian channel with destination cooperation in
Figure~\ref{fig:destinationcoop} is at most the minimum of the following
five quantities and a sum-rate can be achieved within a gap of at most 43
bits of this minimum.
\begin{align}
v_1&=
\left\{\begin{array}{ll}
             \log\left(1 +\left(|h_{1,4}|+|h_C|+
              \left|\frac{h_{1,3}h_C}{h_{2,3}}\right|\right)^2
              +\left|\frac{h_{1,3}}{h_{2,3}}\right|^2\right), 
                &\text{ if } |h_{2,3}|>\max(1,|h_C|)\\
             \log\left(1 +\left(|h_{1,4}|+|h_C|+
              |h_{1,3}|\right)^2\right), &\text{ otherwise}
        \end{array}\right. \notag\\  
    &\quad +   \left\{\begin{array}{ll}
              \log\left(1 +\left(|h_{2,3}|+|h_C|+
              \left|\frac{h_{2,4}h_C}{h_{1,4}}\right|\right)^2
              +\left|\frac{h_{2,4}}{h_{1,4}}\right|^2\right),
                &\text{ if } |h_{1,4}|>\max(1,|h_C|)\\
             \log\left(1 +\left(|h_{2,3}|+|h_C|+
               |h_{2,4}|\right)^2\right), &\text{ otherwise,}
        \end{array}\right.\label{eq:Gaussv1}\\
v_2&=\log(1+(|h_{1,3}|+|h_{1,4}|+|h_C|)^2) +
\log\left(1+\frac{|h_{2,4}|^2}{\max(1,|h_{1,4}|^2)}\right),\label{eq:Gaussv2}\\
v_3 &=\log(1+(|h_{2,4}|+|h_{2,3}|+|h_C|)^2) +
\log\left(1+\frac{|h_{1,3}|^2}{\max(1,|h_{2,3}|^2)}\right),\label{eq:Gaussv3}\\
v_4 &= \log\left(1+(|h_{1,3}|+|h_C|)^2\right) +
    \log\left(1+(|h_{2,4}|+|h_C|)^2\right),\label{eq:Gaussv4}\\
v_5 &= \log\bigg( 1 + 
2 \left(|h_{1,3}|^2+|h_{2,4}|^2+|h_{2,3}|^2+|h_{1,4}|^2\right)\notag\\
& \qquad\qquad+ 4 \left( |h_{1,3}h_{2,4}|^2 + |h_{2,3}h_{1,4}|^2
 - 2|h_{1,3}h_{2,4}h_{2,3}h_{1,4}|\cos\theta\right) \bigg). \label{eq:Gaussv5}
\end{align}
\end{thm}

Notice that the upperbounds $u_4$ and $u_5$, respectively in
Theorem~\ref{thm:sourcecoopG} and identical to $v_4$ and $v_5$,
respectively. We define the quantities
$n_{k_1,k_2},\;k_1\in\{1,2\},k_2\in\{3,4\}$ and $n_C$ as we did in
appendix~\ref{app:Gachieve}.
\begin{align*}
n_{k_1,k_2}&\defineqq [\log|h_{k_1,k_2}|^2]_+,\;k_1\in\{ 1,2\},\,k_2\in\{3,4\},\text{ and}\\
n_C &\defineqq [\log|h_C|^2]_+.
\end{align*}
And in a manner analogous to our definitions of $u_1',u_2',$ and $u_3'$
(repeated below) in
(\ref{eq:u'1})-(\ref{eq:u'3}) from the upperbounds $u_1,u_2,$ and $u_3$ of
Theorem~\ref{thm:sourcecoopG}, we define below $v_1',v_2',$ and $v_3'$ from
$v_1,v_2,$ and $v_3$. 
\begin{align}
v'_1&=\max(n_{1,3}-n_{2,3}+n_C,n_{1,4},n_C) +
\max(n_{2,4}-n_{1,4}+n_C,n_{2,3},n_C),\label{eq:v1'}\\
v_2'&=
\max(n_{2,4},n_{1,4}) + \left(\max(n_{1,3},n_{1,4},n_C)-n_{1,4}\right),
\label{eq:v2'}\\
v_3'&=
\max(n_{1,3},n_{2,3}) + \left(\max(n_{2,4},n_{2,3},n_C)-n_{2,3}\right),
\label{eq:v3'}
\end{align}
\begin{align*}
u'_1&=\max(n_{1,3}-n_{1,4}+n_C,n_{2,3},n_C) +
\max(n_{2,4}-n_{2,3}+n_C,n_{1,4},n_C),\\
u'_2&=
\max(n_{1,3},n_{2,3}) + \left(\max(n_{2,4},n_{2,3},n_C)-n_{2,3}\right),\\
u'_3&=
\max(n_{2,4},n_{1,4}) + \left(\max(n_{1,3},n_{1,4},n_C)-n_{1,4}\right).
\end{align*}
We can show that the respective primed and un-primed quantities are
within a constant number of bits from each other. In particular,
\begin{align*}
 u_k - 7 \leq u_k' \leq u_k,&\qquad k=1,2,3,\text{ and}\\
 v_k - 7 \leq v_k' \leq v_k,&\qquad k=1,2,3.
\end{align*}
It is possible to verify that
\begin{align*}
\min\left(u'_1,u'_2,u'_3\right) = \min\left(v'_1,v'_2,v'_3\right).
\end{align*}
From these two facts, we may conclude that
\begin{align*}
|\min(u_1,u_2,u_3,u_4,u_5) - \min(v_1,v_2,v_3,v_4,v_5)| \leq 7.
\end{align*}
Hence we may conclude from Theorems~\ref{thm:sourcecoopG}
and~\ref{thm:dstcoopG} that the sum-capacities of the two-user Gaussian 
interference channels in Figures~\ref{fig:sourcecoop}
and~\ref{fig:destinationcoop} with source cooperation and destination
cooperation, respectively, are within a constant gap of at most 50 bits.

\end{document}